\documentclass[12pt]{article}
\usepackage{amsfonts}

\catcode`\@=11

\newcount\hour
\newcount\minute
\newtoks\amorpm \hour=\time\divide\hour by 60\minute
=\time{\multiply\hour by 60 \global\advance\minute by-\hour}
\edef\standardtime{{\ifnum\hour<12 \global\amorpm={am}%
        \else\global\amorpm={pm}\advance\hour by-12 \fi
        \ifnum\hour=0 \hour=12 \fi
        \number\hour:\ifnum\minute<10
        0\fi\number\minute\the\amorpm}}
\edef\militarytime{\number\hour:\ifnum\minute<10
0\fi\number\minute}

\def\draftlabel#1{{\@bsphack\if@filesw {\let\thepage\relax
   \xdef\@gtempa{\write\@auxout{\string
      \newlabel{#1}{{\@currentlabel}{\thepage}}}}}\@gtempa
   \if@nobreak \ifvmode\nobreak\fi\fi\fi\@esphack}
        \gdef\@eqnlabel{#1}}
\def\@eqnlabel{}
\def\@vacuum{}
\def\marginnote#1{}
\def\draftmarginnote#1{\marginpar{\raggedright\scriptsize\tt#1}}
\def\draft{
        \pagestyle{plain}
        \overfullrule=2pt
        \oddsidemargin -.5truein
        \def\@oddhead{\sl \phantom{\today\quad\militarytime} \hfil
        \smash{\Large\sl DRAFT} \hfil \today\quad\militarytime}
        \let\@evenhead\@oddhead
        \let\label=\draftlabel
        \let\marginnote=\draftmarginnote
        \def\ps@empty{\let\@mkboth\@gobbletwo
        \def\@oddfoot{\hfil \smash{\Large\sl DRAFT} \hfil}
        \let\@evenfoot\@oddhead}
        \def\@eqnnum{(\theequation)\rlap{\kern\marginparsep\tt\@eqnlabel}%
        \global\let\@eqnlabel\@vacuum}  }

\newcommand{\rf}[1]{(\ref{#1})}
\renewcommand{\theequation}{\thesection.\arabic{equation}}
\renewcommand{\thefootnote}{\fnsymbol{footnote}}
\newcommand{\newsection}{    
\setcounter{equation}{0}\section}
\def\appendix#1{\addtocounter{section}{1}\setcounter{equation}{0}
\renewcommand{\thesection}{\Alph{section}}
\section*{Appendix \thesection\protect\indent \parbox[t]{11.15cm}{#1}}
\addcontentsline{toc}{section}{Appendix \thesection\ \ \ #1}}

\def\smn{{\scriptscriptstyle (n)}}
\def\smt{{\scriptscriptstyle (2)}}
\def\sm3{{\scriptscriptstyle (3)}}
\def\smf{{\scriptscriptstyle (4)}}

\newcommand{\Po}{\mathbb{P}}

\def\LL{{\cal L}}
\def\PP{{\cal P}}
\def\Tr{{\rm Tr}\,}

\def\be{\begin{equation}}
\def\ee{\end{equation}}
\def\beq{\begin{eqnarray}}
\def\eeq{\end{eqnarray}}

\def\nonequal{={\kern -1em}/ \,\,}
\def\pline{\,p\kern -0.45em /}
\def\bpline{\,{\Po}\kern -0.6em\raise0.3ex \hbox{$/$}\,}
\def\qline{{q \kern-0.5em  / } }
\def\Pline{\,P\kern -0.6em\raise0.3ex \hbox{$/$}\,}

\begin{document}


\begin{flushright}
FIAN/TD/08-04
\\
hep-th/0410239
\end{flushright}
\vspace{.5cm}

\vspace{1cm}

\begin{center}

{\Large\bf Eleven dimensional supergravity in light cone gauge}

\vspace{2cm}
R.R. Metsaev

\vspace{1cm} {\it Department of Theoretical Physics, P.N. Lebedev
Physical Institute, Leninsky prospect 53, 119991, Moscow, Russia}

\vspace{3cm}
{\bf Abstract}
\end{center}

\noindent Light-cone gauge manifestly  supersymmetric formulation
of eleven dimensional supergravity is developed. The formulation
is given entirely in terms of light cone scalar superfield,
allowing us to treat all component fields on an equal footing. All
higher derivative on mass shell manifestly supersymmetric 4-point
functions invariant with respect to linear supersymmetry
transformations and corresponding (in  gravitational bosonic
sector) to  terms constructed from four Riemann tensors and
derivatives are found. Superspace representation for 4-point
scattering amplitudes is also obtained. Superfield representation
of linearized interaction vertex of superparticle and supergravity
fields is presented. All 4-point higher derivative interaction
vertices of ten-dimensional supersymmetric Yang-Mills theory are
also determined.

\newpage
\renewcommand{\thefootnote}{\arabic{footnote}}
\setcounter{footnote}{0}

\newsection{Introduction and Summary}

\subsection{Motivations for light-cone gauge approach}

In view of high degree of symmetry $11d$ supergravity \cite{cjs}
has attracted considerable interest for a long period of time.
Presently  due to conjectured interrelation with superstrings
interest in $11d$ supergravity is renewed and by now $11d$
supergravity is viewed as low energy approximation of
M-theory\cite{Townsend:1995kk,Witten:1995ex}. Because application
of the light cone approach turned out to be fruitful in many
problems of superstrings  one can expect this approach might also
be useful to understand M-theory better. Extensive studies
triggered by a conjecture made in Ref.\cite{Banks:1996vh} support
this expectation. Therefore one can believe that study of various
aspects of $11d$ supergravity in the framework of light cone
approach is a fruitful direction to go. This is what we are doing
in this paper.

So far superfield light cone formalism was explored for $d\leq 10$
supergravity theories (see {\it e.g.}
\cite{GS5,GSB,Bengtsson:1983pg,GS6}). The major goal of this paper
is to develop manifest supersymmetric light cone gauge formulation
of $11d$ supergravity and discuss its various applications. Our
method is conceptually very close to that used in
\cite{GS5}-\cite{bgs} to find ten dimensional supersymmetric
theories and is based essentially on a light cone gauge
description of interaction vertices developed in \cite{met22} (see
also \cite{met20,met13,met16}).

One of motivations of our investigation is to demonstrate
efficiency of light cone formulation in study of various higher
derivative 4-points functions which consist in their gravitational
bosonic sector $\partial^{2n}R^4$ terms, where $R$ stands for
Riemann tensor. We shall show that light cone formulation gives
simple way to derive manifestly supersymmetric expressions for
these vertices. As by-product of this study we find superspace
representation for 4-point scattering amplitude of the generic
$11d$ supergravity \cite{cjs}. An attractive feature of our
approach is that the methods we use are algebraic in nature and
this allows us to extend our result to other supersymmetric
theories in a rather straightforward way. That is to say that we
shall extend our discussion to the case of $10d$, non-Abelian
supersymmetric Yang-Mills (SYM) theory and we shall find
manifestly supersymmetric representation of all 4-point functions,
which consist in their bosonic sector $\partial^n F^4$ terms.

As illustration of our approach we shall begin with study of light
cone gauge cubic interaction vertex of generic $11d$ supergravity.
Our interest in light cone gauge re-formulation of the $11d$
supergravity cubic interaction vertex is motivated by the
following reason. For the case of ten dimensional theories it is
light cone gauge cubic vertices of $10d$ supergravity theories
\cite{GS5} that admit simple and natural extension to superstring
theories \cite{GSB}. Therefore it is reasonable to expect that it
is our light cone gauge cubic interaction vertex of $11d$
supergravity that will have natural extension to M-theory, i.e. we
expect that formulation of M-theory, which is still to be
understood, should be simpler within the framework of light cone
approach. The situation here may be analogous to that in string
theory: a covariant formulation of closed string field theories is
non-polynomial, while light cone formulation restricts the string
action to cubic order in string fields.

Another application of our approach is a derivation of linearized
interaction vertex of superparticle with $11d$ supergravity
fields. As is well known in many cases an evaluation of scattering
amplitudes taken to be in the form of quantum mechanical
correlation function of world line linearized interaction vertices
turns out to be more convenient as compared to evaluation in field
theoretical approach. Therefore it seems highly likely that an
extension of world line approach to the $11d$ supergravity should
be fruitful from the perspective of future applications to
M-theory.

\subsection{3-point and 4-point interaction vertices}

To discuss light cone superspace formulation of $11d$ supergravity
we use a superfield $\Phi(p,\lambda)$ which depends on bosonic
momenta $p^I$, $p^+$, Grassmann momentum $\lambda$, while a
dependence of the superfield  $\Phi(p,\lambda)$ on the light cone
evolution parameter $x^+$ is implicit
\footnote{We decompose momenta $p^I$, $I=1,\ldots,9$ into $p^i$,
$i=1,\ldots,7$ and $p^{R,L}$ where $p^{R,L}\equiv (p^8\pm {\rm
i}p^9)/\sqrt{2}$. Grassmann momentum $\lambda$, whose spinor
indices are implicit, transforms in spin one-half representation
of $so(7)$. For momentum in light-cone direction we use simplified
notation $\beta \equiv p^+$.}.
Light cone gauge action describing dynamics of $11d$ supergravity
fields admits then the following standard  representation:
\be \label{lcact00} S=\int dx^+  d^{10} p\, d^8\lambda\,\,
\Phi(-p,-\lambda){\rm i}\,\beta\partial^-\Phi(p,\lambda) +\int
dx^+ P^-\,, \ee
where $P^-$ is the Hamiltonian and $\partial^-\equiv/\partial
x^+$. In theories of interacting fields the Hamiltonian receives
corrections having higher powers of physical fields and one has
the following expansion
\be\label{hanexp00} P^- = \sum_{n=2}^\infty P^-_\smn\,, \ee
where $P^-_\smn$ stands for $n$ - point contribution (degree $n$
in physical fields) to the Hamiltonian. Dynamics of free fields is
described by the well known free Hamiltonian
$P_{{\scriptscriptstyle (2)}}^-$ to be discussed in Section 2,
while  $n$-point interaction corrections $P^-_\smn$ , $n\geq 3$,
admit the following representation
\be  P_\smn^-  = \int d\Gamma_n \prod_{a=1}^n
\Phi(p_a,\lambda_a)\, p_\smn^-\,. \ee
Density  $p_\smn^-$, sometimes referred to as $n$- point
interaction vertex, depends on light cone momenta $\beta_a$,
transverse momenta $p_a^I$ and Grassmann momenta $\lambda_a$,
where an external line index $a=1,\ldots, n$ labels $n$
interacting fields. Explicit expression for an integration measure
$d\Gamma_n$ is given below in \rf{delfun}-\rf{delfun02}.

We begin with discussion of cubic interaction vertex for generic
$11d$ supergravity theory \cite{cjs}. Expression for the 3-point
interaction vertex we find takes the following form:
\beq\label{exprep00} \frac{3}{\kappa}  p_\sm3^- &=& {\Po }^{L2} -
\frac{{\Po }^L}{2\sqrt{2}\hat{\beta}}\Lambda\bpline\Lambda
+\frac{1}{16\hat{\beta}^2}(\Lambda\bpline\Lambda)^2 - \frac{
|\Po|^2}{9\cdot 16\hat{\beta}^2}(\Lambda\gamma^j\Lambda)^2
\nonumber\\[5pt]
&+&\frac{{\Po }^R}{9\cdot 16\sqrt{2}\hat{\beta}^3}
\Lambda\bpline\Lambda(\Lambda\gamma^j\Lambda)^2 +\frac{{\Po
}^{R2}}{2^7\cdot 63\hat{\beta}^4}
((\Lambda\gamma^i\Lambda)^2)^2\,, \eeq
where we use the notation
\be \label{defpi00} {\Po }^I
=\frac{1}{3}\sum_{a=1}^3\check{\beta}_a p_a^I\,,\qquad \Lambda
=\frac{1}{3}\sum_{a=1}^3\check{\beta}_a \lambda_a\,, \ee
\be \check{\beta}_a\equiv \beta_{a+1}-\beta_{a+2}\,, \qquad
\hat{\beta} \equiv \beta_1\beta_2\beta_3\,,\ee
$\bpline \equiv \gamma^i\Po^i$, $|\Po|^2\equiv \Po^I\Po^I$,
$\gamma^i$ are $so(7)$ Dirac matrices, and we use identification
$\beta_a\equiv \beta_{a+3}$ for 3-point interaction vertices. In
formula \rf{exprep00} $\kappa$ is gravitational constant (see
formula \rf{geact} for normalization). Note that cubic vertices
for generic $11d$ supergravity are the only vertices which can be
constructed in cubic approximation. In Section 4 we demonstrate
that the Poincar\'e supersymmetries forbid supersymmetric
extension of $R^2$ and $R^3$ terms.

We proceed to discussion of 4-point interaction vertices. It
should be emphasized from the very beginning that we do not
consider 4-point interaction vertices of the generic $11d$
supergravity \cite{cjs}. We find 4-point vertices that are
invariant with respect to linear supersymmetry transformations and
do not depend on contributions of exchanges generated by cubic
vertices. These 4-point vertices in their gravitational bosonic
sector involve higher derivative terms that can be constructed
from Riemann tensor and derivatives, i.e. they can be presented
schematically as $\partial^{2n} R^4$. The bosonic bodies of these
supersymmetric vertices appeared in various previous studies and
they are of interest because they are responsible for quantum
corrections to classical action of $11d$ supergravity.

4-point interaction vertices $p_\smf^-$ depend on momenta $p_a^I$
and $\lambda_a$ through the following quantities
\be \label{pablab00} {\Po }_{ab}^I\equiv
p_a^I\beta_b-p_b^I\beta_a\,, \qquad \Lambda_{ab}\equiv
\lambda_a\beta_b - \lambda_b\beta_a\,, \ee
where external line indices $a,b$ take values $a,b=1,2,3,4$.
Solution to 4-point interaction vertices we find admits the
representation
\be\label{p4fin0300}  p_\smf^-  =  - \Bigl( ({\bf J}_{12}{\bf
J}_{34})^2 ut E_s + ({\bf J}_{13}{\bf J}_{24})^2 st E_u + ({\bf
J}_{14}{\bf J}_{23})^2 u s E_t\Bigr) g(s,t,u) \,, \ee
where we use the notation
\be\label{bfJdef00}  {\bf J}_{ab} \equiv \Po_{ab}^L \exp\Bigl(
\frac{\Lambda_{ab}\, \qline_{ab}
\Lambda_{ab}}{4\sqrt{2}\beta_a\beta_b \beta_{ab}}\Bigr)\,, \ee
\beq \label{eu00} && E_u \equiv \exp\Bigl( - \frac{
u\Lambda^L\qline_{_L} \Lambda^L}{2\sqrt{2}\beta_{13}q_{_L}^2 ({\Po
}_{13}^L{\Po }_{24}^L)^2}\Bigr)\,, \eeq
\be q_{ab}^i\equiv \frac{\Po^i_{ab}}{\Po^L_{ab}}\,,\qquad
\beta_{ab} \equiv \beta_a + \beta_b\,,\ee
\be \label{qi00}  q_{_L}^i\equiv q_{13}^i-q_{24}^i\,, \qquad \
\Lambda^L \equiv  \Lambda_{13} {\Po }_{24}^L - \Lambda_{24} {\Po
}_{13}^L\,, \ee
$\qline \equiv \gamma^i q^i$, $q_{_L}^2\equiv q_{_L}^iq^i_{_L}$.
In above-given expressions the quantities $s$, $t$, $u$ are the
standard Mandelstam variables (normalization we use for these
variables may be found in \rf{studef},\rf{stulcdef}), while the
quantities $E_s$ and $E_t$ are obtainable from $E_u$ by
appropriate interchange of external line indices 1,2,3,4:
\be \label{EsEtdef00} E_s \equiv E_u|_{2\leftrightarrow
3}\,,\qquad E_t \equiv E_u|_{3\leftrightarrow 4} \,.\ee
Explicit form of a function $g(s,t,u)$ \rf{p4fin0300}, which
should be symmetric in Mandelstam variables, cannot be fixed by
exploiting restrictions imposed by global symmetries alone. This
function is freedom of our solution. Assuming that $g(s,t,u)$
admits Taylor series expansion we get the following lower order
terms in infinite series expansion of $g(s,t,u)$:
\beq  \label{gexp00} g(s,t,u) & = & g_0 + g_2 (s^2+t^2+u^2) + g_3
stu + g_4 (s^4 +t^4 +u^4)
\nonumber\\[7pt]
& + & g_5 stu(s^2 + t^2 +u^2) + g_{6;1} (s^6 + t^6 +u^6) +
g_{6;2}(stu)^2 +  \ldots\,. \eeq
Sign minus in r.h.s. of formula \rf{p4fin0300} is of no physical
significance and is chosen for later convenience.

Our approach allows us to obtain surprisingly compact superspace
representation for 4-point scattering amplitude of generic $11d$
supergravity theory \cite{cjs} which we exhibit here:

\be\label{t4poi00} {\cal A}_\smf = 2 \kappa^2 \Bigl(\frac{({\bf
J}_{12}{\bf J}_{34})^2}{s}E_s + \frac{({\bf J}_{13}{\bf
J}_{24})^2}{u}E_u + \frac{({\bf J}_{14}{\bf
J}_{23})^2}{t}E_t\Bigr)\,.\ee
%

\subsection{Contents of the rest of the  paper}

The rest of the paper contains derivation of above-mentioned
3-point and 4-point interaction vertices, superfield description
of linearized interaction of free superparticle with $11d$
supergravity and related explanations and technical details. The
paper is organized as follows.

In section \ref{DESsec} we introduce a notation and describe the
free level $so(7)$ covariant formulation of $11d$ supergravity in
terms of unconstrained light cone scalar superfield.

In section \ref{GENVERsec}  we discuss arbitrary $n$-point
interaction vertices and find constraints for these vertices
imposed by symmetries of the  Poincar\'e superalgebra.

In section \ref{CUBVERsec} we study a cubic interaction vertex of
$11d$ supergravity and find manifest supersymmetric light cone
representation for this vertex. To do that we use method of
\cite{met22}, which allows us to find simple and compact
representation for the vertex in question. Because this vertex
describes generic $11d$ supergravity it involves terms having the
second power of derivatives.  The formalism we use is algebraic in
nature and this  allows us to study on an equal footing the
vertices involving arbitrary powers of derivatives and describing
in their gravitational bosonic sector $R^2 $ and $R^3$ terms. We
demonstrate explicitly that the Poincar\'e supersymmetries forbid
supersymmetric extension of $R^2$ and $R^3$ terms.

In section \ref{FOUVERsec} we study 4-point vertices that are
invariant with respect to linear Poincar\'e supersymmetry
transformations and involve arbitrary powers of derivatives. These
4-point supersymmetric vertices consist in their gravitational
bosonic sectors $\partial^{2n} R^4$ terms. We find all constraints
imposed by Poincar\'e supersymmetries on such vertices and find
all possible solutions to these constraints. We present explicit
and simple form for these  4-point vertices. Also, as a by-product
of our investigation we present superspace representation for
4-point scattering amplitude of the generic $11d$ supergravity.

In section \ref{PARsec} we develop world line representation for
interaction vertex of $11d$ supergravity. Namely, we obtain
superfield description  of linearized interaction of free
superparticle with $11d$  supergravity.

In section \ref{SYMsec} we extend out formalism to discuss 4-point
vertices for $10d$ nonabelian SYM theory. Because our formalism is
algebraic in nature it allows us,  starting with above mentioned
4-point vertices of $11d$ supergravity, to write down in a
straightforward way  all higher derivatives 4-point interaction
vertices for SYM theory. In their bosonic sector these
supersymmetric vertices correspond to $\partial^n F^4$ terms.

Section \ref{CONsec} summarizes our conclusions and suggests
directions for future research. Appendices contain some
mathematical details and useful formulae.

\newsection{Free $11d$ supergravity in $so(7)$
light cone basis}\label{DESsec}

Method suggested in Ref.\cite{dir} reduces the problem of finding
a new (light cone gauge) dynamical system  to the problem of
finding a new solution of commutation relations of an defining
symmetry algebra\footnote{This method is Hamiltonian version of
the Noether method of finding new dynamical system. Interesting
recent discussion of the Noether method may be found in
\cite{Hurth:1998nq}.}.  Because in our case the defining
symmetries are generated by $11d$ Poincar\'e superalgebra we begin
our investigation with description of the $so(7)$ form of this
superalgebra, which is most convenient for our purposes. The
conventional light cone formalism in eleven dimension based on
$so(9)$ symmetries requires complicated superfield constraints
which we prefer to avoid. Fortunately it turns out that reducing
manifest symmetries to the $so(7)$ symmetries allows one to
develop superfield light cone formulation of $11d$ supergravity in
terms of unconstrained scalar superfield. Another reason why do we
prefer to use the $so(7)$ light cone formulation is that it is the
$so(7)$ symmetries that are manifest symmetries of the general
method of constructing cubic interaction vertices developed in
\cite{met22} (see section \ref{CUBVERsec}). In this section we
focus on free fields.

Poincar\'e superalgebra of $11d$ Minkowski spacetime consists of
translation generators $P^\mu$, rotation generators $J^{\mu\nu}$,
which span $so(10,1)$ Lorentz algebra, and 32 Majorana
supercharges $Q$. The Lorentz covariant form of (anti)commutation
relations is
\beq \label{pj1} &{}
[P^\mu,\,J^{\nu\rho}]=\eta^{\mu\nu}P^\rho-\eta^{\mu\rho}P^\nu\,,
\qquad {} [J^{\mu\sigma},\,J^{\nu\rho}]
=\eta^{\sigma\nu}J^{\mu\rho}+3\hbox{ terms}, &
\\
\label{lorcovcomrel} & [J^{\mu\nu},\,
Q]=-\frac{1}{2}\gamma_{32}^{\mu\nu}Q\,, \qquad \{Q,\, Q\} = -
\gamma_{32}^\mu C_{32}^{-1} P_\mu\,, & \eeq
where $\gamma_{32}^\mu$ are $so(10,1)$ Dirac matrices and we use
mostly positive flat metric tensor $\eta^{\mu\nu}$. The generators
$P^\mu$ are chosen to be hermitian, while the $J^{\mu\nu}$ to be
antihermitian. The supercharges $Q$ satisfy Majorana condition
$Q^\dagger \gamma_{32}^0=Q^t C_{32}$. To develop light cone
formulation we introduce instead of the Lorentz basis coordinates
$x^\mu$ the light cone basis coordinates $x^\pm$, $x^R$, $x^L$,
$x^i$ defined by\footnote{ $\mu,\nu = 0,1,\ldots 10$ are
$so(10,1)$ vector indices, $\alpha=1,\ldots, 8$ is $so(7)$ spinor
index, $I,J,K=1,\ldots 9$ are $so(9)$ vector indices,
$i,j,k=1,\ldots,7$ are $so(7)$ vector indices.}
\be x^\pm \equiv \frac{1}{\sqrt{2}}(x^{10}  \pm x^0), \qquad x^R
\equiv \frac{1}{\sqrt{2}}(x^8+{\rm i}x^9), \qquad x^L\equiv
\frac{1}{\sqrt{2}}(x^8-{\rm i}x^9) \ee and treat $x^{+}$ as an
evolution parameter. In this notation Lorentz basis $11d$ vector
$X^\mu$ is decomposed as $(X^+,X^-,X^I)$, where
$X^I=(X^R,X^L,X^i)$. A scalar product of two $11d$ vectors is
decomposed then as
\be \eta_{\mu\nu}X^\mu Y^\nu = X^+Y^- + X^-Y^+ +X^IY^I\,, \qquad
X^IY^I=X^iY^i+X^RY^L+X^LY^R\,, \ee where the covariant and
contravariant components of vectors are related as $X^+=X_-$,
$X^R=X_L=(X^L)^*$. In the light cone formalism Poincar\'e
superalgebra splits into generators
\be \label{kingen} P^+,\quad P^I,\quad J^{+I},\quad Q^{+R},\quad
Q^{+L},\quad J^{+-},\quad J^{IJ}, \ee which we refer to as
kinematical generators and
\be \label{dyngen} P^-,\quad J^{-I},\quad Q^{-R},\quad Q^{-L}\,,
\ee which we refer to as dynamical generators. For $x^+=0$ the
kinematical generators in the superfield realization are quadratic
in the physical fields\footnote{Namely, for $x^+=\!\!\!\!\!\!/\, 0
$ they have a structure $G= G_1 + x^+ G_2$, where $G_1$ is
quadratic in fields, while $G_2$ contains higher order terms in
fields.}, while the dynamical generators receive higher-order
interaction-dependent corrections.

The $so(7)$ form of Poincar\'e algebra commutators can be obtained
from \rf{pj1} by using the light cone metric having the following
non vanishing elements $\eta^{+-}=\eta^{-+}=1$, $\eta^{RL}=1$,
$\eta^{ij}=\delta^{ij}$. Now we describe the so(7) form of the
remaining (anti)commutators given in \rf{lorcovcomrel}. The
supercharges with superscript $+$ ($-$) have positive (negative)
$J^{+-}$ charge
\be \label{jq1} [J^{+-},Q^{+R,L}] =\frac{1}{2}Q^{+R,L}\,, \quad
[J^{+-},Q^{-R,L}] =-\frac{1}{2}Q^{-R,L} \ee and the superscripts
$R$ and $L$ are used to indicate $J^{RL}$ charge:
\be \label{JRLQRL}[J^{RL},Q^{\pm R}]=\frac{1}{2}Q^{\pm R}\,,
\qquad [J^{RL},Q^{\pm L}]=-\frac{1}{2}Q^{\pm L}\,. \ee
Transformation properties of supercharges with respect to $so(7)$
algebra are given by
\be [J^{ij},Q^{\pm R,L}]=-\frac{1}{2}\gamma^{ij}Q^{\pm R,L}\,. \ee
Remaining commutation relations between supercharges and even part
of superalgebra take the following form
\be\label{210} [J^{Li},Q^{\pm
R}]=-\frac{1}{\sqrt{2}}\gamma^iQ^{\pm L}\,, \qquad [J^{Ri},Q^{\pm
L}]=\frac{1}{\sqrt{2}}\gamma^iQ^{\pm R}\,, \ee

\be
[J^{\pm R},\, Q^{\mp L}]=\pm Q^{\pm R}\,,
\qquad
[J^{\pm L},\, Q^{\mp R}]=\pm Q^{\pm L}\,,
\ee

\be\label{212} [J^{\pm i},Q^{\mp
R}]=\mp\frac{1}{\sqrt{2}}\gamma^iQ^{\pm R}\,, \qquad [J^{\pm
i},Q^{\mp L}]=\pm\frac{1}{\sqrt{2}}\gamma^iQ^{\pm L}\,. \ee
In Eqs.\rf{210},\rf{212} and below $\gamma^i$ stands for $so(7)$
Dirac matrices. Anticommutation relations between supercharges are
\be \label{qq1} \{Q^{\pm R},\, Q^{\pm L}\}=\pm P^\pm, \qquad
\{Q^{+R},\, Q^{-R}\}= P^R, \qquad \{Q^{+L},\, Q^{-L}\}= P^L\,, \ee

\be \label{qq2} \{Q^{\pm L},\, Q^{\mp R}\} =
\frac{1}{\sqrt{2}}\gamma^i P^i\,. \ee
Hermitian conjugation rules in the $so(7)$ basis take the form
$$ P^{\pm \dagger}=P^\pm, \quad P^{i\dagger} = P^i, \quad
P^{R\dagger }=P^L, \quad Q^{\pm R\dagger}=Q^{\pm L}\,, $$ \be
J^{ij\dagger}=-J^{ij}, \quad J^{\pm R\dagger}=-J^{\pm L}, \quad
J^{RL\dagger}=J^{RL}\,. \ee Next step is to find  a realization of
Poincar\'e superalgebra on the space of $11d$ supergravity fields.
To do that we use light cone superspace formalism. First, we
introduce light cone superspace that is based on position
coordinates $x^\mu$ and Grassmann position coordinates
$\theta^\alpha$. Second,  on this light cone superspace we
introduce a scalar superfield $\Phi(x^\mu,\theta)$. In the
remainder of this paper we find it convenient to Fourier
transform\footnote{ Normalization of the Fourier transformation we
use is given in formula \rf{r4ter0}.} to momentum space for all
coordinates except for the time  $x^+$. This implies using $p^+$,
$p^R$, $p^L$, $p^i$, $\lambda^\alpha$, instead of $x^-$, $x^L$,
$x^R$, $x^i$, $\theta^\alpha$ respectively. Thus we consider the
scalar superfield $\Phi(x^+, p^+,p^R,p^L,p^i,\lambda)$ with the
following expansion in powers of the Grassmann momenta $\lambda$
\beq \Phi(p,\lambda)&=&\beta^2 A + \beta\lambda^\alpha\psi^\alpha
+\beta\lambda^{\alpha_1}\lambda^{\alpha_2}A^{\alpha_1\alpha_2}
\nonumber\\
&&+\lambda^{\alpha_1}\lambda^{\alpha_2}\lambda^{\alpha_3}
\psi^{\alpha_1\alpha_2\alpha_3}
+\lambda^{\alpha_1}\ldots\lambda^{\alpha_4} A^{\alpha_1\ldots \alpha_4}
+\frac{1}{\beta}
(\epsilon\lambda^5)^{\alpha_1\alpha_2\alpha_3}
\psi^{\alpha_1\alpha_2\alpha_3*}
\nonumber\\
\label{supfield} &&-\frac{1}{\beta}
(\epsilon\lambda^6)^{\alpha_1\alpha_2} A^{\alpha_1\alpha_2*}
-\frac{1}{\beta^2}(\epsilon\lambda^7)^\alpha\psi^{\alpha*}
+\frac{1}{\beta^2}(\epsilon\lambda^8) A^*\,, \eeq where we use the
notation\footnote{ In what follows a momentum $p$ as argument of
the superfield $\Phi$ and $\delta$- functions designates the set
$\{p^I\,,\beta\}$. Also we do not show explicitly the dependence
of the superfield on evolution parameter $x^+$. Expansion like
\rf{supfield} was introduced for the first time in \cite{GS5,GSB}
to study the light cone formulation of $IIB$ supergravity.}
\be \beta\equiv p^+\,, \qquad
(\epsilon\lambda^{8-n})^{\alpha_1\ldots
\alpha_{n}}\equiv\frac{1}{(8-n)!} \epsilon^{\alpha_1\ldots
\alpha_{n}\alpha_{n+1}\ldots \alpha_8}
\lambda^{\alpha_{n+1}}\ldots\lambda^{\alpha_8} \ee
and $\epsilon^{\alpha_1\ldots \alpha_8}$ is the Levi-Civita
symbol. In \rf{supfield} the fields and their Hermitian conjugated
are related as $(A^*(p))^*=A(-p)$. The superfield $\Phi$ satisfies
the reality constraint\footnote{Integration measure w.r.t.
Grassmann variables is normalized to be $\int d^8\lambda
(\epsilon\lambda^8)=1$. This implies that Grassmann
$\delta$-function is given by $\delta(\lambda)=(\epsilon
\lambda^8)$.}
\be \Phi(-p,\lambda) =\beta^4\int d^8\lambda^\dagger
e^{\lambda\lambda^\dagger/\beta} (\Phi(p,\lambda))^\dagger\,, \ee
where for odd variables $F$, (i.e. Grassmann variables and
fermionic fields) we use the convention $(F_1 F_2)^\dagger
=F_2^\dagger F_1^\dagger$. In \rf{supfield} the component fields
carrying even number of spinor indices are bosonic fields
\beq && A^{\alpha_1\ldots
\alpha_4}(70)=\{h^{ij}(27^0),h^{RL}(1^0),
C^{ijk}(35^0),C^{RLi}(7^0)\}\,,
\\[5pt]
&& A^{\alpha_1\alpha_2}(28)=\{h^{Li}(7^{-1}),
C^{Lij}(21^{-1})\}\,, \eeq while the fields with odd number of
spinor indices are responsible for gravitino field. Explicitly
these fields are related as follows\footnote{$[\alpha_1 \ldots
\alpha_n]$ stands for antisymmetrization in $\alpha_1, \ldots
,\alpha_n$ involving $n!$ terms with overall normalization factor
equal to $\frac{1}{n!}$.  Graviton field $h^{IJ}$ being $so(9)$
tensor field decomposes into $so(7)$ fields as: $h^{ij}$,
$h^{RL}$, $h^{Ri}$, $h^{Li}$. We assume $h^{II}=0$, $h^{ii}=0$.
Antisymmetric $so(9)$ tensor field $C^{IJK}$ decomposes into
$so(7)$ fields in obvious way: $C^{ijk}$, $C^{Rij}$, $C^{Lij}$,
$C^{RLi}$. Interrelation of gravitino field components in
\rf{inter3},\rf{inter4} with those of the so(9) basis is discussed
in Appendix A.}
\beq && A_{\alpha_1\ldots \alpha_4}
=\frac{1}{2^4\sqrt{2}}\gamma^i_{[\alpha_1\alpha_2}
\gamma^j_{\alpha_3\alpha_4]}h^{ij} -\frac{1}{7\cdot 2^3\sqrt{2}}
\gamma_{[\alpha_1\alpha_2}^i\gamma_{\alpha_4\alpha_4]}^ih^{RL}
\nonumber\\[4pt]
\label{inter1} &&\hspace{1.5cm} + \frac{1}{3\cdot
2^4\sqrt{2}}\gamma^i_{[\alpha_1\alpha_2}
\gamma^{jk}_{\alpha_3\alpha_4]} C^{ijk} +\frac{1}{3\cdot
2^4\sqrt{2}}\gamma^i_{[\alpha_1\alpha_2}
\gamma^{ij}_{\alpha_3\alpha_4]} C^{RLj}\,,
\\[6pt]
&& A_{\alpha_1\alpha_2}
=-\frac{1}{4}\gamma^i_{\alpha_1\alpha_2}h^{Li}
-\frac{1}{8}\gamma_{\alpha_1\alpha_2}^{ij}C^{Lij}\,,
\\[6pt]
&& A \ \ \ = \ \ \frac{1}{\sqrt{2}}h^{LL}\,,
\\[6pt]
\label{inter3} &&
\psi_{\alpha_1\alpha_2\alpha_3}=\frac{1}{2\sqrt{2}}\gamma_{[\alpha_1\alpha_2}^i
\psi_{\alpha_3] i}^{\oplus L}\,,
\\[6pt]
\label{inter4}&& \psi_{\alpha} \ \ \ \ = \ \ \
\psi_{R\alpha}^{\oplus L}\,. \eeq Now a representation of the
kinematical generators in terms of differential operators acting
on the superfield $\Phi$ is given by\footnote{Throughout this
paper without loss of generality we analyze generators of
Poincar\'e superalgebra and their commutators for $x^+=0$.}
\be \label{pp} P^+=\beta\,, \qquad P^I=p^I\,, \qquad
Q^{+R}=\beta\theta\,, \qquad Q^{+L}=\lambda\,, \ee

\be J^{+I}=\partial_{p^I}\beta\,, \ee
\beq && \label{jpm}
J^{+-}=\partial_\beta\beta-\frac{1}{2}\theta\lambda+2\,,
\\
\label{jij} && J^{ij}=p^i\partial_{p^j} - p^j\partial_{p^i}
+\frac{1}{2}\theta\gamma^{ij}\lambda\,,
\\
\label{jrl} && J^{RL}=p^R\partial_{p^R} - p^L\partial_{p^L}
+\frac{1}{2}\theta\lambda-2\,,
\\
\label{jri} && J^{Ri}=p^R\partial_{p^i} - p^i\partial_{p^L}
-\frac{1}{2\sqrt{2}}\beta\theta\gamma^i\theta\,,
\\
\label{jli} && J^{Li}=p^L\partial_{p^i} - p^i\partial_{p^R}
+\frac{1}{2\sqrt{2}\beta}\lambda\gamma^i\lambda\,. \eeq Here and
below we use the notation
$$
\partial_\beta\equiv \partial/\partial \beta\,,
\quad
\partial_{p^i}\equiv \partial/\partial p^i\,,
\quad
\partial_{p^R}\equiv \partial/\partial p^R \,,
\quad
\partial_{p^L}\equiv \partial/\partial p^L\,,
\qquad \pline \equiv \gamma^i p^i\,, $$
\be\gamma^{ij}\equiv \frac{1}{2}\gamma^i\gamma^j -
(i\leftrightarrow j)\,. \ee
Representation of the dynamical generators in terms of
differential operators acting on the superfield $\Phi$ is given by
\be\label{dyn1} P^- = p^-\,, \qquad p^- \equiv
-\frac{p^Ip^I}{2\beta}\,, \ee

\beq  \label{jmr} && J^{-R}=\partial_{p^L} p^--\partial_\beta p^R
-\frac{1}{2\sqrt{2}}\theta\pline \theta
+\frac{1}{\beta}p^R\theta\lambda -\frac{4}{\beta}p^R\,,
\\
\label{jml} && J^{-L}=\partial_{p^R} p^--\partial_\beta p^L
+\frac{1}{2\sqrt{2}\beta^2}\lambda\pline\lambda\,,
\\
\label{jmi} && J^{-i}=\partial_{p^i} p^- - \partial_\beta p^i
+\frac{1}{2\beta}\theta\gamma^i\pline \lambda
-\frac{1}{2\sqrt{2}\beta^2}p^R\lambda\gamma^i\lambda
+\frac{1}{2\sqrt{2}}p^L\theta\gamma^i\theta -\frac{2}{\beta}p^i\,,
\ \ \ \ \eeq

\beq   \label{qmr} &&
Q^{-R}=\frac{1}{\sqrt{2}}\theta\pline+\frac{1}{\beta}p^R\lambda\,,
\\
\label{qml} && Q^{-L} = p^L \theta + \frac{1}{\sqrt{2}\beta}\pline
\lambda \,. \eeq
The Grassmann coordinates $\theta$ and momenta $\lambda$ satisfy
the following anticommutation and hermitian conjugations rules
\be \{\lambda^{\alpha_1}, \theta^{\alpha_2}\} =
\delta^{\alpha_1\alpha_2}\,, \qquad \lambda^\dagger = p^+\theta\,,
\qquad \theta^\dagger = \frac{1}{p^+}\lambda\,. \ee The
above-given  expressions provide realization of Poincar\'e
superalgebra in terms of  differential operators acting on the
physical superfield $\Phi$. Now let us write down a field
theoretical realization of this algebra in terms of the physical
superfield $\Phi$.  As we mentioned above the kinematical
generators $\hat{G}^{kin}$ are realized quadratically in $\Phi$,
while the dynamical generators $\hat{G}^{dyn}$ are realized
non-linearly. At a quadratical level both $\hat{G}^{kin}$ and
$\hat{G}^{dyn}$ admit the following representation
\be \label{fierep} \hat{G}=\int \beta d^{10}p\, d^8\lambda
\Phi(-p,-\lambda) G \Phi(p,\lambda)\,, \qquad d^{10}p \equiv
d\beta d^9p\,,\ee where $G$ are the differential operators given
above in \rf{pp}-\rf{qml}. The field $\Phi$ satisfies the
Poisson-Dirac commutation relation
\be {}[\Phi(p,\lambda), \,\Phi(p^\prime,
\lambda^\prime)]\Bigl|_{equal\, x^+} =
\frac{\delta^{10}(p+p^\prime)}{2\beta}
\delta^8(\lambda+\lambda^\prime)\,. \ee With these definitions one
has the standard commutation relation
\be \label{phig} [\Phi,\hat{G}\,]=G\Phi\,. \ee Note that our
normalization of the component fields in expansion of the
superfield $\Phi$ (see \rf{supfield} and \rf{inter1}-\rf{inter4})
is chosen so that contributions of component fields to the
generators, say for $P^+$, are weighted as follows
\be\label{norm} P^+=\int  d^{10}p\,\,\beta^2
(\frac{1}{2}h^{IJ}(-p)h^{IJ}(p)+\frac{1}{3!}C^{IJK}(-p)C^{IJK}(p)
+\psi_I^{\oplus}(-p)C_{16}\psi_I^\oplus(p))\,, \ee where we used a
notation of the $so(9)$ basis. Light-cone gauge action takes then
the following standard form
\be \label{lcact} S=\int dx^+  d^{10} p\, d^8\lambda\,\,
\Phi(-p,-\lambda){\rm i}\, \beta \partial^-\Phi(p,\lambda) +\int
dx^+ P^-\,. \ee
This representation for the light cone action is valid both for
free and interacting theory. Hamiltonian of free theory can be
obtained from Eqs.\rf{dyn1},\rf{fierep}.

\newsection{General structure of $n$-point interaction vertices }\label{GENVERsec}

We begin with discussion of the general structure of Poincar\'e
superalgebra dynamical generators \rf{dyngen}. In theories of
interacting fields the dynamical generators receive corrections
having higher powers of physical fields and one has the following
expansion for them
\be\label{GDYN01} G^{dyn}=\sum_{n=2}^\infty G^{dyn}_\smn\,, \ee
where $G^{dyn}_\smn$ stands for $n$ - point contribution (degree
$n$ in physical fields) to the dynamical generators. The
generators $G^{dyn}$ of classical SYM theories do not receive
corrections higher than fourth order in fields
\cite{BLN,MAN,britol}, while the generators $G_\smn^{dyn}$ for
supergravity theories are nontrivial for all $n\geq 2$
\cite{gors,hor,ara}\footnote{ Generators of closed string field
theories, which involve graviton field, terminates at cubic
correction $G_3^{dyn}$ \cite{GSB,GS6}. On the other hand it is
natural to expect that generators of general covariant theory
should involve all powers of graviton field $h_{\mu\nu}$. The fact
that closed string field theories do not involve higher than third
order vertices in $h_{\mu\nu}$ implies that the general covariance
in closed string field theories is realized in a highly nontrivial
way.  In string theory the general covariance manifests itself
upon integration out of massive string modes and going to the low
energy expansion. See \cite{tsecov} for interesting discussion of
this theme.}.

The `free' generators $G_{(2)}^{dyn}$ \rf{GDYN01}, which are
quadratical in fields, were discussed in the preceding section
(see \rf{fierep}). In this section we discuss general structure of
`interacting' dynamical generators $G_{(n)}^{dyn}$, $n\geq 3$.
Namely, we describe those properties of the dynamical generators
$G_\smn^{dyn}$, $n\geq 3$ that can be obtained from commutation
relations between $G^{kin}$ and $G^{dyn}$. In other words we find
restrictions imposed by kinematical symmetries on the dynamical
`interacting' generators. We proceed in the following way.

({\bf i}) {}First of all we consider restrictions imposed by
kinematical symmetries on the following dynamical generators
\be\label{dyngen10} P^-\,,\quad Q^{-R}\,, \quad Q^{-L}\,. \ee As
seen from (anti)commutators \rf{pj1},\rf{jq1}-\rf{qq2} all
kinematical generators \rf{kingen} with exception of $J^{+-}$,
$J^{IJ}$ have the following commutation relations  with dynamical
generators \rf{dyngen10}: $[G^{dyn},G^{kin}]=G^{kin}$. Because
$G^{kin}$ are quadratic in fields we get from this the
(anti)commutation relations
\be \label{gdynngkinr} [G_\smn^{dyn},G^{kin}]=0\,, \qquad n\geq 3
\,. \ee Exploiting \rf{gdynngkinr} for $G^{kin}=(P^I,P^+,Q^{+L})$
we get the following representation for the dynamical generators
\rf{dyngen10}:
\beq \label{pm1} && P_\smn^-  = \int d\Gamma_n\Phi_\smn
p_\smn^-\,,
\\[6pt]
\label{qrl1} && Q_\smn^{-R,L}  = \int d\Gamma_n\Phi_\smn
q_\smn^{-R,L}\,, \eeq where we use the notation
\beq \label{Phin} && \Phi_\smn\equiv \prod_{a=1}^n
\Phi(p_a,\lambda_a)\,,
\\[8pt]
&& \label{delfun} d\Gamma_n \equiv d\Gamma_n(p)
d\Gamma_n(\lambda)\,,
\\
&& \label{delfun01} d\Gamma_n(p) \equiv (2\pi)^{d-1}
\delta^{d-1}(\sum_{a=1}^np_a) \prod_{a=1}^n \frac{d^{d-1}
p_a}{(2\pi)^{(d-1)/2}}\,, \qquad d=11\,,
\\
&& \label{delfun02} d\Gamma_n(\lambda) \equiv
\delta^8(\sum_{a=1}^n\lambda_a) \prod_{a=1}^n d^8\lambda_a\,. \eeq
Densities  $p_\smn^-$, $q_\smn^{-R,L}$ in \rf{pm1},\rf{qrl1}
depend on light cone momenta $\beta_a$, transverse momenta $p_a^I$
and Grassmann momenta $\lambda_a$. The $\delta$- functions in
\rf{delfun01},\rf{delfun02} respect conservation laws for these
momenta. Here and below the indices $a,b=1,\ldots n$ label $n$
interacting fields.

\medskip
({\bf ii}) Making use of \rf{gdynngkinr} for
$G^{kin}=(J^{+I},Q^{+R})$ we find that the densities $p_\smn^-$,
$q_\smn^{-R,L}$ depend on momenta $p_a^I$ and $\lambda_a$ through
the following quantities
\be \label{pablab} {\Po }_{ab}^I\equiv
p_a^I\beta_b-p_b^I\beta_a\,, \qquad \Lambda_{ab}\equiv
\lambda_a\beta_b - \lambda_b\beta_a\,, \ee
{\it i.e.} the densities $p_\smn^-$, $q_\smn^{-R,L}$ turn out to
be functions of ${\Po }_{ab}^I$, $\Lambda_{ab}$\footnote{Note that
due to momentum conservation laws not all ${\Po }_{ab}^I$, (and
$\Lambda_{ab}$) are independent. It easy to check that $n$-point
vertex involves $n-2$ independent `momenta' $\Po_{ab}^I$ (and
$\Lambda_{ab}$).} instead of $p_a^I$, $\lambda_a$:
\beq \label{pmpm} && p_\smn^-(p_a,\lambda_a,\beta_a)
=p_\smn^-({\Po }_{ab},\Lambda_{ab},\beta_a)\,,
\\[5pt]
&& \label{qmqm} q^{-R,L}_\smn (p_a,\lambda_a,\beta_a)
=q_\smn^{-R,L}({\Po }_{ab},\Lambda_{ab},\beta_a)\,. \eeq

\medskip
({\bf iii}) Commutators between $G^{dyn}$ and remaining
kinematical generators $G^{kin}=(J^{+-}, J^{IJ})$ have the form
$[G^{kin},G^{dyn}]=G^{dyn}$. Because $G^{kin}$ are quadratic in
physical fields, {\it i.e.} $G_\smn^{kin}=0$ for $n>2$, we get the
important commutation relations between above mentioned $G^{kin}$
and all $G^{dyn}$ to be written schematically as
\be \label{gdynngkin} [G_\smn^{dyn},G^{kin}]=G_\smn^{dyn}\,,
\qquad n\geq 2\,. \ee
Before to proceed we introduce $j^{+-}$- and $j^{RL}$- charges by
relations\footnote{
$j^{+-}$- and $j^{RL}$- charges can be easily obtained from
commutators \rf{pj1},\rf{jq1} and \rf{JRLQRL}.}
\be [J^{+-},G]=j^{+-}G\,, \qquad [J^{RL},G]=j^{RL}G\,. \ee
It is straightforward to check that commutators \rf{gdynngkin}
taken for $G^{kin}=J^{+-},J^{RL}$ lead to the following respective
equations for densities $g_\smn =p_\smn^-$, $q_\smn^{-R,L}$:
\beq \label{jmpgn2} && \sum_{a=1}^n (\beta_a\partial_{\beta_a}
+\frac{1}{2}\lambda_a\partial_{\lambda_a})g_\smn =(2n +
j^{+-}-3)g_\smn\,,
\\
\label{jrlgn3} && \sum_{a=1}^n
(p_a^L\partial_{p_a^L}-p_a^R\partial_{p_a^R}
+\frac{1}{2}\lambda_a\partial_{\lambda_a})g_\smn =(2n -
j^{RL}-4)g_\smn\,.\eeq
Remaining equations we will need below can be obtained from
commutators \rf{gdynngkin} taken for $G^{dyn}= P^-$ and
$G^{kin}=J^{Ri},J^{Li}$. For these generators the commutators
\rf{gdynngkin} take a form $[P^-,G^{kin}]=0$ and the latter
commutators lead to the following equations for density
$p_\smn^-$:
\beq && \sum_{a=1}^n(p_a^i\partial_{p_a^L} -p_a^R\partial_{p_a^i}
-\frac{1}{2\sqrt{2}}\beta_a\theta_a\gamma^i\theta_a)p_\smn^- =0\,,
\\
\label{jrign3} && \sum_{a=1}^n(p_a^i\partial_{p_a^R}
-p_a^L\partial_{p_a^i}
+\frac{1}{2\sqrt{2}\beta_a}\lambda_a\gamma^i\lambda_a)p_\smn^-
=0\,. \eeq

\medskip
({\bf iv}) To complete a description of the dynamical generators
we should consider the dynamical generator $J^{-I}$. Making use of
commutation relations of $J^{-I}$ with the kinematical generators
we get the following representation for $J^{-I}$:
\beq \label{jmrd} J^{-R}_{\smn} &=&\int d\Gamma_n
\Bigl(\Phi_{\smn} j_{\smn}^{-R}
+\frac{1}{n}(\sum_{a=1}^n\partial_{p_a^L} \Phi_{\smn})p_\smn^-
+\frac{1}{n}(\sum_{a=1}^n\theta_a \Phi_{\smn})q_\smn^{-R}\Bigr)\,,
\\
\label{jmld} J^{-L}_{\smn} &=&\int d\Gamma_n \Bigl(\Phi_{\smn}
j^{-L}_{\smn} +\frac{1}{n}(\sum_{a=1}^n\partial_{p_a^R}
\Phi_{\smn})p_\smn^-
+\frac{1}{n}\Phi_{\smn}\sum_{a=1}^n\frac{\lambda_a}{\beta_a}
q_\smn^{-L} \Bigr)\,,
\\
\label{jmid} J^{-i}_{\smn}\! &=&\!\int d\Gamma_n \Bigl(\Phi_{\smn}
j^{-i}_{\smn} +\frac{1}{n}(\sum_{a=1}^n\partial_{p_a^i}
\Phi_{\smn})p_\smn^-
\nonumber\\
{}\!&+&\!\frac{1}{\sqrt{2}n}(\sum_{a=1}^n \theta_a\gamma^i
\Phi_{\smn}) q_\smn^{-L} -\frac{1}{\sqrt{2}n}\Phi_{\smn}
\sum_{a=1}^n \frac{\lambda_a}{\beta_a}\gamma^i
q_\smn^{-R}\Bigr)\,. \eeq Here we introduce new densities
$j_\smn^{-I}$. Commutation relations of $J^{-I}$ with the
kinematical generators tell us that $j_\smn^{-I}$ depend on
$\beta_a$ and momenta ${\Po }_{ab}^I$, $\Lambda_{ab}$ \rf{pablab}.

To summarize, commutation relations between the kinematical and
dynamical generators give us the expressions for dynamical
generators \rf{pm1},\rf{qrl1},\rf{jmrd}-\rf{jmid}, where all
densities $p_\smn^-$, $q_\smn^{-R,L},j_\smn^{-I}$  depend on ${\Po
}_{ab}^I$, $\Lambda_{ab}$ and the densities satisfy the equations
\rf{jmpgn2}-\rf{jrign3}.

In order to fix the densities $p_\smn^-$,
$q_\smn^{-R,L},j_\smn^{-I}$ we should consider commutation
relations between respective dynamical generators and a general
strategy of finding these densities consists basically of the
following two steps:

\begin{itemize}

\item First step is to find restrictions imposed by commutation
relations of Poincar\'e superalgebra between dynamical generators.
Usually from these commutation relations one learns that not all
densities are independent. It turns out that density $p_\smn^-$ is
still to be independent, while all remaining densities, {\it i.e.}
$q_\smn^{-R,L},j_\smn^{-I}$,  are expressible in terms of
$p_\smn^-$.

\item Second step is to find solution to the independent density
$p_\smn^-$. The solution is found, as we will demonstrate below,
from the requirement that all densities, {\it i.e.} $p_\smn^-$,
$q_\smn^{-R,L}$, $j_\smn^{-I}$,  be polynomial in the transverse
momenta $p_a^I$ (and Grassmann momenta $\lambda_a$ as well). This
requirement we shall refer to as a locality condition.

\end{itemize}

Below we apply this strategy to study 3- and 4-point interaction
vertices.

\newsection{Cubic interaction vertices}\label{CUBVERsec}

Although many examples of cubic vertices are known in the
literature, constructing cubic vertices for concrete field
theoretical models is still a challenging procedure. The general
method essentially simplifying the procedure of obtaining cubic
interaction vertices was discovered in \cite{met20}, developed in
\cite{met13,met16} and formulated finally  in \cite{met22}. One of
the characteristic features of this method is reducing manifest
transverse $so(d-2)$ invariance (which is $so(9)$ for $11d$
supergravity) to $so(d-4)$ invariance (which is $so(7)$ in this
paper)\footnote{ In the preceding studies \cite{GS6}, reducing the
manifest $so(d-2)$ symmetry to $so(d-4)$ was used to formulate
superfield theory of $IIA$ superstrings. In the latter reference
the reducing was motivated by desire to get the unconstrained
superfield formulation. In \cite{met22} the main motivation for
reducing was desire to get the most general solution for cubic
vertex for arbitrary spin fields of (super) Poincar\'e invariant
theory. Discussion of the $so(7)$ formalism in the context of
M(atrix) theory can be found in \cite{pw}.}. On the other hand, it
is the $so(7)$ symmetries that are manifest symmetries of the
unconstrained superfield formulation of $11d$ supergravity. In
other words the manifest symmetries of our method and those of the
unconstrained superfield formulation of $11d$ supergravity match.
In this section we would like to demonstrate how the method of
Ref.\cite{met22} works for the case of $11d$ supergravity.

As was explained above (see \rf{pmpm}) cubic vertex $p_{\sm3}^-$
depends on the `momenta' ${\Po }_{ab}$, $\Lambda_{ab}$ and
$\beta_a$, where $a,b=1,2,3$ label three interacting fields in
cubic vertex. The variables ${\Po }_{12}^I$, ${\Po }_{23}^I$,
${\Po }_{31}^I$ however are not independent: all of them are
expressible in terms of ${\Po }^I$ defined by\footnote{By using
momentum conservation laws for $p_a^I$ and $\beta_a$ it is easy to
check that ${\Po }_{12}^I={\Po }_{23}^I={\Po }_{31}^I={\Po }^I$.}
\be \label{defpi} {\Po }^I =\frac{1}{3}\sum_{a=1}^3\check{\beta}_a
p_a^I\,, \qquad \check{\beta}_a\equiv \beta_{a+1}-\beta_{a+2}\,,
\quad \beta_a\equiv \beta_{a+3}\,. \ee
The same holds for Grassmann variables $\Lambda_{ab}$, {\it i.e.}
due to momentum conservation laws for $\beta_a$ and Grassmann
momenta $\lambda_a$ the variables $\Lambda_{12}$, $\Lambda_{23}$
$\Lambda_{31}$ (see \ref{pablab}) are also expressible in terms of
the variable $\Lambda$ defined by
\be\label{Lambda} \Lambda =\frac{1}{3}\sum_{a=1}^3\check{\beta}_a
\lambda_a\,. \ee
The usage of ${\Po }^I$ and $\Lambda$ is advantageous since they
are manifestly invariant under cyclic permutations of indices
$1,2,3$, which label three interacting fields. Thus the vertex
$p_\sm3^-$ is eventually function of ${\Po }^I$, $\Lambda$ and
$\beta_a$:
\be \label{p2v} p_\sm3^- = p_\sm3^-({\Po }, \Lambda,\beta_a)\,.
\ee In general the vertex $p_\sm3^-$ is a monomial of degree $k$
in ${\Po }^I$. As is well know the generic $11d$ supergravity is
described by vertex $p_\sm3^-$, which is monomial of degree two in
transverse `momentum' ${\Po }^I$, i.e. have to set $k=2$. However
for flexibility we keep $k$ to be arbitrary. By doing this we will
be able to demonstrate explicitly the fact that cubic vertices
corresponding to $k=4,6$, which are responsible for the higher
derivative $R^2$ and $R^3$- terms ($R$ stands for Riemann tensor),
do not admit supersymmetric extension.

The method of finding cubic vertices suggested in \cite{met22}
consists of the following steps:

\medskip
({\bf i}) {}First we find dependence of the vertex $p_\sm3^-$ on
`momentum' ${\Po }^I$. To this end we use commutation relations
$[P^-,J^{IJ}]=0$, which lead to the following equations for
$p_\sm3^-$:
\be \label{JIJp3} J^{IJ}(\Po,\Lambda) p_\sm3^-=0\,,\qquad
J^{IJ}(\Po,\Lambda) \equiv L^{IJ}(\Po)+M^{IJ}(\Lambda)\,, \ee
where orbital part of angular momentum is given by
\be\label{LIJ} L^{IJ}(\Po)\equiv
\Po^I\partial_{\Po^J}-\Po^J\partial_{\Po^I}\,, \ee while spin
operators are given by
\beq \label{MIJ1} &&
M^{RL}(\Lambda)=\frac{1}{2}\theta_\Lambda\Lambda-2\,,
\\[5pt]
\label{MIJ4} && M^{ij}(\Lambda)
=\frac{1}{2}\theta_\Lambda\gamma^{ij}\Lambda\,,
\\[5pt]
\label{MIJ2} && M^{Ri}(\Lambda)
=-\frac{1}{2\sqrt{2}}\hat{\beta}\theta_\Lambda\gamma^i\theta_\Lambda\,,
\\[5pt]
\label{MIJ3} && M^{Li}(\Lambda)
=\frac{1}{2\sqrt{2}\hat{\beta}}\Lambda\gamma^i\Lambda\,. \eeq
Here and below we use the notation
\be\label{hatbet} \hat{\beta}\equiv\beta_1\beta_2\beta_3\,, \ee
and $\theta_\Lambda$ is a derivative with respect to $\Lambda$:
\be \label{tldef} \theta_\Lambda\equiv \partial_\Lambda\,,\qquad
\{\theta_\Lambda, \Lambda\}=1\,. \ee In what follows we prefer to
exploit instead of `momenta' ${\Po }^I =(\Po^L,\Po^R,\Po^i)$ a
dimensionfull `momentum' ${\Po }^L$ and dimensionless variables
$q^i$, $\rho$ defined by
\be \label{newvar} q^i\equiv \frac{{\Po }^i}{{\Po }^L}\,, \qquad
\rho\equiv \frac{{\Po }^i{\Po }^i+2{\Po }^R{\Po }^L} {2({\Po
}^L)^2}\,, \qquad \frac{{\Po }^R}{{\Po }^L}= \rho -
\frac{q^2}{2}\,. \ee In terms of the new variables the cubic
interaction vertex can be cast into the form
\be \label{p3v} p_\sm3^-=({\Po }^L)^k
V(q\,,\rho\,,\beta\,,\Lambda)\,, \ee which demonstrates explicitly
that the vertex $p_\sm3^-$ is a monomial of degree $k$ in
transverse `momentum' ${\Po }^I$. In terms of new variables
various components of the orbital momentum operator \rf{LIJ} take
the following form
\beq \label{Lrl} && L^{RL} =q\partial_q+2\rho\partial_\rho -{\Po
}^{L}
\partial_{{\Po }^L}\,,
\\
\label{Lij}
&&
L^{ij}
=q^i\partial_{q^j}-q^j\partial_{q^i}\,,
\\
\label{Lli}
&&
L^{Li}
=\partial_{q^i}\,,
\\
\label{Lri} && L^{Ri} =(\rho-\frac{q^2}{2})\partial_{q^i} +q^i
(q\partial_q+2\rho\partial_\rho -{\Po }^L\partial_{{\Po }^L})\,.
\eeq To demonstrate main idea of introducing the variable $q^i$
let us focus on the $Li$ part of Eqs.\rf{JIJp3}. Plugging in the
$Li$ part of Eqs.\rf{JIJp3} the respective representations for
$p_\sm3^-$ and $L^{Li}$ in \rf{p3v} and \rf{Lli} we get
\be \label{Li2} (\partial_{q^i} + M^{Li}(\Lambda))V=0\,, \ee where
$M^{Li}$ is given in \rf{MIJ3}. Solution of \rf{Li2} is easily
found to be
\be \label{veq} V(q\,,\rho\,,\beta\,,\Lambda)
=E_q\tilde{V}(\rho\,,\beta\,,\Lambda)\,, \ee where an operator
$E_q$ is defined to be
\be \label{eq} E_q \equiv \exp(-q^j M^{Lj}(\Lambda))\,. \ee Thus
collecting above-given expressions we get the following
intermediate representation for the vertex $p_\sm3^-$:
\be \label{p3int} p_\sm3^-=({\Po
}^L)^kE_q\tilde{V}(\rho\,,\beta\,,\Lambda)\,. \ee Next step is to
find dependence on variable $\rho$.  To do that we use $LR$, $Ri$
and $ij$ parts of Eqs.\rf{JIJp3}. Details of derivation may be
found in Appendix B and our result is given by
\be \label{tvrho} \tilde{V}(\rho,\beta,\Lambda)= E_\rho
\tilde{V}_0(\beta,\Lambda)\,, \ee where an operator $E_\rho$ is
defined by relation
\be \label{erho} E_\rho\equiv \sum_{n=0}^k (-\rho)^n
\frac{\Gamma(\frac{7}{2}+k-n)}{2^n n!\Gamma(\frac{7}{2}+k)}
(M^{Lj}(\Lambda)M^{Lj}(\Lambda))^n \ee and new vertex
$\tilde{V}_0$ satisfies the following equations
\beq \label{mrltv0} && (M^{RL}(\Lambda)-k)\tilde{V}_0=0\,,
\\[5pt]
\label{mritv0} && M^{Ri}(\Lambda)\tilde{V}_0=0\,,
\\[5pt]
\label{mijtv0} && M^{ij}(\Lambda)\tilde{V}_0=0\,. \eeq As seen
from \rf{tvrho} the vertex $\tilde{V}_0$ depends only on Grassmann
`momentum' $\Lambda$ and light cone momenta $\beta_a$. The
dependence on the transverse space `momentum' ${\Po }^I$ is thus
fixed explicitly and we get the following representation for the
cubic vertex
\be \label{p3v0} p_\sm3^-({\Po },\Lambda,\beta_a) =({\Po }^L)^k
E_q E_\rho \tilde{V}_0(\Lambda,\beta_a)\,, \ee where $\tilde{V}_0$
satisfies Eqs.\rf{mrltv0}-\rf{mijtv0}. We note that while deriving
this representation we used general form of the orbital momentum
$L^{IJ}$ \rf{JIJp3}, which is valid for arbitrary Poincar\'e
invariant theory. Therefore the representation for the vertex
$p_\sm3^-$ given in \rf{p3v0} is universal and is valid for
arbitrary Poincar\'e invariant theory. Various theories  differ
by: (i) spin operators in angular momentum (for the case under
consideration these spin operators are given in
\rf{MIJ1}-\rf{MIJ3}); (ii) the vertex $\tilde{V}_0$, which for
case under consideration depends on Grassmann `momentum' $\Lambda$
and light cone momenta $\beta_a$. Now we proceed to the second
step of our method.

\medskip
({\bf ii}) At this stage we find dependence on Grassmann
`momentum' $\Lambda$. To this end we use
Eqs.\rf{mrltv0}-\rf{mritv0}, which turn out to be very simple to
analyze. Indeed, making use of expression for $M^{RL}(\Lambda)$
given in \rf{MIJ1} we find the equation
\be \label{dohl} \Lambda\theta_\Lambda
\tilde{V}_0=2(2-k)\tilde{V}_0\,. \ee An operator
$\Lambda\theta_\Lambda$  counts power of Grassmann `momentum'
$\Lambda$ involved in the vertex $\tilde{V}_0$ that cannot involve
terms having negative power of $\Lambda$, {\it i.e.} eigenvalues
of the operator $\Lambda\theta_\Lambda$ must be non-negative.
Eq.\rf{dohl} implies then that vertices with terms higher than
second order in ${\Po }^I$, i.e. with $k>2$, are forbidden. Note
that the values $k=4$ and $k=6$ correspond to the $R^2$- and
$R^3$- terms. Therefore, the fact that vertices with $k=4$ and
$k=6$ are forbidden implies that the $R^2$ and $R^3$- terms do not
allow supersymmetric extension. One important thing to note is
that we proved absence of supersymmetric extension of the $R^2$
and $R^3$- terms by using only commutation relations between
Hamiltonian $P^-$ and the kinematical generators. It is reasonable
to think that the kinematical generators do not  receive quantum
corrections. If this indeed would be the case then our result
could be considered as light cone proof of non-renormalization of
$R^2$ and $R^3$ terms in $11d$ supergravity \footnote{Discussion
of the $R^2$ and $R^3$- terms in string theory effective action
may be found in \cite{MTO}.}.

The remaining case of $k=2$ corresponds to cubic vertex of generic
$11d$ supergravity\footnote{Note that Eq.\rf{dohl} does not
formally rule out the cases of $k=0,1$. It is easy to demonstrate
however that these cases are ruled out by
Eqs.\rf{mritv0},\rf{mijtv0}.} and Eq.\rf{dohl} tells us that for
$11d$ supergravity  the $\tilde{V}_0$ does not depends on
$\Lambda$ at all, {\it i.e.} entire dependence of the cubic vertex
$p_\sm3^-$ \rf{p3v0} on `momenta' $\Po^i$ and $\Lambda$ is
governed by $E$-operators $E_q$, $E_\rho$ \rf{eq},\rf{erho}, which
are purely algebraic in nature.

\medskip
({\bf iiii}) Last step is to find dependence of $\tilde{V}_0$ on
three momenta $\beta_a$. Because of conservation law
$\sum_{a=1}^3\beta_a=0$ the vertex $\tilde{V}_0$ depends on two
light cone momenta. Therefore we need two equations to fix
$\tilde{V}_0$. One of equations is obtainable from commutator
$[P^-,J^{+-}]=P^-$ and was given in \rf{jmpgn2}, where we have to
set $j^{+-}=-1$, $n=3$, $g_\sm3=p_\sm3^-$. Exploiting then the
representation \rf{p2v} we obtain the following equation
\be
(\sum_{a=1}^3\beta_a\partial_{\beta_a}+\frac{3}{2}\Lambda\theta_\Lambda
+k-2)p_\sm3^-=0\,, \ee which in terms of $\tilde{V}_0$ takes the
following form
\be  \label{jmnhom} \sum_{a =1}^3 \beta_a
\partial_{\beta_a}^{\vphantom{5pt}}\tilde{V}_0=0\,.
\ee  The second equation for  $\tilde{V}_0$ can be found from
commutation relations between dynamical generators and requirement
we call a locality condition to be formulated precisely below.
Namely, making use of commutator $[P^-,J^{-I}]=0$ and expression
for $J^{-I}$ given in \rf{jmrd}-\rf{jmid} one can find the
following relation (for details see Appendix C)
\be \label{j3p3} j_\sm3^{-I}=-\frac{2{\Po }^I}{3|{\Po
}|^2}\sum_{a=1}^3
\check{\beta}_a\beta_a\partial_{\beta_a}p_\sm3^-\,. \ee This
expression tells us that the density $j_\sm3^{-I}$ is not
independent quantity but expressible in terms of the interacting
vertex $p_\sm3^-$. Remaining commutators between the dynamical
generators also do not fix the vertex $p_\sm3^-$ uniquely. This
implies that restrictions imposed by commutation relations of
Poincar\'e superalgebra by themselves are not sufficient to fix
the interaction vertex $p_\sm3^-$ uniquely. To choose physical
relevant vertices $p_\sm3^-$ and $j_\sm3^{-I}$, i.e. to fix them
uniquely, we impose the requirement we refer to as the locality
condition: demand the vertices $p_\sm3^-$, $j_\sm3^{-I}$ be
monomial in ${\Po }^I$. As to the vertex $p_\sm3^-$ we demand this
vertex be local ({\it i.e.} monomial in ${\Po }^I$) from the very
beginning. However from Eq.\rf{j3p3} it is clear that local
$p_\sm3^-$ does not lead automatically to local density
$j_\sm3^{-I}$. From the expressions for $\rho$ (see \ref{newvar})
and formulas \rf{erho},\rf{p3v0}  on can demonstrate that the
locality condition amounts to requiring the $\tilde{V}_0$
satisfies the equation
\beq \label{V0loc1} \sum_{a=1}^3 \check{\beta}_a \beta_a
\partial_{\beta_a}^{\vphantom{5pt}}\tilde{V}_0=0\,.
\eeq
This equation reflects simply the fact that in order to cancel
denominator $|\Po|^2$ in r.h.s of Eq.\rf{j3p3} we have to cancel
the contribution of $n=0$ term to the expansion \rf{erho}.
Equations \rf{jmnhom} and \rf{V0loc1} tell us that $\tilde{V}_0$
does not depend on momenta $\beta_a$ at all and therefore
$\tilde{V}_0$ is fixed to be
\be \label{tv0fin} \tilde{V}_0 = \frac{\kappa}{3 }\,, \ee
where $\kappa$ is gravitational constant (see formula \rf{geact}
below). Thus our final result for the cubic vertex is
\be \label{finres} p_\sm3^-({\Po },\Lambda,\beta) =
\frac{\kappa}{3}\, {\Po }^{L\,2}E_qE_\rho\,, \ee where the $E$-
operators $E_q$ and $E_\rho$ are given by \rf{eq} and \rf{erho}.
Note that in expression for $E_\rho$ corresponding to $11d$
supergravity vertex we have to set $k=2$ and this gives the
expansion
\be E_\rho = 1 -\frac{\rho}{9} M^{Li}(\Lambda) M^{Li}(\Lambda)
+\frac{\rho^2}{7\cdot 18 }(M^{Li}(\Lambda) M^{Li}(\Lambda))^2\,.
\ee Making use of these expressions and formula for
$M^{Li}(\Lambda)$ given in \rf{MIJ3} we can work out an explicit
representation for the cubic vertex in a rather straightforward
way
\beq\label{exprep} \frac{3}{\kappa}  p_\sm3^- &=& {\Po }^{L2} -
\frac{{\Po }^L}{2\sqrt{2}\hat{\beta}}\Lambda\bpline\Lambda
+\frac{1}{16\hat{\beta}^2}(\Lambda\bpline\Lambda)^2 - \frac{
|\Po|^2}{9\cdot 16\hat{\beta}^2}(\Lambda\gamma^j\Lambda)^2
\nonumber\\[5pt]
&+&\frac{{\Po }^R}{9\cdot 16\sqrt{2}\hat{\beta}^3}
\Lambda\bpline\Lambda(\Lambda\gamma^j\Lambda)^2 +\frac{{\Po
}^{R2}}{2^7\cdot 63\hat{\beta}^4}
((\Lambda\gamma^i\Lambda)^2)^2\,, \eeq
where throughout this paper we use the notation
\be \bpline \equiv \Po^i\gamma^i\,,\qquad |\Po|^2 \equiv \Po^I
\Po^I\,.\ee Thus the action in cubic approximation is given by
expression \rf{lcact}, where we have to insert Hamiltonian $P^-$
given by \rf{pm1} and \rf{finres}. Note also that formulas
\rf{j3p3} and \rf{exprep} imply the relation
\be j_\sm3^{-I}=0\,.\ee

We choose normalization \rf{tv0fin} so that the cubic vertex for
graviton field obtainable from our action \rf{lcact} coincides
with that of Einstein-Hilbert action taken in the normalization
\be\label{geact} S_{EH} = \int d^{11}x\, {\cal L}_{EH}\,, \qquad
{\cal L}_{EH} \equiv \frac{1}{2\kappa^2} \sqrt{-g}\, R\,. \ee
Let us make comment how the normalization of our vertex can be
related with that of the action \rf{geact} in a most simple way.
Making use an expansion for metric tensor\footnote{We adopt the
conventions $R^\mu{}_{\nu\lambda\rho} \equiv
\partial_\lambda \Gamma_{\nu\rho}^\mu +\ldots $, $R_{\nu\rho}
\equiv R^\mu{}_{\nu\mu\rho}$.}
\be\label{gmnexp} g_{\mu\nu}=\delta_{\mu\nu}+\sqrt{2}\,\kappa
h_{\mu\nu}\ee
we get the following expansion (up to total derivatives) for
Einstein-Hilbert Lagrangian \rf{geact} in cubic approximation
\be \label{r3hhh} \frac{1}{\sqrt{2}\kappa } {\cal
L}_{EH}\,|_{_{(3)}} = \frac{1}{4}h_{\mu\nu}\partial_\mu
h_{\rho\lambda}\partial_\nu h_{\rho\lambda} +
\frac{1}{2}h_{\mu\nu}\partial_\lambda h_{\mu\rho}
\partial_\rho h_{\nu\lambda}
+ \frac{1}{2}h_{\mu\nu} \partial_\rho h_{\nu\lambda}\partial_\rho
h_{\lambda\mu}\,. \ee
To derive this formula we use the constraints $h_\mu^\mu=0$ and
$\partial_\mu h^{\mu\nu}=0$ as we are going to treat cubic
vertices in light cone gauge
\be\label{lcgau} h^{++}=0\,, \qquad h^{+-}=0\,, \qquad
h^{+I}=0\,,\ee
which leads to such constraints.  Making use of light cone gauge
\rf{lcgau} gives the following solution to non-physical degrees of
freedom ($h^{II}=0$):
\be\label{nondyn} h^{-I}(p)=- \frac{p^J}{\beta}\,h^{IJ}(p)\,,
\qquad h^{--}(p) = \frac{p^Ip^J}{\beta^2}\,h^{IJ}(p) \,. \ee
Plugging \rf{lcgau},\rf{nondyn} into \rf{geact} and keeping in
mind \rf{lcact} we get Hamiltonian
\be \label{gemet} P_\sm3^-  = \sqrt{2}\kappa \int d\Gamma_3(p)
\Bigl(h_1^{KN}h_2^{KN} h_3^{IJ}\frac{{\Po }^I{\Po }^J}{4\beta_3^2}
+ h_1^{IK}h_2^{JN}h_3^{KN}\frac{{\Po }^I{\Po }^J}{2\beta_1\beta_2}
\Bigr)\,, \ee
where $h^{IJ}_a\equiv h^{IJ}(p_a)$. An integration measure
$d\Gamma_3(p)$ is given by \rf{delfun01} in which we set $n=3$.
Now in order to fix the normalization \rf{tv0fin} it suffices to
compare the ${\Po }^{L2}h^{RR} h^{RR} h^{LL}$ terms in \rf{finres}
and \rf{gemet}.

To summarize we got two equivalent representations for the cubic
interaction vertex given in \rf{finres},\rf{exprep}. The
representation \rf{exprep} being written manifestly in terms of
the bosonic `momentum' $\Po^I$ and Grassmann `momentum' $\Lambda$
is not convenient, however, in calculations. In contrast to this,
the representation \rf{finres} does not show explicitly a
dependence on the `momenta' $\Po^i$ and $\Lambda$. However the
remarkable feature of representation \rf{finres} is that it is
expressed entirely in terms of $E$- operators which have simple
algebraic properties. For this reason the representation
\rf{finres} is most convenient in various practical calculations.
It is representation \rf{finres} that is universal and can
therefore be extended in a straightforward way to the cases of
$10d$ $IIA$ supergravity and SYM theories.

\subsection{Supercharges in cubic approximation}\label{QchaCUB}

As was said above in supersymmetric theories of interacting fields
the dynamical supercharges also receive interaction - dependent
corrections. In order to complete our study we have to fix
dynamical supercharges too. To derive the supercharges we could
use the procedure we exploited for evaluation of Hamiltonian in
the preceding paragraph. For the case of supercharges there is
however a shorter way, which is based on exploiting the expression
for Hamiltonian density $p_\sm3^-$ above obtained. We proceed as
follows. From commutation relations
\be [P^-,Q^{-R,L}]=0 \ee we get the following representations for
the supercharge densities $q^{-R,L}$:
\be\label{qQp} q_\sm3^{-R} \equiv -\frac{2\hat{\beta}}{|\Po|^2}
Q^{-R}(\Lambda) p_\sm3^-\,, \qquad
q_\sm3^{-L}=-\frac{2\hat{\beta}}{|\Po|^2} Q^{-L}(\Lambda)
p_\sm3^-\,, \ee
where we introduce operators $Q^{-R,L}(\Lambda)$ defined by
\be Q^{-R}(\Lambda)=\frac{1}{\sqrt{2}}\theta_\Lambda\bpline
+\frac{1}{\hat{\beta}}{\Po }^R\Lambda\,, \qquad
Q^{-L}(\Lambda)\equiv  {\Po }^L\theta_\Lambda +
\frac{1}{\sqrt{2}\hat{\beta}}\bpline \Lambda
 \ee
and $\hat{\beta}$ is defined in \rf{hatbet}. By using the
representation for $p_\sm3^-$ given in \rf{exprep} we find from
\rf{qQp} the following manifest representation for the
supercharges
\beq \label{qlfin} && \frac{3}{\kappa} q_\sm3^{-L} =\frac{{\Po
}^L}{18\hat{\beta}}\gamma^i\Lambda\Lambda\gamma^i\Lambda
-\frac{1}{36\sqrt{2}\hat{\beta}^2}\gamma^i\Lambda\Lambda\gamma^i\Lambda
\Lambda\bpline\Lambda -\frac{{\Po }^R}{16\cdot
63\hat{\beta}^3}\gamma^i\Lambda \Lambda\gamma^i\Lambda
(\Lambda\gamma^j\Lambda)^2\,,\hspace{1cm}
\\[5pt]
\label{qrfin} &&  \frac{3}{\kappa} q_\sm3^{-R} =-{\Po }^L\Lambda
+\frac{{\Po }^i}{18\sqrt{2}\hat{\beta}}
(8\Lambda\Lambda\gamma^i\Lambda
-\gamma^{ij}\Lambda\Lambda\gamma^j\Lambda) +\frac{{\Po
}^R}{72\hat{\beta}^2}\Lambda(\Lambda\gamma^i\Lambda)^2\,. \eeq
Because $Q^{-L}$ commutes with $J^{Li}$ the supercharge density
$q^{-L}_\sm3$ admits the representation similar to \rf{finres} (or
\rf{p3v0}):
\be\label{qeetq} q_\sm3^{-L}({\Po },\Lambda,\beta) ={\Po
}^LE_qE_\rho \tilde{q}_0^{-L}(\Lambda,\beta)\,. \ee where the
$\tilde{q}_0^{-L}$ can be considered to some extent as
superpartner of $\tilde{V}_0$ \rf{tv0fin} and is given by
\be \frac{3}{\kappa}\,\tilde{q}_0^{-L}(\Lambda,\beta) =
\frac{1}{18\hat{\beta}}\gamma^i\Lambda \Lambda\gamma^i\Lambda\,.
\ee Note that $E_\rho$ in \rf{qeetq} is given by \rf{erho}, where
we have to set $k=1$. The expressions for $q_\sm3^{-R,L}$
\rf{qlfin},\rf{qrfin} can be used to rederive the $p_\sm3^-$
\rf{exprep} by using anticommutator $\{Q^{-R},Q^{-L}\}=-P^-$. This
provides additional check to our calculations.

\newsection{4-point interaction vertices}\label{FOUVERsec}

In this section we extend our analysis to 4-point interaction
vertices. It should be emphasized from the very beginning that we
do not consider 4-point interaction vertices of the generic $11d$
supergravity \cite{cjs,brihow}. Here we study 4-point vertices
that are invariant with respect to linear supersymmetry
transformations and do not depend on contributions of exchanges
generated by cubic vertices. These 4-point vertices in their
gravitational bosonic sector involve higher derivative terms that
can be constructed from Riemann tensor and derivatives, i.e. they
can be presented schematically as $\partial^{2n} R^4$. We will be
able to find these vertices with arbitrary powers of derivatives,
i.e. for arbitrary value $n$. The bosonic bodies of these
supersymmetric vertices appeared in various previous studies and
they are of interest because they are responsible for quantum
corrections to classical action of $11d$ supergravity. Light-cone
formulation we promote here allows us to find simple and compact
expressions for these 4-point vertices, which might be useful in
various future studies. One of interesting results of our study is
that it is the supersymmetric formulation based on unconstrained
light cone superfield  that allows us to develop these simple and
compact expressions for higher derivative vertices. Note also that
a by-product of our study will be a derivation of superspace tree
level 4-point scattering amplitude of the generic $11d$
supergravity theory.

Thus we are interested in 4-point Hamiltonian
\be \label{pm122}   P_\smf^-  = \int d\Gamma_4\Phi_\smf
p_\smf^-\,,\ee where $\Phi_\smf$ and $d\Gamma_4$ are given by
formulas \rf{Phin},\rf{delfun} in which we set $n=4$. According to
\rf{pmpm} 4-point interaction vertex $p_\smf^-$ depends on the
variables
\be {\Po }_{ab}^I, \quad \Lambda_{ab}, \quad \beta_a\,, \qquad
a,b=1,2,3,4, \ee where indices $a,b$ label four interacting
fields. The `momenta' ${\Po }_{ab}^I$ and $\Lambda_{ab}$ however
are not independent. Indeed making use of conservation laws it is
straightforward to get  the following relations
\be\label{bas} {\Po }_{12}^I=\frac{\beta_2}{\beta_{13}}{\Po
}_{13}^I +\frac{\beta_1}{\beta_{13}}{\Po }_{24}^I\,, \qquad
\Lambda_{12}=\frac{\beta_2}{\beta_{13}}\Lambda_{13}
+\frac{\beta_1}{\beta_{13}}\Lambda_{24}\,,\ee
\be \beta_{ab}\equiv \beta_a+\beta_b\,. \ee
Using these relations and those obtainable from them by making
cyclic permutations of indices $1,2,3,4$ it is easy to see that
all $\Po_{ab}^I$ and $\Lambda_{ab}$ can be expressed in terms of
`momenta' $\Po_{13}^I$, $\Po_{24}^I$ and Grassmann `momenta'
$\Lambda_{13}$ $\Lambda_{24}$. Therefore we can choose the
representation in which the vertex depends only on ${\Po
}_{13}^I$, ${\Po }_{24}^I$, $\Lambda_{13}$, $\Lambda_{24}$ and
$\beta_a$. The usage of ${\Po }_{13}^I$, ${\Po }_{24}^I$ and
$\Lambda_{13}$, $\Lambda_{24}$ is advantageous because these
variables transform into each other under cyclic permutations of
indices $1,2,3,4$. By analogy with \rf{newvar} we introduce then
new dimensionless variables $q_{ab}$, $\rho_{ab}$:
\be \label{newvar4} q_{ab}^i\equiv \frac{{\Po }_{ab}^i}{{\Po
}_{ab}^L}\,, \qquad \rho_{ab}\equiv \frac{{\Po }_{ab}^i{\Po
}_{ab}^i+2{\Po }_{ab}^R{\Po }_{ab}^L} {2{\Po }_{ab}^{L2}}\,,
\qquad \frac{{\Po }_{ab}^R}{{\Po }_{ab}^L}=
\rho_{ab}-\frac{q_{ab}^iq_{ab}^i}{2}\,. \ee Note that we drop the
summation rule for repeated four external line indices $a,b$. With
this notation it is clear that the most general 4-point
interaction vertex depending on ${\Po }_{13}^I$, ${\Po }_{24}^I$,
$\Lambda_{13}$, $\Lambda_{24}$ and $\beta_a$ can be cast into the
form
\be \label{p4dep}
p_{(4)}^-=p_{(4)}^-(q_{13},q_{24},\Lambda_{13},\Lambda_{24}, {\Po
}_{13}^L,{\Po }_{24}^L,\rho_{13},\rho_{24},\beta_a)\,. \ee Thus by
using \rf{newvar4} we have replaced the variables ${\Po }_{13}^i$,
${\Po }_{13}^R$ and ${\Po }_{24}^i$, ${\Po }_{24}^R$ by the
respective variables $q_{13}^i$, $\rho_{13}$ and $q_{24}^i$,
$\rho_{24}$.

As before the vertex $p_\smf^-$ can be found by exploiting
commutation relations of Poincar\'e superalgebra and requirement
of locality with respect to transverse momenta.  In 4-point
approximation the commutation relations between Hamiltonian $P^-$
and remaining generators $G$ take the following form in general
\be \label{pg4} [P^-,G]_\smf=[P_\smt^-,G_\smf]+[P_\smf^-,G_\smt] +
[P_\sm3^-,G_\sm3]\,. \ee Here we are interested in Hamiltonian
that has no interaction corrections to cubic order i.e. we
restrict our attention to the case $P_\sm3^-=0$. This implies that
in \rf{pg4} we can drop the commutator $[P_\sm3^-,G_\sm3]$. Note
that it is the latter commutator that is responsible for exchange
contributions. The restriction $P_\sm3^-=0$ simplifies analysis
significantly. Because we have demonstrated that in cubic
approximation there is only the vertex of generic $11d$
supergravity  theory our analysis does not involve only the
4-point interaction vertex of that theory. In other words it is
the assumption of $P_\sm3^-=0$ that leaves aside the 4-point
vertex of the generic $11d$ supergravity. Thus taking into account
the restriction $P_\sm3^-=0$ the 4-point commutator \rf{pg4}
simplifies as
\be \label{pg4s}
[P^-,G]_\smf=[P_\smt^-,G_\smf]+[P_\smf^-,G_\smt]\,. \ee It turns
out that in order to find $P_\smf^-$ it suffices to consider the
following commutation relations $[P^-,J^{+-}]=P^-$, $[P^-,
J^{IJ}]=0$ and $[P^-, J^{-L}]=0$, $[P^-,Q^{-R,L}]=0$, which lead
in 4-point approximation to the equations
\be \label{pg4ss}  [P_\smf^-,J^{+-}] = P_\smf^-\,,\qquad
[P_\smf^-, J^{IJ}] = 0\,,  \ee

\be \label{pg4ss1} {}[P_\smt^-, J_\smf^{-L}]+[P_\smf^-,
J_\smt^{-L}]=0\,,\qquad  [P_\smt^-, Q_\smf^{-R,L}]+[P_\smf^-,
Q_\smt^{-R,L}]=0\,.  \ee Moreover we are interested in vertices
$p_\smf^-$ whose momenta $p_a^I$ satisfy the relation
\be \label{enesur} \sum_{a=1}^4 p_a^-=0\,, \qquad p_a^-\equiv
-\frac{p_a^Ip_a^I}{\beta_a}\,. \ee Because this relation expresses
simply the energy conservation law we shall refer the hypersurface
defined by \rf{enesur} to as the energy surface. Obviously the
condition \rf{enesur} decreases number of independent variables of
vertex $p_\smf^-$ shown in r.h.s. of Eq.\rf{p4dep}. Indeed making
use of relation
\be\label{enesur02} \sum_{a=1}^4 \frac{p_a^Ip_a^I}{\beta_a}
=\frac{{\Po }_{13}^I{\Po }_{13}^I}{\beta_1\beta_2\beta_{13}}
+\frac{{\Po }_{24}^I{\Po }_{24}^I}{\beta_2\beta_4\beta_{24}} \ee
we are going to demonstrate that on the energy surface the
variables $\rho_{13}$ and $\rho_{24}$, which were independent so
far, can be now expressed in terms of remaining variables shown in
r.h.s. of Eq.\rf{p4dep} and Mandelstam variable $u$
\be\label{rhoman} \rho_{13}=\frac{\beta_1\beta_3 u}{2{\Po
}_{13}^{L2}}\,, \qquad \rho_{24}=\frac{\beta_2\beta_4 u}{2{\Po
}_{24}^{L2}}\,. \ee To this end and in order to fix our notation
we use Mandelstam variables defined by relations
\be\label{studef} s\equiv -(p_1+p_2)^2,\qquad t \equiv
-(p_1+p_4)^2,\qquad u\equiv -(p_1+p_3)^2\,, \ee
\be \label{stu0} s+t+u=0\,.\ee Making use of  on mass-shell
equation for massless particles (2nd relation in \rf{enesur}) it
is easy to get a representation of Mandelstam variables in light
cone basis:

\be\label{stulcdef} s = \frac{{\Po }_{12}^I{\Po
}_{12}^I}{\beta_1\beta_2}\,,\qquad t = \frac{{\Po }_{14}^I{\Po
}_{14}^I}{\beta_1\beta_4}\,,\qquad u=\frac{{\Po }_{13}^I{\Po
}_{13}^I}{\beta_1\beta_3}\,. \ee By using then the relations given
in \rf{newvar4} and representation for $u$ given in \rf{stulcdef}
one can make sure that on the energy surface \rf{enesur} (see also
\rf{enesur02}) the variables $\rho_{13}$, $\rho_{24}$ admit the
representation \rf{rhoman} indeed.

Now we consider commutation relations \rf{pg4ss},\rf{pg4ss1} in
which the vertex $p_\smf^-$ is restricted to the energy surface.
Exploiting the condition \rf{enesur} in \rf{pg4ss},\rf{pg4ss1}
leads to the following equation for 4-point Hamiltonian
\be\label{pg4sss} [P_\smf^-,J^{+-}] = P_\smf^-\,,\qquad [P_\smf^-,
J^{IJ}]=0\,, \qquad [P_\smf^-, Q_\smt^{-R,L}]=0\,, \qquad
[P_\smf^-, J_\smt^{-L}]=0\,. \ee
These equations in terms of the vertex $p_\smf^-$ defined by
\rf{pm122} take the following form
\be \label{defeqs0}  (\Po_{13}^L\partial_{\Po_{13}^L}
+\Po_{24}^L\partial_{\Po_{24}^L}
+\frac{3}{2}\Lambda_{13}\theta_{\Lambda_{13}}
+\frac{3}{2}\Lambda_{24}\theta_{\Lambda_{24}}
+\sum_{a=1}^4\beta_a\partial_{\beta_a}-4)p_{(4)}^- = 0\,, \ee
\beq \label{defeqs1} && (J_{13}^{IJ}+J_{24}^{IJ})p_{(4)}^-=0\,,
\\[7pt]
\label{defeqs2} && (Q_{13}^{-R,L}+Q_{24}^{-R,L})p_\smf^-=0\,,
\\[7pt]
\label{defeq3} && (J_{13}^{-L}+J_{24}^{-L})p_{(4)}^-=0\,, \eeq
where the differential operators $J_{ab}^{IJ}$, $Q_{ab}^{-R,L}$
are defined by
\be J_{ab}^{IJ}\equiv {\Po }_{ab}^I\partial_{{\Po }_{ab}^J} - {\Po
}_{ab}^J\partial_{{\Po }_{ab}^I}+M_{ab}^{IJ}\,, \ee

\beq && M_{ab}^{RL} \equiv \frac{1}{2}\theta_{
 \Lambda_{ab} } \Lambda_{ab}-2\,,
\\[6pt]
&& M_{ab}^{ij}\equiv
\frac{1}{2}\theta_{\Lambda_{ab}}\gamma^{ij}\Lambda_{ab}\,,
\\[6pt]
&& M_{ab}^{Ri} \equiv \frac{\beta_a\beta_b\beta_{ab}}{2\sqrt{2}}
\theta_{\Lambda_{ab}}\gamma^i\theta_{\Lambda_{ab}}\,,
\\[6pt]
\label{opem} && M_{ab}^{Li} \equiv -\frac{\Lambda_{ab} \gamma^i
\Lambda_{ab}}{2\sqrt{2}\beta_a\beta_b \beta_{ab}}\,, \eeq

\beq && Q_{ab}^{-R}\equiv
\frac{1}{\sqrt{2}}\theta_{\Lambda_{ab}}\bpline_{ab} -
\frac{1}{\beta_a\beta_b\beta_{ab}}{\Po }^R_{ab}\Lambda_{ab}\,,
\\
&& Q_{ab}^{-L}\equiv {\Po }_{ab}^L\theta_{\Lambda_{ab}} -
\frac{1}{\sqrt{2}\beta_a\beta_b\beta_{ab}}\bpline_{ab}\Lambda_{ab}\,.
\eeq
We succeeded in finding most general solution to the defining
equations \rf{defeqs0}-\rf{defeq3} and our result is given by
(details may be found in Appendix D):
\be\label{p4fin} p_{(4)}^-= (q_{_L}^2)^2 \frac{({\Po }_{13}^L{\Po
}_{24}^L)^4}{\beta_{13}^4} E_{q_{13}} E_{q_{24}} E_u g(s,t,u)\,,
\ee  where operators $E_{q_{ab}}$, $E_u$  are defined to be
\beq\label{eqab} &&E_{q_{ab}}\equiv\exp(-q_{ab}^iM_{ab}^{Li})\,,
\\[7pt]
\label{eu} && E_u \equiv \exp\Bigl( - \frac{ u\Lambda^L\qline_{_L}
\Lambda^L}{2\sqrt{2}\beta_{13}q_{_L}^2 ({\Po }_{13}^L{\Po
}_{24}^L)^2}\Bigr)\,. \eeq The new variables $q_{_L}^i$ and
$\Lambda^L$ which enter the operator $E_u$ \rf{eu} are defined to
be
\beq \label{qi} && q_{_L}^i\equiv q_{13}^i-q_{24}^i\,,
\\[5pt]
\label{defThe} && \Lambda^L \equiv  \Lambda_{13} {\Po }_{24}^L -
\Lambda_{24} {\Po }_{13}^L\,, \eeq while the operators
$M_{ab}^{Li}$ which enter $E_{q_{ab}}$ \rf{eqab} are defined by
\rf{opem}.

{}Explicit form of a function $g(s,t,u)$ \rf{p4fin}, which depends
on Mandelstam variables $s$, $t$, $u$, cannot be fixed by
exploiting restrictions imposed by global symmetries alone. This
function is freedom of our solution. The only requirement is that
the $g(s,t,u)$ be symmetric with respect to $s$, $t$,
$u$\footnote{Because the measure $d\Gamma_4$ and product of four
superfields $\Phi_\smn$ in \rf{pm122} are symmetric upon any
permutations of the four external line indices 1,2,3,4, the
4-point vertex $p_\smf^-$ \rf{pm122} should also be symmetric upon
such permutations. Below we prove that $(q_{_L}^2)^2$- term which
is front of function $g(s,t,u)$ \rf{p4fin} is symmetric upon any
permutations of the indices 1,2,3,4 and this leads to requirement
the $g(s,t,u)$ be symmetric upon any permutations of $s,t,u$.}. If
we assume that $g(s,t,u)$ admits Taylor series expansion then the
lower order terms in infinite series expansion of $g(s,t,u)$ take
the form
\beq  \label{gexp} g(s,t,u) & = & g_0 + g_2 (s^2+t^2+u^2) + g_3
stu + g_4 (s^4 +t^4 +u^4)
\nonumber\\[7pt]
& + & g_5 stu(s^2 + t^2 +u^2) + g_{6;1} (s^6 + t^6 +u^6) +
g_{6;2}(stu)^2 +  \ldots\,. \eeq
Note that all that is important to derive this expansion are the
relation \rf{stu0} and requirement the function $g(s,t,u)$ be
symmetric in $s$, $t$, $u$. Plugging $g_{n}$ terms \rf{gexp} in
\rf{p4fin} gives superinvariants that consist in their
gravitational bosonic body the higher derivative terms constructed
from four Riemann tensors and $2n$ derivatives
\be\label{bossec} g_{n}\partial^{2n} R^4\,, \ee
where $\partial^{2n}$ stands for $2n$ derivatives spread among
four Riemann tensors.

Because in the literature $R^4$ terms usually are considered in
covariant Lagrangian formulation we relate now the coefficient
$g_0$ in expansion \rf{gexp} with that of covariant formulation.
This is to say that $g_0$ is given by
\be \label{g0k4} g_0 =\frac{8}{3}\kappa^4 \kappa_{_{(4)}}\,, \ee
where $\kappa_{_{(4)}}$ is a coupling constant that appears in
front of $R^4$ terms in covariant Lagrangian\footnote{It is clear
that the factor $\sqrt{-g}$ in \rf{covlagr4} is not important for
our analysis as we restricted ourselves to the 4-point vertices.}
\be\label{covlagr4} {\cal L}_{R^4} = \kappa_{_{(4)}} \sqrt{-g}\,\,
W_{R^4} \,, \qquad W_{R^4} \equiv W_1 + \frac{1}{16} W_2\,, \ee
and we use the notation\footnote{We exploit the basis of $R^4$
terms \rf{r42def}-\rf{r46def}, which was introduced in Appendix B2
of Ref.\cite{Peeters:2000qj}. From that Appendix one can learn
that $W_{R^4}$ admits the representation  $ W_{R^4} =
\frac{1}{16\cdot 4!} t_8 t_8 R^4 $. }
\beq\label{w1def} && W_1 \equiv R_{42} +\frac{1}{2}R_{41} -
\frac{1}{4}R_{45} - \frac{1}{8} R_{46}\,,
\\[6pt]
\label{w2def} && W_2 \equiv  R_{43} +\frac{1}{2}R_{44} - 4 R_{45}
- 2 R_{46}\,,\eeq
\beq \label{r42def} && R_{42} = \Tr\ R_{\mu\nu}R_{\nu\rho}
R_{\mu\sigma} R_{\sigma\rho}\,,
\\[8pt]
&&  R_{41} = \Tr\ R_{\mu\nu}R_{\nu\rho} R_{\rho\sigma}
R_{\sigma\mu}\,,
\\[8pt]
\label{r43def} && R_{43} = \Tr\ R_{\mu\nu} R_{\rho\sigma}\, \Tr
R_{\mu\nu} R_{\rho\sigma}\,,
\\[8pt]
&& R_{44} = (\Tr\ R_{\mu\nu} R_{\mu\nu})^2\,,
\\[8pt]
&& R_{45} = \Tr\ R_{\mu\nu} R_{\mu\nu} R_{\rho\sigma}
R_{\rho\sigma}\,,
\\[8pt]
 \label{r46def} && R_{46} = \Tr\ R_{\mu\nu}  R_{\rho\sigma} R_{\mu\nu}
R_{\rho\sigma}\,. \eeq In these formulas an matrix $R_{\mu\nu}$
stands for Riemann tensor $R_{\mu\nu}^{AB}$ and $\Tr$ indicates
trace over Lorentz indices $A,B$.

In other words, if gravitational bosonic body of covariant
supersymmetric Lagrangian is given by the expression \rf{covlagr4}
then the corresponding light cone gauge supersymmetric Hamiltonian
is given by expressions \rf{pm122},\rf{p4fin} with $g_0$ given by
\rf{g0k4}. Remaining $g_{n}$- terms with $n=1,2\ldots $ in
interaction vertex \rf{p4fin} corresponding to Lagrangian
\rf{covlagr4} should be set equal to zero. Derivation of the
relation \rf{g0k4} may be found in Appendix E.

The formula \rf{g0k4} linking constants of bosonic body of
covariant Lagrangian and corresponding light cone gauge
supersymmetric Hamiltonian can easily be generalized to the higher
derivative terms. This is to say that if $\partial^{2n} R^4$-
terms of covariant Lagrangian are given by
\be\label{covlagr4f} {\cal L}_{f\,R^4} =   \sqrt{-g}\,\, f(s,t,u)
W_{R^4} \,,\ee
then the corresponding 4-point light cone gauge supersymmetric
Hamiltonian is given by formulas \rf{pm122},\rf{p4fin}, where the
function $g(s,t,u)$ \rf{gexp} is expressible in terms of the
function $f(s,t,u)$ as follows
\be  g(s,t,u) =\frac{8}{3}\kappa^4 f(s,t,u)\,. \ee
The function $f(s,t,u)$ being symmetric in $s$ $t$, $u$ has the
expansion similar to that of the function $g(s,t,u)$
\rf{gexp}\footnote{The coefficient $f_0$ for effective M-theory
action is given in \cite{Green:1997as,Russo:1997mk} (see also
related discussion in \cite{Frolov:2001jh}). For the case of $11d$
supergravity the function $f(s,t,u)$ describes quantum correction
to the classical action of $11d$ supergravity. Calculation of
various loop corrections to the coefficients $f_0$, $f_4$, $f_6$
may be found in \cite{Bern:1998ug,Deser:1998jz}. Review of this
theme and extensive list of references may be found in
\cite{Deser:1999mh}-\cite{Bern:2000fm}.}:
\beq  \label{fexp} f(s,t,u) & = & f_0 + f_2 (s^2+t^2+u^2) + f_3
stu + f_4 (s^4 +t^4 +u^4)+\ldots \,. \eeq

A few remarks are in order.

({\bf i}) Rewriting the interaction vertex \rf{p4fin} in the form
\be  \label{p4finalt} p_{(4)}^-= K g(s,t,u)\,,\qquad K\equiv
(q_{_L}^2)^2 \frac{({\Po }_{13}^L{\Po }_{24}^L)^4}{\beta_{13}^4}
E_{q_{13}} E_{q_{24}} E_u \,, \ee
we remind that the quantity $K$ is usually referred to as
kinematical factor. Our solution to interaction vertex implies
that there is unique kinematical factor. The fact that there is
unique on shell kinematical factor consisting in gravitational
bosonic sector $R^4$- terms has been previously
argued\footnote{For a discussion of covariant superfield
description of $R^4$ terms in $10d$ see {\it e.g.}
\cite{Nilsson:1986rh,deHaro:2002vk}. For a collection papers
devoted to superspace description of $R^4$ terms in $10d$, $11d$
theories see {\it e.g.} \cite{Cederwall:2004cg}.} by lifting the
unique kinematical factor of $IIA$ $10d$ supergravity to eleven
dimensions. Our study demonstrates this fact directly in eleven
dimensions, i.e. our approach does not rely upon procedure of
lifting which should be used with some care in supersymmetric
theories\footnote{Arguments beyond the lifting procedure may be
found in \cite{Peeters:2000qj}.}. Thus a complete set of on shell
4-point interaction vertices is given by product of unique
kinematical factor and arbitrary function $g(s,t,u)$. It is
interesting to note that for $n=0,2,3,4,5$ the expansion \rf{fexp}
involves only one symmetric monomial of degree $n$ in $s,t,u$.
This implies simply that for these values $n$ there is only one on
shell supersymmetric 4-point vertex.

\medskip
({\bf ii}) The expression for the vertex $p_\smf^-$ given in
\rf{p4fin} is not manifestly polynomial in ${\Po }_{13}^L$, ${\Po
}_{24}^L$ and $q_{_L}^2$ because these quantities appear sometimes
in denominators. One can check however that making use of various
Fierz identities for gamma matrices leads to cancellation of some
nonlocal expressions in ${\Po }_{13}^L$, ${\Po }_{24}^L$ and
$q_{_L}^2$. Remaining nonlocal expressions can be removed then by
exploiting the relations
\be \Po_{13}^R = \frac{\beta_1\beta_3}{2\Po_{13}^L}u
-\frac{\Po_{13}^L}{2}q_{13}^2\,, \qquad  \Po_{24}^R =
\frac{\beta_2\beta_4}{2\Po_{24}^L}u
-\frac{\Po_{24}^L}{2}q_{24}^2\,, \ee which imply that some
nonlocal expressions in $\Po_{13}^L$, $\Po_{24}^L$ can be traded
for local expressions in terms of $\Po_{13}^R$, $\Po_{24}^R$. In
fact the price we paid to get simple representation for the
4-point interaction vertex \rf{p4fin} is a loss of a manifest
locality with respect to ${\Po }_{13}^L$, ${\Po }_{24}^L$ and
$q_{_L}^2$.

\medskip
({\bf iii}) The interaction vertex \rf{p4fin} (or \rf{p4finalt})
should be symmetric under any permutations of four external line
indices 1,2,3,4. The symmetry properties of function $g(s,t,u)$
are not fixed by supersymmetries and therefore we simply demand
this function be symmetric under any permutations of indices
1,2,3,4, i.e. $s,t,u$ variables. As to the kinematical factor $K$
\rf{p4finalt} its form is fixed uniquely by supersymmetries and it
turns out that $K$ is indeed symmetric under any permutations of
1,2,3,4. Let us demonstrate this important feature of the
kinematical factor explicitly. Kinematical factor $K$ is
explicitly symmetric under {\it cyclic} permutations of indices
1,2,3,4 and two permutations $1\leftrightarrow 3$ and
$2\leftrightarrow 4$. Thus all that remains is to prove that $K$
is symmetric under permutation $2\leftrightarrow 3$. All remaining
permutations can be presented as combination of the
above-mentioned permutations. We proceed with analysis of
$(q_{_L}^2)^2$- term that is in front of $E$-operators. This
$(q_{_L}^2)^2$- term can be cast into the form that is manifestly
symmetric under any permutations of 1,2,3,4. Indeed making use of
relations
\beq\label{plplstdec}
\frac{(\Po_{13}^L\Po_{24}^L)^2}{\beta_{13}^2} q_{_L}^2 & = &
\Po_{12}^L\Po_{34}^L t - \Po_{14}^L\Po_{23}^L s
\nonumber\\[12pt]
&  = & -\Po_{13}^L\Po_{24}^L s - \Po_{12}^L\Po_{34}^L u
\nonumber\\[12pt]
&  = & \Po_{13}^L\Po_{24}^L t + \Po_{14}^L\Po_{23}^L u\,. \eeq it
is easy to see that $(q_{_L}^2)^2$- term of $K$ can be cast into
the manifestly symmetric form
\be\label{q4stu} \frac{(\Po_{13}^L\Po_{24}^L)^4}{\beta_{13}^4}
(q_{_L}^2)^2 =-(\Po_{12}^L\Po_{34}^L)^2 ut -
(\Po_{13}^L\Po_{24}^L)^2 st -(\Po_{14}^L\Po_{23}^L)^2 su\,,\ee
which demonstrate that $(q_{_L}^2)^2$- term of $K$ is indeed
symmetric under any permutations of 1,2,3,4. All that remains is
to prove that
\be\label{4poi10} E_{q_{13}}E_{q_{24}} E_u  - \hbox{ is invariant
under permutation }\, 2\leftrightarrow 3\,. \ee This important
property can be proved as follows. Introducing new `momentum'
$P^i$ by relation
\be\label{Pidef} P^i \equiv \Po_{13}^L\Po_{24}^L q_{_L}^i \ee and
making use of relations \rf{bas} one can prove that
$P^i/\beta_{13}$ and $\Lambda^L/\beta_{13}$ are antisymmetric
under any permutations of indices 1,2,3,4. This is to say that
these quantities change sign upon permutation $2\leftrightarrow
3$:
\be\label{Pi23}
\frac{P^i}{\beta_{13}}\,\,\stackrel{2\leftrightarrow
3}{\longrightarrow } \,\, - \frac{P^i}{\beta_{13}}\,, \qquad
\qquad \frac{\Lambda^L }{\beta_{13}} \,\,
\stackrel{2\leftrightarrow 3}{\longrightarrow } \,\, -
\frac{\Lambda^L }{\beta_{13}}\,.\ee
In terms of variable $P^i$ one has the following representation
for $\ln E_u$ (see \rf{eu}):
\be  \ln E_u = - \frac{u \Lambda^L \Pline \Lambda^L
}{2\sqrt{2}\beta_{13}P^2 \Po_{13}^L\Po_{24}^L}\,.  \ee
Taking into account this representation and \rf{Pi23} we get
relations
\be  \ln E_u \ \ \ \stackrel{2 \leftrightarrow 3}{\longrightarrow
} \ \ \ \frac{s \Lambda^L \Pline \Lambda^L  }{
2\sqrt{2}\beta_{13}P^2 \Po_{12}^L\Po_{34}^L}
 = \ln E_u - \frac{ \Lambda^L \qline_{_L} \Lambda^L }{
2\sqrt{2}\beta_{13}^3 \Po_{12}^L\Po_{34}^L}\,, \ee
where we use the formula

\be  s  =  -\frac{\Po_{12}^L\Po_{34}^L}{\Po_{13}^L\Po_{24}^L}u
-\frac{\Po_{13}^L\Po_{24}^L}{\beta_{13}^2} q_{_L}^2\,. \ee
Finally, because of relation
\be \ln E_{q_{13}} + \ln E_{q_{24}} \ \ \ \stackrel{2
\leftrightarrow 3}{\longrightarrow } \ \ \ -q_{12}^iM_{12}^{Li} -
q_{34}^iM_{34}^{Li} \ee
and formula
\be -q_{12}^iM_{12}^{Li} - q_{34}^iM_{34}^{Li} - \frac{ \Lambda^L
\qline_{_L} \Lambda^L  }{ 2\sqrt{2}\beta_{13}^3
\Po_{12}^L\Po_{34}^L} =- q_{13}^i M_{13}^{Li} -
q_{24}^iM_{24}^{Li}\,, \ee
we see that the statement in \rf{4poi10} is indeed true.

Making use of just proved symmetry properties of the $E$-
operators the 4-point vertex can be cast into more symmetric form
with respect to $s$, $t$, $u$ variables. Indeed introducing the
notation
\beq\label{bfJdef} &&  {\bf J}_{ab} \equiv \Po_{ab}^L
\sqrt{E_{q_{ab}}}\,,
\\[7pt]
&&\label{EsEtdef} E_s \equiv E_u|_{2\leftrightarrow 3}\,,\qquad
E_t \equiv E_u|_{3\leftrightarrow 4}\eeq and using formula
\rf{q4stu} the expression for the 4-point vertex \rf{p4fin} can be
cast into the form
\be\label{p4fin03}  p_\smf^-  =  - \Bigl( ({\bf J}_{12}{\bf
J}_{34})^2 ut E_s + ({\bf J}_{13}{\bf J}_{24})^2 st E_u + ({\bf
J}_{14}{\bf J}_{23})^2 u s E_t\Bigr) g(s,t,u) \,. \ee

\medskip
({\bf iv}) The expression for the interaction vertex \rf{p4fin}
(or \rf{p4fin03}) can be used to obtain superspace representation
for tree level 4-point scattering amplitude of the generic $11d$
supergravity theory \cite{cjs}. To this end we introduce standard
representation for the $S$-matrix
\be  S = 1 - 2\pi  {\rm i}\, T\,, \ee where the $T$-matrix has an
expansion
\be T = \sum_{n} T_\smn \,,\ee and $n$ point interaction-dependent
correction to the $T$- matrix takes standard form
\be\label{Tndef} T_\smn  = \int d\Gamma_n\, \delta(\sum_{a=1}^n
p_a^-)\, \prod_{a=1}^n \phi(p_a,\lambda_a)\,\, t_\smn\, \,.\ee In
this formula $p_a^-$ are those of \rf{enesur} and superfield
$\phi(p,\lambda)$ enters a solution to free equations of motion:
\be\label{Phiphirel} \Phi(p,\lambda) = \exp({\rm
i}x^+p^-)\phi(p,\lambda)\,. \ee $n$-point density $t_\smn$ in
\rf{Tndef} depends on light cone momenta $\beta_a$, transverse
momenta $p_a^I$ and Grassmann momenta $\lambda_a$. It is easy to
check that the invariance requirement of the 4-point
$T_\smf$-matrix with respect to $11d$ Poincar\'e supersymmetries
leads to equations for the 4-point density $t_\smf$, which
coincide with equations for 4-point interaction vertex $p_\smf^-$
restricted to the energy surface. Therefore the general solution
obtained for interaction vertex \rf{p4fin} can be used to obtain
solution to the $T$-matrix density $t_\smf$. All that remains is
to find suitable function $g(s,t,u)$. The function $g(s,t,u)$
corresponding to 4-point scattering amplitude is fixed uniquely by
the following two requirements:

\begin{itemize}

\item  The scattering amplitude should has simple poles in
Mandelstam variables;

\item  The superspace representation for the tree level 4-point
scattering amplitudes being restricted to a sector of bosonic
fields should has homogeneity of degree 2 in momenta $p^I_a$ and
$\beta_a$.

\end{itemize}
These two requirements can be easily satisfied by choice $g(s,t,u)
=-\kappa^2/(12stu) $ in \rf{p4fin03}. This leads to the following
compact superspace representation for the tree level 4-point
density $t_\smf$:
\be\label{t4poi} t_\smf =  \frac{\kappa^2}{12}\Bigl(\frac{({\bf
J}_{12}{\bf J}_{34})^2}{s}E_s + \frac{({\bf J}_{13}{\bf
J}_{24})^2}{u}E_u + \frac{({\bf J}_{14}{\bf
J}_{23})^2}{t}E_t\Bigr)\,.\ee
Overall coefficient is fixed so that the 4-point scattering
amplitude for graviton field obtainable from \rf{t4poi} coincides
with the standard graviton 4-point scattering amplitude of
Einstein-Hilbert action \rf{geact}.

Above-given expression for 4-point $T$-matrix can be used for
deriving representation for 4-point scattering amplitude in
various bases of in-states. For example consider representation of
quantized superfield \rf{Phiphirel} in terms of creation and
annihilation operators:
\be\label{phiaabar} \phi(p,\lambda) =
\frac{\epsilon(\beta)}{\sqrt{2\beta}}\, \bar a(p,\lambda) +
\frac{\epsilon(-\beta)}{\sqrt{-2\beta}}\, a(-p,-\lambda)\,, \ee
where $\epsilon(\beta)=1(0)$ for $\beta
>0(\beta < 0)$ and the creation $a(p,\lambda)$ and the annihilation
operators
$\bar a(p,\lambda)$ satisfy the standard commutator
\be [\bar a(p,\lambda),a(p',\lambda')]
=\delta^{10}(p-p')\delta^8(\lambda - \lambda')\,. \ee
Introducing then the basis of ingoing superparticles
$N(\beta)a(p,\lambda)|0\rangle$ we get for the tree level 4-point
amplitude ${\cal A}_\smf$ defined by formula\footnote{We use the
normalization $N(\beta)\equiv (2\pi)^5\sqrt{2\beta}$. Basis of
outgoing superparticles is defined by $\langle 0|\bar
a(-p,-\lambda)N(-\beta)$, where $\beta < 0$.}
\be\label{ampdef} \langle 3,4\,|\, T_\smf \, | 1,2\rangle =
(2\pi)^{10}\delta^{11,8} {\cal A}_\smf \ee the following
superspace representation
\be {\cal A}_\smf =  4!\, t_\smf \,,\ee where $t_\smf$ is given in
\rf{t4poi} and we use the notation
\be\label{d11d8} \delta^{11,8} \equiv \delta^{11}(\sum_{a=1}^4
p_a)\delta^8(\sum_{a=1}^4 \lambda_a)\,.\ee

Conventional scattering amplitude for component fields can be
obtained from the above-given superspace amplitude in a
straightforward way. Consider explicit representation of quantized
fields in terms of polarizations which, say for graviton field,
takes the form
\be\label{hijpoldec} h^{ij}(p) =
\frac{\epsilon(\beta)}{\sqrt{2\beta}}\,\zeta_{_A}^{ij}(p) \bar
a_{_A}(p) + \frac{\epsilon(-\beta)}{\sqrt{-2\beta}}\,
\zeta_{_A}^{ij*}(-p)a_{_A}(-p)\,, \ee
where $\zeta_{_A}^{ij}$ is a basis of graviton polarization states
and summation over polarization states counted by subindex $A$ is
assumed. Using for the quantized $11d$ supergravity fields a
representation similar to that given in \rf{hijpoldec} we get the
conventional 4-point scattering amplitude
\be A_4 = \int \delta^8(\sum_{a=1}^4 \lambda_a) \prod_{a=1}^4
d^8\lambda_a\,\,\prod_{a=1}^4 \phi_{pol}(p_a,\lambda_a)\, {\cal
A}_4\,,\ee
where a superfield $\phi_{pol}(p,\lambda)$ is obtainable from
\rf{supfield} by replacing the quantized fields by appropriate
polarization vectors.

({\bf v}) Because the procedure of our derivation is algebraic in
nature the result of this section can be easily extended to other
supersymmetric theories, which have light cone formulation with
manifest $so(d-4)$ transverse symmetry. In section \ref{SYMsec} we
discuss such extension to the case of $10d$ SYM theory.

\newsection{Superfield form of vertex operators}\label{PARsec}

In the preceding sections we have treated {\it field theoretical}
description of $11d$ supergravity. Alternative approach to study
various aspects of interacting fields is based on usage of
technique of vertex operators. Because such a technique turns out
to be fruitful and sometimes is preferable in some
applictaions\footnote{For instance, world line representation for
interaction vertex of particle with $11d$ supergravity in terms of
components fields \cite{gre1} was used to analyze loop corrections
of $11d$ supergravity \cite{gre1,gre2}. Applications of world line
approach (sometimes referred to as string inspired formalism) to
discussion of UV divergences in gauge theories was discussed in
\cite{mettse,Bern:1987tw}. Review and extensive list of references
may be found in \cite{Bern:1992ad}-\cite{Avramis:2002xf}.} we
would like to reformulate our result in terms of vertex operators.
Thus our goal in this section is to find linearized interaction
vertices of superparticle with fields of $11d$ supergravity. As
before we prefer to formulate our results entirely in terms of the
{\it unconstrained scalar superfield}\footnote{Discussion of light
cone world line representation for interaction vertices of
particles with $11d$ supergravity  in terms of {\it components
fields} may be found in \cite{gre1}. Thus as compared to this
references we formulate our results entirely in terms of
superfield. Also we do not exploit the widely adopted constraint
$k^+=0$.}.

In order to explain the setup we are going to use to study
superparticle vertices let us start with discussion of an
interaction vertex of bosonic particle with Maxwell field. In
phase space approach an action of free particle takes the form
\be S_{free} = \int d\tau \LL_{free}\,,\qquad  \LL_{free}
=\PP_\mu\dot X^\mu -\frac{1}{2}e \PP_\mu\PP^\mu\,, \ee where
$X^\mu(\tau)$ and $\PP_\mu(\tau)$ are coordinates and momenta of
particle, while $e$ is 1d metric tensor of the particle world
line. Phase space equations of motion take then the form
\be\label{parcovequmot} \dot X^\mu = e \PP^\mu\,, \qquad
\dot\PP^\mu =0\,, \qquad \PP^2 =0\,. \ee In light cone gauge
\be\label{parlcgau} X^+(\tau) = \tau \ee we get solution to
equations of motion
\be\label{Xsolequ1} X^I(\tau)= x^I+\frac{p^I}{p^+}\tau\,, \qquad
X^-(\tau)= x^-+\frac{p^-}{p^+}\tau\,, \ee

\be\label{Psolequ1} \PP^I(\tau)= p^I\,, \qquad  \PP^+(\tau) =
p^+\,, \qquad \PP^-(\tau) = p^-\,,\ee

\be\label{esolequ1} e(\tau)=\frac{1}{p^+}\,, \qquad p^- \equiv
-\frac{p^Ip^I}{2p^+}\,.\ee

In covariant approach an interaction of particle with spin 1
massless field is described by action
\be\label{Sint} S_{int}=\int d\tau\,
\phi_\mu(X)\dot{X}^\mu(\tau)\,. \ee Exploiting on mass shell
condition for spin 1 massless field $\phi^\mu$ taken in light cone
gauge
\be \Box \phi^I=0\,, \qquad \phi^+=0\,, \qquad
\phi^-=-\frac{\partial_{x^I}}{\partial_{x^-}}\phi^I \ee we get the
standard representation for solution to equations of motion
\be \phi^I(X^+,X)=\int \frac{d^{d-1}k}{(2\pi)^{(d-1)/2}}\, e^{{\rm
i}k^\mu X^\mu}\, \phi^I(k)\,, \ee where on mass-shell condition in
momentum representation takes the form
\be\label{onsh1} k^-=-\frac{k^Ik^I}{2k^+}\,. \ee Plugging above
given solutions of particle and field equations of motion into
$S_{int}$ we get the following light cone Hamiltonian (sometimes
to be referred to as interaction vertex) describing interaction of
particle with spin 1 massless field:
\be\label{vecvlc} P_{int}^-= \int
\frac{d^{d-1}k}{(2\pi)^{(d-1)/2}}\, e^{{\rm i}k^\mu X^\mu}
\phi^I(k) \frac{\Po^I}{p^+k^+}\,, \ee where `momenta' $\Po^I$ is
defined by (cf. \rf{defpi})
\be\label{Popar} \Po^I \equiv p^Ik^+ - k^I p^+\,. \ee

\medskip
Now let us turn to $11d$ supergravity. Our goal is to find of
analog of $P_{int}^- $ \rf{vecvlc} for the case of $11d$
superparticle interacting with $11d$ supergravity fields. To this
end we could use covariant approach and find an interaction vertex
by considering a particle approximation of world volume action of
membrane interacting with fields of $11d$ supergravity. Such
approach however is very complicated and is not useful in
practical calculations because of basically the following two
reasons: i) covariant $11d$ supergravity superfields involve terms
of 32 powers in fermionic coordinates and have therefore very
complicated structure\footnote{Progress in explicit description of
expansion $11d$ supergravity superfields in fermionic coordinates
was achieved very recently in \cite{Tsimpis:2004gq}.}; ii)
tractable quantization of Green-Schwartz superparticle action is
available only in light cone gauge. Because light cone approach
allows us to avoid these troubles of covariant approach it seems
reasonable to use light cone approach from the very beginning.
This is what we are doing below. We will start directly with light
cone representation and find the interaction vertex by exploiting
requirement of invariance with respect to Poincar\'e superalgebra.

Bosonic body of $11d$ superparticle light cone phase
space\footnote{Discussion of interrelation of this light cone
superspace and covariant superspace may be found in \cite{gre1}.}
consists of  coordinates $X^-(\tau)$, $X^I(\tau)$ and momenta
$\PP^I(\tau)$, $\PP^+(\tau)$ given in \rf{Xsolequ1},\rf{Psolequ1}.
Odd part of the light cone phase space of $11d$ superparticle
involves eight fermionic Grassmann coordinates $\theta(\tau)$ and
eight fermionic momenta $\lambda(\tau)$, which satisfy the
equations of motion
\be\label{parsol} \dot{\theta}(\tau)=0\,,\qquad
\dot{\lambda}(\tau)=0\,. \ee Obvious solution to these equations
is fixed to be
\be\label{parsol1} \theta(\tau)= \theta\,,\qquad
\lambda(\tau)=\lambda\,. \ee

Before going into details of deriving interaction vertex of
superparticle with $11d$ supergravity fields we present our final
result.

Hamiltonian describing interaction of the superparticle with the
supergravity fields is found to be\footnote{Without loss of
generality we consider the interaction vertex and generators of
Poincar\'e superalgebra for $\tau = 0$. Interaction vertex for
arbitrary value of $\tau$ can be easily obtained by using solution
to equations of motion for superfield \rf{Phiphirel} and
superparticle \rf{Xsolequ1}-\rf{esolequ1},\rf{parsol1}.}
\be\label{parvlc1} P_{int}^- = \int\, d\Gamma\, e^{\Omega}\,
\Phi(k,\chi)\, p_{int}^- \,, \ee where $\Phi(k,\chi)$ is the light
cone superfield with expansion in components fields of $11d$
supergravity given in \rf{supfield}. Measure $d\Gamma$ and
quantity $\Omega$ in \rf{parvlc1} are defined by formulas
\be\label{Omedef} \Omega \equiv {\rm i} (k x)-\chi\theta\,, \qquad
(kx)\equiv k^+ x^- + k^Ix^I\,,\ee

\be d\Gamma = \frac{dk^+d^9k}{(2\pi)^5} d^8\chi \ee and the
interaction vertex $p_{int}^-$ is fixed to be
\beq \label{pintfin} \frac{p^+}{\kappa} p_{int}^- &=& \Po^{L2}
-\frac{\Po^L}{2\sqrt{2}\hat{\beta}}\Lambda\bpline\Lambda
+\frac{1}{16\hat{\beta}^2}(\Lambda\bpline\Lambda)^2
-\frac{|\Po|^2}{9\cdot 16\hat{\beta}^2}(\Lambda\gamma^i\Lambda)^2
\nonumber\\
&+&\frac{\Po^R}{9\cdot 16\sqrt{2}\hat{\beta}^3}
\Lambda\bpline\Lambda(\Lambda\gamma^j\Lambda)^2
+\frac{\Po^{R2}}{2^7\cdot 63\hat{\beta}^4}
((\Lambda\gamma^i\Lambda)^2)^2\,, \eeq where the `momentum'
$\Po^I$ is given in \rf{Popar}, while the quantities $\Lambda$ and
$\hat\beta$ are defined by relations (cf. \rf{Lambda},\rf{hatbet})
\be\label{Lampar} \Lambda \equiv \lambda k^+ - \chi p^+\,,\ee

\be\label{hatbpar} \hat{\beta} \equiv -p^{+2}k^+\,.\ee  In formula
\rf{pintfin} $\kappa$ is gravitational constant and we choose
normalization so that the interaction vertex for the graviton
obtainable from \rf{parvlc1} coincides with that of the standard
action of particle interacting with graviton
\be\label{parGRA} S_{int} = \int d\tau\, \frac{1}{2e}\,
g_{\mu\nu}(X)\dot X^\mu \dot X^\nu\,,\ee where the expansion for
metric tensor given in \rf{gmnexp} and the light cone gauge (see
\rf{lcgau}, \rf{nondyn}, \rf{parlcgau}-\rf{Psolequ1}) should be
used.

The expression for the vertex \rf{pintfin} coincides with that of
field theoretical approach given in \rf{exprep} by module of
definitions of the quantities $\Po^I$, $\Lambda$, $\hat\beta$. In
field theoretical vertex \rf{exprep} we should exploit expressions
for $\Po^I$, $\Lambda$, $\hat\beta$ given in
\rf{defpi},\rf{Lambda},\rf{hatbet}, while appropriate quantities
for vertex \rf{pintfin} are defined in
\rf{Popar},\rf{Lampar},\rf{hatbpar}. Therefore the vertex
\rf{pintfin} can be also written in terms of $E$-operators:
\be \label{finrespar} p_{int}^-  = \frac{\kappa}{p^+}\, {\Po
}^{L\,2}E_qE_\rho\,. \ee As compared to \rf{exprep} the vertex
\rf{finrespar} involves an extra factor $p^+$. Appearance of this
factor is related to light cone gauge on superparticle coordinate
$X^+(\tau)$ given in \rf{parlcgau}.

\subsection{Restrictions imposed by kinematical symmetries}

As before to derive above-given linearized interaction vertex of
superparticle with supergravity fields \rf{pintfin} we use the
general approach of Ref.\cite{dir}. This is to say that both the
particle phase variables and supergravity fields are considered to
be dynamical variables and we require this dynamical system
respects symmetries of $11d$ Poincar\'e superalgebra\footnote{In
widely used alternative approach the superparticle variables are
treated as the dynamical variables, while the supergravity fields
are considered as external (background) fields. In this approach
variation of vertices under supersymmetry transformation of
superparticle variables is replaced by supersymmetry
transformation of the supergravity fields. Various application of
this approach may be found in \cite{Green:1983sg,gre1,ple2}.}. As
before following setup of Ref.\cite{dir} we should find
realization of commutation relations of Poincar\'e superalgebra
for the dynamical system involving superparticle and supergravity
fields. Below following this setup we demonstrate that
restrictions imposed by Poincar\'e supersymmetries and requirement
of light cone locality allows us to fix linearized interaction
vertex uniquely.

Let $G_{part}$ and $G_{field}$ be respective generators of the
Poincar\'e superalgebra for the free superparticle and
supergravity fields, while $G_{int}$ be generators responsible for
linearized interaction of particle with supergravity fields. The
generators of the system superparticle plus supergravity are given
then by
\be \label{gtot} G\equiv G_{particle} + G_{field} + G_{int}\,. \ee
The generators of free supergravity fields $G_{field}$ are given
in \rf{fierep},\rf{pp}-\rf{qml}. To apply our method we need
explicit expressions for free superparticle generators
$G_{particle}$. These generators can be obtained in a rather
straightforward way by using standard methods of the Noether
procedure. We present therefore expressions for the generators
without derivation\footnote{The superparticle generators are
normalized so that if we plug a representation for
(super)coordinates implied by \rf{xpbra} into the superparticle
generators then these generators coincide with those of Section 2.
The coincidence is true by module of some terms generated by
ordering procedure of coordinates and momenta operators in angular
generators $J^{\mu\nu}$ of Section 2.}:
\be P^-=p^-,\qquad P^+=p^+\,, \qquad P^I=p^I\,, \qquad
p^-=-\frac{p^Ip^I}{2p^+} \ee
\be J^{+I}=-{\rm i}x^I p^+\,, \ee
\beq && J^{+-}=-{\rm i}x^-p^+-\frac{1}{2}\theta\lambda\,,
\\
&& J^{ij}={\rm i}(x^ip^j-x^jp^i)
+\frac{1}{2}\theta\gamma^{ij}\lambda\,,
\\
&& J^{RL}={\rm i}(x^R p^L-x^L p^R) +\frac{1}{2}\theta\lambda\,,
\\
&& J^{Ri}={\rm i}(x^R p^i-x^i p^R) -\frac{p^+}{2\sqrt{2}}
\theta\gamma^i\theta\,,
\\
&& J^{Li}={\rm i}(x^L p^i-x^i p^L)
+\frac{1}{2\sqrt{2}p^+}\lambda\gamma^i\lambda\,, \eeq
\beq  && J^{-R}={\rm i}(x^- p^R - x^R p^-)
-\frac{1}{2\sqrt{2}}\theta\pline \theta
+\frac{p^R}{p^+}\theta\lambda\,,
\\[5pt]
&& J^{-L}= {\rm i}(x^- p^L  - x^L p^-)
+\frac{1}{2\sqrt{2}p^{+2}}\lambda\pline\lambda\,,
\\[5pt]
&& J^{-i}={\rm i}(x^- p^i - x^i p^-)
+\frac{1}{2p^+}\theta\gamma^i\pline \lambda
-\frac{p^R}{2\sqrt{2}p^{+2}}\lambda\gamma^i\lambda
+\frac{p^L}{2\sqrt{2}}\theta\gamma^i\theta\,, \ \ \ \ \eeq
\beq && Q^{+R}=p^+\theta\,,
\\[5pt]
&& Q^{+L}=\lambda\,,
\\[5pt]
&& Q^{-R}=\frac{1}{\sqrt{2}}\theta\pline+\frac{p^R}{p^+}\lambda\,,
\\[5pt]
&& Q^{-L} = p^L\theta + \frac{1}{\sqrt{2}p^+}\pline\lambda\,. \eeq

The superparticle coordinates and momenta satisfy Poisson brackets
that are normalized to be\footnote{Note that in spite of
appearance of imagine unity ${\rm i}$ the brackets \rf{xpbra}
stand for the {\it classical} Poisson brackets. Such `quantum'
normalization of the classical Poisson brackets is convenient for
us.}
\be\label{xpbra} [x^I,p^J]={\rm i}\delta^{IJ},\qquad
[x^-,p^+]={\rm i}\,,\qquad \{\theta,\lambda\}= 1\,.\ee

Now we are going to derive restrictions imposed by the kinematical
symmetries. By definition, these restrictions are obtainable from
commutation relations of Hamiltonian $P_{int}^-$ with kinematical
generators of $11d$ Poincar\'e superalgebra \rf{kingen}. To
elucidate the procedure of constructing interaction vertex let us
analyze the commutation relations  of $P_{int}^-$ with the various
kinematical generators in turn.

i) Firstly, making use of the commutation relations of $P_{int}^-$
with $P^I$, $P^+$, and $Q^{+L}$ in linearized approximation we
find that a dependence of interaction vertex \rf{parvlc1} on the
bosonic coordinates $x^I$, $x^-$ and Grassmann coordinates
$\theta$ enters throughout the quantity $\Omega$ \rf{Omedef}.

ii) Secondly, we analyze restrictions imposed by commutation
relations of $P_{int}^-$ with generators $J^{+I}$, $Q^{+R}$. To
this end we evaluate the commutators
\beq\label{pintjcom} &&  [P_{int}^-,J^{+I}] = - \int d\Gamma\,
e^\Omega \,\Phi(k,\chi)\, \Bigl(k^+\partial_{k^I} +
p^+\partial_{p^I}\Bigr) p_{int}^-\,,
\\[6pt]
&&\label{pintqcom}  [P_{int}^-,Q^{+R}] =-\int d\Gamma\, e^\Omega\,
\Phi(k,\chi)\Bigl(k^+\partial_\chi+p^+\partial_\lambda\Bigr)p_{int}^-\,.
\eeq
Because kinematical generators $P^I$ do not get interaction
corrections commutators of $11d$ Poincar\'e superalgebra
$[P^-,J^{+I}]=P^I$, $[P^-,Q^{+R}]=0$ leads to commutators
$[P_{int}^-,J^{+I}]=0$, $[P_{int}^-,Q^{+R}]=0$. The latter
commutators and formulas \rf{pintjcom},\rf{pintqcom} give then
equations
\be (k^+\partial_{k^I} + p^+\partial_{p^I}) p_{int}^- =0\,,\qquad
(k^+\partial_\chi+p^+\partial_\lambda)p_{int}^-=0\,. \ee
These equations tell us that the vertex $p_{int}^-$ can be
presented as
\be \label{vlc2} p_{int}^-= p_{int}^-({\Po }, \Lambda,p^+,k^+)\,.
\ee
In other words $p_{int}^-$ depends on $p^I$, $k^I$, $\lambda$ and
$\chi$ throughout the `momenta' ${\Po }^I$ and $\Lambda$, which
are defined in \rf{Popar},\rf{Lampar}.

iii) Commutator $[P^-,J^{IJ}]=0$ gives equations
\be\label{jijequ1} J^{IJ}(\Po,\Lambda)\, p_{int}^- =0\,,\qquad
J^{IJ}(\Po,\Lambda) \equiv L^{IJ}(\Po)+M^{IJ}(\Lambda)\,, \ee
where the orbital operator $L^{IJ}({\Po })$ and spin operators are
given by expressions \rf{LIJ}-\rf{MIJ3} in which we should use
${\Po }^I$, $\Lambda$, $\hat{\beta}$ given in \rf{Popar},
\rf{Lampar} and \rf{hatbpar} respectively.

Remaining kinematical symmetry related with commutation relation
of $P^-$ with $J^{+-}$ gives equation
\be\label{jpmequ1} (J^{+-}(\Po,\Lambda)+1)p_{int}^- = 0\,,\ee
where we use the notation
\be J^{+-}(\Po,\Lambda) \equiv p^+\partial_{p^+}
+k^+\partial_{k^+} + \Po^I\partial_{\Po^I}
+\frac{3}{2}\Lambda\partial_\Lambda-2\,. \ee

Because Eqs.\rf{jijequ1} take the same form as those given in
\rf{JIJp3} we can apply the same procedure we exploited while
solving Eqs.\rf{JIJp3}. Introducing the variables $q^i$, $\rho$ as
in \rf{newvar} we get from Eqs.\rf{jijequ1} the following solution
\be \label{rep1} p_{int}^-(\Po,\Lambda,p^+,k^+) = {\Po }^{L\,2}
E_qE_\rho \tilde{V}_0(p^+,k^+)\,, \ee where in expressions for
$E_q$ \rf{eq}, $E_\rho$ \rf{erho}, $M^{Li}(\Lambda)$ \rf{MIJ3} we
have to insert appropriate values of $\Po^I$, $\Lambda$,
$\hat{\beta}$ given in \rf{Popar}, \rf{Lampar}, \rf{hatbpar}. In
Eq.\rf{rep1} we have extracted dimesionfull factor ${\Po }^{L2}$
which is appropriate for supergravity theories.

One can check that Eq.\rf{jpmequ1} leads to the following equation
for vertex $\tilde{V}_0(p^+,k^+)$
\be\label{remequ1}
(p^+\partial_{p^+}+k^+\partial_{k^+}+1)\tilde{V}_0(p^+,k^+)
=0\,.\ee Thus as in the case of field theoretical supergravity
interaction vertices we see that the kinematical symmetries admit
to fix dependence of particle interaction vertex on the `momenta'
$\Po^I$ and $\Lambda$ completely, while the dependence on two
light cone momenta $p^+$, $k^+$ is restricted only by one equation
\rf{remequ1}, i.e. to fix dependence on $p^+$, $k^+$ completely we
need one additional equation. Such equation can be obtained by
exploiting commutation relations between the dynamical generators
and using the locality requirement. This will be done in next
section.

\subsection{Nonlinear symmetries and locality requirement}

Systematic procedure of fixing dependence of interaction vertex on
light cone momenta $p^+$, $k^+$ is based on study commutation
relations between dynamical generators \rf{dyngen}. Another reason
for study of the dynamical generators is that the light cone gauge
breaks manifest Lorentz symmetries and in order to check that
these symmetries still present one needs to construct all
generators and make sure that all commutation relations are
satisfied. This is that what we are doing in this section.

The kinematical generators \rf{kingen} do not receive interaction
corrections, while the dynamical generators depend on interaction.
Making use of commutation relations between the dynamical
generators and kinematical generators one can make sure that
interaction corrections to the dynamical generators take the
following form
\beq \label{dyngen52}&& Q_{int}^{-R} =\int d\Gamma\, e^\Omega
\Phi(k,\chi)q_{int}^{-R}\,,
\\[6pt]
\label{dyngen51} && Q_{int}^{-L} =\int d\Gamma\, e^\Omega
\Phi(k,\chi)q_{int}^{-L}\,,
\\[6pt]
\label{dyngen54}&& J_{int}^{-R} =-{\rm i}x^RP_{int}^- +\theta
Q_{int}^{-R}+\int d\Gamma\, e^\Omega \Phi(k,\chi)j_{int}^{-R}\,,
\\[6pt]
\label{dyngen53}&& J_{int}^{-L} =-{\rm i}x^LP_{int}^-
+\frac{\lambda}{p^+}Q_{int}^{-L}+\int d\Gamma\, e^\Omega
\Phi(k,\chi)j_{int}^{-L}\,,
\\[6pt]
\label{dyngen55}&& J_{int}^{-i} =-{\rm i}x^iP_{int}^- +
\frac{1}{\sqrt{2}}\theta\gamma^i
Q_{int}^{-L}-\frac{\lambda}{\sqrt{2}\,p^+}\gamma^i
Q_{int}^{-R}+\int d\Gamma\, e^\Omega \Phi(k,\chi)j_{int}^{-i}\,,
\eeq
where densities ${\cal
X}=(q_{int}^{-R},q_{int}^{-L},j_{int}^{-I})$ depend on `momenta'
$\Po^I$, $\Lambda$ and light cone momenta $p^+$, $k^+$:
\be {\cal X}={\cal X}(\Po,\Lambda; p^+,k^+)\,.\ee

With expression for dynamical generators
\rf{dyngen52}-\rf{dyngen55} at our hands we are ready to study
commutators between dynamical generators. We proceed as follows.

i) Making use of commutators $[P^-,Q^{-L}]=0$, $[P^-,Q^{-R}]=0$ we
get the following relations for the supercharge densities
\be\label{for56} q_{int}^{-L} =
-\frac{2\hat{\beta}}{|\Po|^2}Q^{-L}(\Lambda)p_{int}^-\,, \qquad
q_{int}^{-R} = -\frac{2\hat{\beta}}{|\Po|^2}Q^{-R}(\Lambda)
p_{int}^-\,,\ee
where we use the notation for differential operators:
\beq \label{for57a} && Q^{-L}(\Lambda)\equiv \Po^L\partial_\Lambda
+ \frac{1}{\sqrt{2} \hat{\beta}}\bpline\Lambda\,,
\\[8pt]
\label{for57b}  && Q^{-R}(\Lambda) \equiv
\frac{1}{\sqrt{2}}\partial_\Lambda\bpline
+\frac{1}{\hat{\beta}}\Po^R\Lambda \,.\eeq

ii) Making use of these formulas we get from the commutators
$[P^-,J^{-I}]=0$ the following expression for the density
$j_{int}^{-I}$:
\be \label{for58}
 j_{int}^{-I} = -p^+\partial_{\Po^I}p_{int}^- -
\frac{2p^+\Po^I}{|\Po|^2}(p^+\partial_{p^+}+\Lambda\partial_\Lambda+1)p_{int}^-\,.
\ee {}From this formula we see that for the density $j_{int}^{-I}$
to satisfy the locality requirement we should impose on the
interaction vertex the following equation
\be\label{for59}
(p^+\partial_{p^+}+\Lambda\partial_\Lambda+1)p_{int}^-=0\,. \ee
Taking into account the formula \rf{rep1} we find that equation
\rf{for59} leads to the following equation for the vertex
$\tilde{V}_0(p^+,k^+)$:
\be \label{for60}(p^+\partial_{p^+} +
1)\tilde{V}_0(p^+,k^+)=0\,.\ee {}Solution to this equation and
Eq.\rf{remequ1}  is fixed to be

\be \tilde{V}_0(p^+,k^+)= \frac{\kappa}{p^+}\,,\ee where $\kappa$
is the gravitational constant. In this formula a normalization is
chosen so that to respect the normalization of the standard action
of particle interacting with gravity \rf{parGRA}. Note that
Eq.\rf{for59} leads to the following simplified  expression for
$j_{int}^{-I}$ \rf{for58}:
\be j_{int}^{-I} = -p^+\partial_{\Po^I}p_{int}^-\,. \ee To
complete light cone description we should write down expressions
for the supercharge densities $q_{int}^{-R}$, $q_{int}^{-L} $. The
expression for densities given in \rf{for56} are still to be
formal because they are nonlocal in $\Po^I$. However these
expressions become local in transverse `momenta' $\Po^I$ once we
insert explicit solution to interaction vertex $p_{int}^-$.
Comparison with calculations of field theory densities made in
Section \ref{QchaCUB} allows us to write down expressions for
densities $q_{int}^{-R}$, $q_{int}^{-L}$ in a rather
straightforward way. Indeed confronting Eqs.\rf{for56} with
Eqs.\rf{qQp} and formula \rf{pintfin} with \rf{exprep} we see that
$q_{int}^{-R}$, $q_{int}^{-L}$ can be obtained by making the
following substitutions in formulas \rf{qlfin} \rf{qrfin}: (i) in
l.h.s. of expressions \rf{qlfin} \rf{qrfin} the factor $3/\kappa$
should be replaced by $p^+/\kappa$; (ii) in r.h.s. of expressions
\rf{qlfin} \rf{qrfin} we should insert the expressions for
$\Po^I$, $\Lambda$, $\hat\beta$ given in \rf{Popar},\rf{Lampar},
\rf{hatbpar} respectively.

\newsection{Cubic and 4- point interaction
vertices of $10d$  SYM theory}\label{SYMsec}

In this section we would like to extend our analysis to the case
of ten dimensional SYM theory. There are two approaches to
superfield light cone description of this theory. One of them
keeps manifest $so(8)$ symmetries and is based on {\it constrained
vector} superfield \cite{bgs}. In alternative approach, we use
below, the $so(8)$ symmetries are reduced to the manifest $so(6)$
symmetries and the action is formulated in terms of {\it
unconstrained scalar} superfield \cite{GS6}. We prefer to use the
latter approach because in this approach a representation of
Poincar\'e superalgebra generators is similar to that of $11d$
supergravity. Due to that we can straightforwardly generalize our
results to the SYM case.

Our derivation for cubic and 4-point supergravity vertices was
essentially algebraic because this derivation was based
significantly on the usage of Poincar\'e superalgebra generators
given in \rf{pp}-\rf{qml}. Note that similar generators appear in
$IIA$, $10d$ supergravity and, with some minor modification,  in
$10d$ SYM theory. As to $IIA$ supergravity the generalization of
our results to this theory is trivial and can be achieved simply
by using dimensional reduction. All that we have to do is to set
one of transverse momenta equal zero in expressions for cubic
vertices \rf{finres} (or \rf{exprep}) and \rf{p4fin}  Now let us
turn to $10d$ SYM theory.

To extend our supergravity results to the case of SYM theory we
should make the following modifications. First of all we have to
set one of transverse momenta, say $p^7$, equal to zero. Next step
is to replace odd part of light cone superspace, i.e. $\lambda$
(and $\theta$) by appropriate Grassmann variables of SYM theory.
Namely, instead of $\lambda$ (and $\theta$) transforming in spinor
representation of the $so(7)$ algebra we introduce $\lambda_A$
(and $\theta^A$), $A=1,2,3,4$, which transform in
(anti)fundamental, i.e. vector, representation of the $su(4)$
algebra. This modification reflects the fact that $10d$ SYM theory
involves $16$ supercharges instead of $32$ supergravity
supercharges. Appropriate unconstrained light cone scalar
superfield has the following expansion in powers of Grassmann
momenta $\lambda_A$
\beq \Psi(p,\lambda)=\beta \phi^L + \lambda_A\psi^A +
\lambda_A\lambda_B\phi^{AB} +\frac{1}{\beta}(\epsilon \lambda^3)^A
\psi_A  - \frac{1}{\beta}(\epsilon \lambda^4)\phi^R\,, \eeq where
we use the notation
\be\label{epsl3} (\epsilon \lambda^3)^A \equiv
\frac{1}{3!}\epsilon^{AA_1A_2A_3}\lambda_{A_1}\lambda_{A_2}\lambda_{A_3}\,,
\qquad (\epsilon \lambda^4) \equiv
\frac{1}{4!}\epsilon^{A_1A_2A_3A_4}\lambda_{A_1}\ldots
\lambda_{A_4}\,. \ee All fields have indices of the suitable  Lie
algebra, which we do not show explicitly. The self-dual field
$\phi^{AB}$ of the $su(4)$ algebra can be related with vector
field $\phi^i$ of the $so(6)$ algebra in a standard way
\be\label{phiabphii} \phi^{AB} = \frac{1}{2\sqrt{2}}\rho^{i\, AB}
\phi^i \,,\ee where normalization of the corresponding
Clebsch-Gordan coefficients (or Dirac matrices\footnote{These
matrices satisfy the standard relation $\rho^{iAC}\rho_{CB}^j +
(i\leftrightarrow j) = 2\delta^{ij}\delta^A_B$ and we adopt
complex conjugation rule $\rho^{iAB*}=-\rho^i_{AB}$. Note that our
sign convention in \rf{rhonor1} differs from that adopted in
formulas (A.7), (A.9) in Ref.\cite{GS6}. In this section we use
$so(6)$ vector indices $i,j=1,\ldots,6$, $so(8)$ vector indices
$I,J = 1,\ldots, 6, R,L$ and $su(4)$ vector indices
$A,B=1,\ldots,4$.}) is chosen so that the following relations are
true
\be\label{rhonor1} \rho_{[AB}^i\rho_{CD]}^j =
\frac{1}{3}\epsilon_{ABCD}\delta^{ij}\,,\qquad \rho_{{AB}}^i
=-\frac{1}{2}\epsilon_{ABCD}\rho^{i\, CD}\,.\ee Reality condition
for vector field $\phi^i(-p) = \phi^{i*}(p)$ and relation
\rf{phiabphii} lead to self-duality condition for $\phi^{AB}$:
\be \phi^{AB}(p) =\frac{1}{2}\epsilon^{ABCD} (\phi^{CD}(-p))^*\,.
\ee Fields $\phi^R$, $\phi^L$ ($\phi^L(-p)=\phi^{R*}(p)$) and
$\phi^i$ describe eight bosonic physical d.o.f of SYM,  while
$\psi^A$ are fermionic fields subject to hermitian conjugation
rule $\psi^A(-p)=(\psi_A(p))^\dagger$. Reality condition in terms
of superfield $\Psi$ takes the form
\be \Psi(-p,\lambda) = \beta^2 \int d\lambda^\dagger\,
e^{\lambda\lambda^\dagger/\beta}(\Psi(p,\lambda))^\dagger\,. \ee
Superfield $\Psi$ satisfies the following commutation
relation\footnote{Grassmann delta function is chosen to be
$\delta^4(\lambda)\equiv (\epsilon \lambda^4)$ (see \rf{epsl3}).
Accordingly an integral over Garssman variables is normalized to
be $\int d^4\lambda \delta^4(\lambda)=1$.}
\be [\Psi(p,\lambda),\Psi(p',\lambda')]|_{_{equal\, x^+}} =
\frac{1}{2\beta} \delta^{9}(p + p')\delta^4(\lambda + \lambda')\ee
which implies that the component fields should satisfy the
commutation relations
\beq &&   [\phi^I(p),\phi^J(p')] = \frac{1}{2\beta} \delta^{9}(p +
p')\delta^{IJ}\,,
\\[6pt]
&& \{ \psi_A(p),\psi^B(p') \} = \frac{1}{2} \delta^{9}(p +
p')\delta_A^B\,. \eeq

In order to get representation of Poincar\'e superalgebra on the
superfield $\Psi$ we should make the following replacements in
expressions for generators given in \rf{pp}-\rf{qml}:

({\bf i}) Scalar product $\theta\lambda$ in \rf{jpm},\rf{jmr}
should be replaced by $\theta\lambda\equiv \theta^A\lambda_A$.

({\bf ii}) Expressions  $\theta\gamma^i\theta$ and
$\lambda\gamma^i\lambda$ in \rf{jri}-\rf{jmi} should be replaced
by $\theta \rho^i \theta \equiv \theta^A\rho^i_{AB}\theta^B$ and
$\lambda \rho^i\lambda \equiv \lambda_A(\rho^i)^{AB}\lambda_B$
respectively.

({\bf iii}) Expression $\theta\gamma^{ij}\lambda$ in \rf{jij}
should be replaced by $\theta^A(\rho^{ij})_A{}^B\lambda_B$, where
\be (\rho^{ij})_A{}^B\equiv \frac{1}{2}\rho^i_{AC} \rho^{j\,CB}
-(i \leftrightarrow j)\,. \ee

({\bf iv}) Last terms in \rf{jpm},\rf{jrl},\rf{jmr},\rf{jmi}, {\it
i.e.} $2$, $-2$, $-4p^R/\beta$, $-2p^i/\beta$ should  be replaced
by $1$, $-1$, $-2p^R/\beta$, $-p^i/\beta$ respectively.

After this we can extend our supergravity calculations to the case
of SYM theory in a rather straightforward way. We note that it is
usage of $E$-operators that allows us to do such straightforward
extension. Below we present interaction vertices of SYM theory
without derivation. Let us first consider the cubic vertices.

The superspace representation of $10d$ SYM theory Hamiltonian in
cubic approximation is given by
\be\label{Ham3fin} P_{(3)}^- = \int d\Gamma_3 \Tr
(\prod_{a=1}^3\Psi(p_a,\lambda_a)) p_\sm3^-({\Po
},\Lambda,\beta)\,, \ee where the $d\Gamma_3$ is obtainable from
\rf{delfun}-\rf{delfun02} by setting $n=3$, $d=10$ and replacing
$d^8\lambda$ by $d^4\lambda$. The cubic vertex is found to
be\footnote{The representation for the cubic vertex in terms of
$E$-operators \rf{ymcub} is result of this paper. Explicit
representation in terms of `momenta' $\Po^I$ and $\Lambda$ (see
Eq.\rf{ymcubexp} below) was found in Ref.\cite{GS6}.}
\be\label{ymcub} p_\sm3^-({\Po },\Lambda,\beta) = -
\frac{2g_{_{YM}}}{3}\, {\Po }^L E_qE_\rho\,, \ee where operators
$E_q$ and $E_\rho$ are given by\footnote{Operator $E_\rho$ can be
obtained  by setting $d=10$ and $k=1$ in the  general solution
given in \rf{erhod}.}
\beq\label{eqym} && E_q=\exp(-q^j M^{Lj} (\Lambda))\,,
\\[5pt]
&& E_\rho\equiv 1 - \frac{\rho}{6}  M^{Lj}(\Lambda)
M^{Lj}(\Lambda)\,,
\\[5pt]
&& M^{Li}(\Lambda) \equiv \frac{1}{2\sqrt{2}\hat{\beta}}\Lambda
\rho^i \Lambda\,. \eeq The quantity $\hat{\beta}$ is given by
\rf{hatbet}, while $\Lambda$ is obtainable from \rf{Lambda}
replacing the $so(7)$ Grassmann momentum $\lambda_\alpha$ by the
$su(4)$ Grassmann momentum $\lambda_A$. The expression for
interaction vertex can be re-written manifestly in terms of
momenta $\Po^I$ and $\Lambda$:
\be\label{ymcubexp}  p_\sm3^-({\Po },\Lambda,\beta) =
-\frac{2g_{_{YM}}}{3}\,\Bigl(\Po^L
-\frac{\Po^i}{2\sqrt{2}\hat{\beta}}\Lambda \rho^i \Lambda -
\frac{\Po^R}{\hat{\beta}^2}(\epsilon \Lambda^4)\Bigr)\,.\ee

Normalization coefficient in the vertex \rf{ymcub} (or
\rf{ymcubexp}) is chosen so that the bosonic body of covariant
action corresponding to the supersymmetric light cone Hamiltonian
\rf{Ham3fin} is given by
\be S_{_{YM}} = \int d^{10}x\, {\cal L}_{_{YM}}\,,\qquad {\cal
L}_{_{YM}}=\Tr -\frac{1}{4} F_{\mu \nu}F_{\mu\nu}\,,\ee where
field strength is defined as
\be F_{\mu \nu} =
\partial_\mu \phi_\nu  -\partial_\nu \phi_\mu + {\rm i} g_{_{YM}}\, [
\phi_\mu, \phi_\nu] \ee and gauge field $\phi_\mu$ is a hermitian
matrix $\phi_\mu^\dagger = \phi_\mu$.

Now let us consider higher derivative 4-point vertices. We are
interested in superspace light cone representation of 4-point
interaction vertices involving in their bosonic sector $\partial^n
F^4$ terms\footnote{Supersymmetric completion of {\it non-abelian}
$F^4$ terms to the second order in fermions was obtained in
\cite{brs}. Full supersymmetric completion of {\it abelian} and
{\it non-abelian} $F^4$ terms was obtained in \cite{met3} and
\cite{Bergshoeff:2001dc} respectively. Supersymmetric action to
all orders in {\it abelian} field strength $F$ and fermions was
found in \cite{Aganagic:1996nn}.}. Following step by step the
procedure we used while deriving supergravity 4-point vertex we
obtain the following 4-point supersymmetric Hamiltonian:
\be \label{Ham4fin} P_{(4)}^- = \int d\Gamma_4 \Tr (\prod_{a=1}^4
\Psi(p_a,\lambda_a)) p_\smf^- ({\Po },\Lambda,\beta)\,, \ee where
the expression for 4-point vertex is fixed to be
\be \label{p4finym} p_{(4)}^-  =  q_{_L}^2 \frac{({\Po
}_{13}^L{\Po }_{24}^L)^2}{\beta_{13}^2}\,   E_{q_{13}}
E_{q_{24}}E_u g(s,t)\,, \ee  and the measure $d\Gamma_4$ is
obtainable from \rf{delfun} by setting $n=4$, $d=10$ and replacing
$d^8\lambda$ by $d^4\lambda$. The variables $q_{_L}^i$ and
$\Lambda^L$ take the same form as in \rf{qi},\rf{defThe}, while
the operators $E_{q_{ab}}$ and $E_u$ are given by
Eqs.\rf{eqab},\rf{eu} in which we have to make replacement above
mentioned in (ii) and (iii). Function $g(s,t)$ depending on
Mandelstam variables $s$, $t$ cannot be fixed by global
supersymmetries. This function should be cyclically symmetric
\be\label{gdua} g(s,t) =g(t,s) \ee  and if we assume that the
function $g(s,t)$ admits Taylor series expansion then lower order
terms of expansion take the form
\be\label{gstexp} g(s,t) = g_0 + g_1 u + g_{2;1} st + g_{2;2} u^2
+ \ldots\,. \ee

As compared to analogous expressions for $11d$ supergravity vertex
\rf{p4fin} we see that $q_{_L}^2$- term which is in front of the
$E$-operators \rf{p4finym} turns out to be a square root of the
corresponding term in the supergravity vertex \rf{p4fin}. This can
be considered to some extent as a sort of Kawai Lewellen Tye
relationship between gravity and gauge theory.

Making use of formulas \rf{plplstdec} it is easy to see that
$q_{_L}^2$- term in \rf{p4finym} is symmetric upon any
permutations of the four external line indices 1,2,3,4. The
product of $E$-operators in \rf{p4finym} is also symmetric upon
such permutations\footnote{This total symmetry of the $q_{_L}^2$-
term and product of the $E$-operators is related to the well known
total symmetry of the kinematic factor $K$ that enters scattering
amplitude of type $I$ superstring theory (see {\it e.g.} section
4.2. in Ref.\cite{Schwarz:1982jn}).}. Making use of these symmetry
properties the 4-point vertex can be cast into more symmetric form
with respect to $s$, $t$ variables. This is to say that
introducing new quantities ${\bf J}_{ab}$:
\be\label{bfJdefym} {\bf J}_{ab} \equiv \Po_{ab}^L
E_{q_{ab}}\,,\ee and exploiting the first relation in
\rf{plplstdec} we can cast the expression \rf{p4finym} into the
form
\beq p_\smf^- & = & \left({\bf J}_{12}{\bf J}_{34} t E_s - {\bf
J}_{14}{\bf J}_{23} s E_t \right)g(s,t)\eeq where operators $E_s$
and $E_t$ are obtainable from $E_u$ in the same way as in
\rf{EsEtdef}.

The function $g(s,t)$ can be expressed in terms of constants that
appear in covariant Lagrangian formulation. Let a bosonic body of
covariant supersymmetric Lagrangian at order $F^4$ is given
by\footnote{Lagrangian \rf{symL} describes $F^4$ corrections to
low energy dynamics of massless spin 1 modes of type I superstring
theory \cite{Tseytlin:1986ti,Gross:1986iv}.}
\beq \label{symL} {\cal L}_{_{F^4}} & = & g_{_{F^4}} W_{_{F^4}}\,,
\\[6pt]
\label{symW}W_{_{F^4}} &\equiv & {\rm Tr}\,\,
F_{\mu\rho}F_{\nu\rho} F_{\mu\sigma}F_{\nu\sigma} +
\frac{1}{2}F_{\mu\rho}F_{\nu\rho} F_{\nu\sigma}F_{\mu\sigma}
\nonumber\\[6pt]
&-& \frac{1}{4}F_{\mu\nu}F_{\mu\nu} F_{\rho\sigma}F_{\rho\sigma} -
\frac{1}{8}F_{\mu\nu}F_{\rho\sigma} F_{\mu\nu} F_{\rho\sigma}\,,
\eeq where $g_{_{F^4}}$ is some constant. Then the corresponding
light cone gauge supersymmetric Hamiltonian is given by
Eqs.\rf{Ham4fin},\rf{p4finym}, where the function $g(s,t)$
\rf{gstexp} is fixed to be

\be\label{g0grel} g(s,t)  = \frac{1}{2}g_{_{F^4}} \,. \ee Details
of derivation this relationship may be found at the end of
Appendix E. The overall constant $g_{_{F^4}}$ in \rf{symL} is
dynamical-dependent and cannot be fixed by global symmetries. For
the case of tree level superstring effective Lagrangian for
massless spin 1 fields this constant can be expressed in terms of
string tension \cite{Tseytlin:1986ti,Gross:1986iv}.

The formula \rf{g0k4} linking constants of bosonic body of
covariant Lagrangian and corresponding light cone gauge
supersymmetric Hamiltonian can easily be generalized to the higher
derivative $\partial^{n} F^4$ terms. Namely, if  bosonic body of
covariant Lagrangian is given by
\beq\label{LfF4} && {\cal L}_{_{f\,F^4}} =  f(s,t) W_{_{F^4}}\,,
\\[7pt]
&&\label{fstexp} f(s,t) = f_0 + f_1 u + f_{2;1} st + f_{2;2} u^2 +
\ldots\,, \eeq then the corresponding light cone gauge
supersymmetric Hamiltonian is given by Eqs.\rf{Ham4fin},
\rf{p4finym}, where the function $g(s,t)$ \rf{gstexp} is fixed to
be\footnote{Note that for establishing a relation \rf{gstfstrel}
we do not need concrete form of expansion for the function
$f(s,t)$ given in \rf{fstexp}. All that is required for derivation
the relation \rf{gstfstrel} is `crossing symmetry' of the function
$f$: $f(s,t)=f(t,s)$.}
\be\label{gstfstrel} g(s,t) = \frac{1}{2 } f(s,t)\,. \ee Concrete
form of the function $f(s,t)$ (and hence $g(s,t)$) is dynamical -
dependent. For the case of superstring theory the coefficients
$f_0,f_1,\ldots$ describe  tree level effective Lagrangian for
massless modes as well as quantum string loop corrections. For
$10d$ SYM theory these coefficients are responsible for quantum
loop corrections. Evaluation of contribution of the SYM theory
one-loop UV divergencies to the coefficients $f_0$, $f_1$ may be
found in \cite{mettse} (the case of constant abelian $F$ was
considered in \cite{Fradkin:1982kf}).

As in the case of $11d$ supergravity the expression for
interaction vertex \rf{p4fin} can be used to obtain superspace
representation for tree level 4-point scattering amplitude of the
generic $10d$ SYM theory. As before to get 4-point $T$- matrix
from $P_{(4)}^-$ \rf{Ham4fin} we should multiply the 4-point
interaction vertex $p_\smf^-$ \rf{p4finym} by delta function
$\delta(\sum_{a=1}^4 p_a^-)$ that respects energy conservation
law. Beyond this the function $g(s,t)$ should be chosen so that to
respect the following two requirements: (i) the scattering
amplitude should has simple poles in Mandelstam variables. (ii)
the amplitude being restricted to the sector of bosonic fields
should has homogeneity of degree 0 in momenta $p^I_a$, $\beta_a$.
These two requirements can be fulfilled by the following choice of
$g(s,t)$:
\be g(s,t) = \frac{g_{_{YM}}^2}{st}\,.\ee This leads to the
following compact superspace representation for tree level 4-point
$T$-matrix of $10d$ SYM theory:
\be T_\smf  = \int d\Gamma_4\, \delta(\sum_{a=1}^4 p_a^-)\,
\Tr\prod_{a=1}^4 \psi(p_a,\lambda_a)\,\, t_\smf\, \,.\ee where the
expression for density $t_\smf$ is given by
\beq\label{t4sym} t_\smf  =  g_{_{YM}}^2\left(\frac{{\bf
J}_{12}{\bf J}_{34}}{s} E_s - \frac{{\bf J}_{14}{\bf J}_{23}}{t}
E_t \right) \eeq and a superfield $\psi(p,\lambda)$ enters a
solution to free equations of motion:
\be \Psi(p,\lambda) = \exp({\rm i}x^+p^-)\psi(p,\lambda)\,. \ee We
introduce then the standard decomposition similar to that given in
\rf{phiaabar}
\be \psi(p,\lambda) =  \frac{\epsilon(\beta)}{\sqrt{2\beta}}\,
\sigma \bar a (p,\lambda) +
\frac{\epsilon(-\beta)}{\sqrt{-2\beta}}\,\sigma a (-p,-\lambda)\,,
\ee
where $\sigma$ are Lie algebra matrices and we define a superspace
4-point amplitude by relation
\be\label{ampdefym} \langle 3,4\,|\, T_\smf \, | 1,2\rangle =
(2\pi)^9\delta^{10,4} {\cal A}_\smf \ee in which
$\delta^{10,4}$-function is an analog of \rf{d11d8}.  These
relations lead to the following 4-point amplitude
\be\label{symamp} {\cal A}_4 = {\cal A}^{abel}_4\Tr(
\sigma_1\ldots \sigma_4) + \hbox{ non-cyclic perms. of }
1,2,3,4\,,\ee where we introduce `abelian' part of the amplitude
defined by
\be\label{symampabe} {\cal A}_4^{abel} =   4 t_\smf\,. \ee

Superspace light cone representation for 4-point scattering
amplitude of SYM theory was obtained in \cite{GS6} (see formula
(5.48) in Ref.\cite{GS6}). Attractive feature of our alternative
representation for 4-point scattering amplitude
\rf{symamp},\rf{symampabe} is that the complicated dependence of
the amplitude on the bosonic `momenta' $\Po_{ab}^i$ and the
Grassmann `momenta' $\Lambda_{ab}$ is entirely collected in the
$E$-operators. It is usage of the $E$- operators that allows us to
find compact representation for amplitude given in
\rf{t4sym},\rf{symamp}.

\section{Conclusions}\label{CONsec}

We developed the superspace light cone formalism for $11d$
supergravity. In this paper we applied this formalism to study of
the superspace representation of the higher derivative 4-point
interaction vertices. We found also superfield representation for
cubic interaction vertex and for the tree level 4-point scattering
amplitude of the generic $11d$ supergravity \cite{cjs}. By analogy
with the fact that it is light cone gauge cubic vertices of the
$10d$ supergravity theories that admit natural extension to
superstring field theories we expect that our vertices will admit
natural extension to M-theory. Because the formalism we presented
is algebraic in nature it allows us to find various interaction
vertices in a relative straightforward way. Comparison of this
formalism with other approaches available in the literature leads
us to  the conclusion that this is a very efficient formalism.

Long term motivation for our study is related to conjectured
supergravity theory in $AdS_{11}$ spacetime \cite{gun}. As is well
known the standard $11d$ supergravity \cite{cjs} does not admit an
extension with a cosmological constant, i.e. does not have
$AdS_{11}$ vacuum \cite{des,des2}(see also \cite{nic,sag}). On the
one hand, in Ref.\cite{gun} certain massless $AdS_{11}$ graviton
supermultiplet was found\footnote{This novel graviton
supermultiplet is to transform in representation of
orthosymplectic superalgebra $osp(32,1)$. Recent interesting
discussion of unitary representations of $osp(32,1)$ superalgebra
and gravity theories based on such superalgebra may be found in
\cite{Dobrev:2004km} and \cite{Nastase:2003wb,Bilal:2001an}
respectively.}. This novel supermultiplet contains fields of the
usual $11d$ supergravity plus additional ones. One can expect that
these additional fields may allow one to overcome no-go theorem
and construct a consistent supergravity admitting $AdS_{11}$
ground state. Certain massless $AdS_{11}$ graviton multiplet is
also predicted by eleven dimensional version of $AdS_{10}$ higher
spin gauge theories discovered in \cite{vas1}. These theories
allow more or less straightforward generalization to $AdS_{11}$.
Usually a tower of infinite higher spin fields contains of
supergravity multiplet and therefore one expects that an extension
to $d=11$ of ten dimensional theories discussed in
\cite{vas2,Vasiliev:2003ev} also describes some $AdS_{11}$
graviton multiplet. On the other hand in \cite{met08} it was
demonstrated that, under certain assumption about {\it spontaneous
breaking of AdS symmetries}, totally symmetric massless higher
spin field in $AdS_{d+1}$ spacetime leads to a massive field in
$d$ dimensional Minkowski spacetime whose mass and spin are
related in the same manner as for a massive field belonging to the
leading Regge trajectory of string theory. {\it This suggests that
superstrings in 10 dimensional Minkowski space-time, viewed as a
boundary, could be related to higher spin massless fields theory
living in $AdS_{11}$ spacetime, viewed as a bulk}. Because
superstring theory admits simple and elegant formulation in light
cone gauge one can expect that eleven dimensional theories should
also be formulated within the light cone gauge. In this
perspective the study of this paper can be considered as a warm-up
for generalization to $AdS_{11}$ spacetime. Light-cone form of
field dynamics in $AdS$ spacetime developed in \cite{metlc}
implies that such generalization  is possible in principle.

The results presented here should have a number of interesting
applications and generalizations, some of which are:

(i) generalization to eleven dimensional anti-de Sitter spacetime
$AdS_{11}$ and study of interaction vertices for massless
$AdS_{11}$ graviton supermultiplet found in \cite{gun}.

(ii) generalization to 3- and 4- point interaction vertices of
type IIB supergravity in $AdS_5\times S^5$ background and
application to study of superspace form of AdS/CFT correspondence
at the level of 3- and 4-points correlation functions.

(iii) application of manifestly supersymmetric light-cone
formalism to the study of the various aspects of M-theory along
the lines \cite{ple3,gr}.

Another interesting application, which triggered our
investigation, is related to certain massless (nonsupersymmetric)
triplets in $d=11$, the dimension of M-theory. It was found in
\cite{Pengpan:1998qn} that some irreps of $so(9)$ algebra
naturally group together into triplets to be referred to as Euler
triplets which are such that bosonic and fermionic degrees of
freedom match up the same way as in $11d$ supergravity. Later on
Euler triplets were studied in
Refs.\cite{Brink:2002zq}-\cite{Brink:1999te} and it was
conjectured that these triplets might be organized in a
relativistic theory so that this theory would presumably be
finite. The methods we developed and used in this paper for study
$11d$ supergravity admit straightforward generalization to study
of higher spin Euler triplets. On the other hand a world line
approach used in Refs.\cite{gre1,gre2} for evaluation of quantum
loop corrections to $11d$ supergravity can also be generalized to
study loop corrections for higher spin fields in a relative
straightforward way. In principle just mentioned methods and
approaches should make it possible to address question of UV
finiteness of higher spin fields theory based on Euler triplets.
We hope to return to these problems in future publications.

\setcounter{section}{0} \setcounter{subsection}{0}
\begin{center}
{\bf Acknowledgments}
\end{center}

This work was supported by the INTAS project 00-00254, by the RFBR
Grant 02-02-16944 and RFBR Grant for Leading Scientific Schools No
1578-2003-2.

\setcounter{section}{0} \setcounter{subsection}{0}

\appendix{$11d$ Poincar\'e superalgebra
and gravitino field in $so(7)$  basis}

In order to cast $11d$ Poincar\'e superalgebra commutators
\rf{lorcovcomrel} into the $so(7)$ basis we first transform them
to the $so(9)$ basis. To this end we use the following
decomposition of $32\times 32$ gamma matrices and charge
conjugation matrix
\be \gamma_{32}^\mu=\left(
\begin{array}{cc}
\delta_I^\mu \gamma_{16}^I & \sqrt{2} \delta_+^\mu
\\[6pt]
\sqrt{2} \delta_-^\mu  & -\delta_I^\mu \gamma_{16}^I
\end{array}\right)\,,
\qquad
C_{32}=\left(
\begin{array}{ll}
0 & C_{16}\\[6pt]
-C_{16} & 0
\end{array}
\right)\,, \ee $ \{\gamma_{32}^\mu,\gamma_{32}^\nu\}
=2\eta^{\mu\nu}$,  $\gamma_{32}^{\mu\,t }=-C_{32}\gamma_{32}^\mu
C_{32}^{-1}$ and $\eta^{\mu\nu}$ is mostly positive flat metric
tensor. $\gamma_{16}^I$ and $C_{16}$ are $16\times 16$ gamma
matrices and charge conjugation matrix respectively:
\be \{\gamma_{16}^I,\gamma_{16}^J\}=2\delta^{IJ}\,, \qquad
\gamma_{16}^{I\,t }=C_{16}\gamma_{16}^I C_{16}^{-1}\,, \qquad
C_{16}^t=C_{16}\,. \ee Decomposition of the 32-component
supercharges $Q$ into two 16-component supercharges $Q^\pm$ and
using Majorana condition $Q^\dagger \gamma_{32}^0=Q^tC_{32}$ gives
\be\label{qdec} Q=2^{1/4}\left(
\begin{array}{l}
Q^-\\[4pt]
Q^+
\end{array}
\right)\,, \qquad (Q^\pm)^\dagger=(Q^\pm)^t C_{16}\,. \ee In the
$so(9)$ basis (anti)commutators of Poincar\'e superalgebra given
in \rf{lorcovcomrel} take then the form
\beq \label{so91} & \{Q^\pm, Q^\pm\}=\pm C_{16}P^\pm\,, \qquad
\{Q^+,Q^-\}=\frac{1}{\sqrt{2}}\gamma_{16}^IC_{16} P^I\,, &
\\[8pt]
\label{so92} & [J^{+-},Q^\pm]=\pm \frac{1}{2}Q^\pm, \qquad
[J^{IJ},Q^\pm]=-\frac{1}{2}\gamma_{16}^{IJ}Q^\pm, \qquad [J^{\pm
I},\, Q^\mp]=\pm\frac{1}{\sqrt{2}}\gamma_{16}^I Q^\pm. & \eeq
To convert these commutators into those of $so(7)$ basis we use
the following decomposition of $\gamma_{16}^I$ and $C_{16}$
matrices
\be \gamma_{16}^I=\left(
\begin{array}{cc}
\delta_i^I \gamma^i & \sqrt{2} \delta_R^I
\\[6pt]
\sqrt{2} \delta_L^I  & -\delta_i^I \gamma^i
\end{array}\right)\,,
\qquad
C_{16}=\left(
\begin{array}{ll}
0 & 1\\
1 & 0
\end{array}
\right)\,, \ee
where $\gamma^i$ are $8\times 8$ gamma matrices that are
antisymmetric and hermitian: $\gamma^{i t}=-\gamma^i$, $\gamma^{i
\dagger}=\gamma^i$. The decomposition for supercharges $Q^\pm $ we
use is
\be Q^\pm=2^{1/4}\left(
\begin{array}{l}
Q^{\pm L}\\[4pt]
Q^{\pm R}
\end{array}
\right)\,. \ee In terms of the supercharges $Q^{\pm R,L}$ the
Majorana condition \rf{qdec} takes the form $Q^{\pm
R\dagger}=Q^{\pm L}$. By exploiting these formulas in
(anti)commutators of the Poincar\'e superalgebra taken in the
$so(9)$ basis \rf{so91},\rf{so92} we arrive at the $so(7)$ basis
(anti)commutators given in Section 2. The reader interested in the
$so(7)$ $\gamma^i$- matrices identities is advised to consult the
appendices of Ref.\cite{Cremmer:1979up}.

Now let us describe relationship of the $so(7)$ basis physical
components of gravitino field, which enter superfield expansion
\rf{supfield}, with the nomenclature of covariant approach.
Lorentz covariant equations of motion and constraints for
gravitino field take the form
\be \gamma_{32}^\mu p_\mu \psi_\nu=0\,, \qquad p^\mu \psi_\mu=0\,,
\qquad \gamma_{32}^\mu \psi_\mu=0 \ee Making use of gauge
$\psi_-=0$, one proves that the components given by $\psi_I^\oplus
\equiv \frac{1}{2}\gamma_{32}^-\gamma_{32}^+\psi_I$ are physical
fields, while the components $\psi_I^\ominus \equiv
\frac{1}{2}\gamma_{32}^+\gamma_{32}^-\psi_I$ are non-physical
d.o.f. The physical gravitino field satisfies algebraic constraint
which by exploiting the 16-component notation can be written in
the $so(9)$ basis as

\be \gamma_{16}^I\psi_I^\oplus=0\,. \ee We solve this constraint
by exploiting $so(7)$ basis. Namely, introducing
\be \psi_I^{\oplus\,R} \equiv
\frac{1}{2}\gamma_{16}^L\gamma_{16}^R\psi_I^\oplus\,,\qquad
\psi_I^{\oplus\, L} \equiv
\frac{1}{2}\gamma_{16}^R\gamma_{16}^L\psi_I^\oplus\,, \ee it is
easily seen that the components $\psi_i^{\oplus\,R }$,
$\psi_L^{\oplus\,R }$ (and their hermitian conjugated partners
$\psi_i^{\oplus\,L }$, $\psi_R^{\oplus\, L }$) are independent,
while the remaining components  $\psi_R^{\oplus\,R }$ (and
$\psi_L^{\oplus\,L }$) are expressible in terms of that
independent components
\be \psi_R^{+R}=-\frac{1}{\sqrt{2}}\gamma^i\psi_i^{+L}\,, \qquad
\psi_L^{+L}=\frac{1}{\sqrt{2}}\gamma^i\psi_i^{+R}\,. \ee It is the
components $\psi_i^{\oplus\,R }$, $\psi_L^{\oplus\,R }$ (and
$\psi_i^{\oplus\,L }$, $\psi_R^{\oplus\, L }$) that enter
expansion of the superfield $\Phi$ in \rf{supfield}.

\appendix{Derivation of representation for the cubic vertex \rf{tvrho}.}

To derive the representation \rf{tvrho} we use $RL$, $Ri$ and $ij$
parts of Eqs.\rf{JIJp3}. Acting with angular momenta
$J^{IJ}(\Po,\Lambda)$ on the vertex $p_\sm3^-$ \rf{p3int} we find
the expressions
\beq && J^{RL}(\Po,\Lambda) p_\sm3^- = ({\Po }^L)^k
E_q\Bigl(M^{RL}(\Lambda) + 2\rho\partial_\rho -k\Bigr)\tilde{V}\,,
\\
&& J^{Ri}(\Po,\Lambda) p_\sm3^- =q^i J^{RL}(\Lambda) p_\sm3^-
+({\Po }^L)^k E_q \Bigl(M^{Ri}(\Lambda) -\rho M^{Li}(\Lambda) +q^j
M^{ij}(\Lambda)\Bigr)\tilde{V}\,, \ \ \
\\
&& J^{ij}(\Po,\Lambda) p_\sm3^- =({\Po }^L)^k E_q
M^{ij}(\Lambda)\tilde{V}\,. \eeq From these expressions it is
easily seen that the $RL$, $Ri$ and $ij$  parts of Eqs.\rf{JIJp3}
lead to the following equations for $\tilde{V}$
\beq \label{tv1} && (M^{RL}(\Lambda) + 2\rho\partial_\rho
-k)\tilde{V}=0\,,
\\[4pt]
\label{tv2} && (M^{Ri}(\Lambda) -\rho
M^{Li}(\Lambda))\tilde{V}=0\,,
\\[4pt]
\label{tv3}&& M^{ij}(\Lambda)\tilde{V}=0\,. \eeq {}From the
relations in \rf{newvar} and the fact that the vertex $p_\sm3^-$
is a monomial of degree $k$ in ${\Po }^I$ it follows that
$\tilde{V}$ should be polynomial of degree $k$ in $\rho$, {\it
i.e.} we can use the expansion
\be\label{tvexp} \tilde{V}(\rho,\Lambda,\beta) =\sum_{n=0}^k
\rho^n \tilde{V}_n(\Lambda,\beta)\,. \ee  Plugging this expansion
in Eq.\rf{tv2} we get the following equations\footnote{In addition
to Eqs.\rf{vnvn-1},\rf{vnvn-2} one has extra equation
$M^{Li}(\Lambda)\tilde{V}_k =0$. Because $\tilde{V}_k$ turns out
to be monomial of degree $4k$ in Grassmann momentum $\Lambda$
(note that $\Lambda^9=0$ as $\Lambda$ has eight components) this
extra equation amounts to the equation $\Lambda^{4k+2}=0$ which
satisfies automatically for supergravity theories $k\geq 2$. }
\beq &&\label{vnvn-1} M^{Ri}(\Lambda)\tilde{V}_n =
M^{Li}(\Lambda)\tilde{V}_{n-1}\,,\qquad n=1,\ldots, k\,,
\\
\label{vnvn-2} && M^{Ri}(\Lambda)\tilde{V}_0=0\,, \eeq while
Eqs.\rf{tv1},\rf{tv3} lead to the respective equations for
$\tilde{V}_0$ given in \rf{mrltv0},\rf{mijtv0}.

Now we focus on Eqs.\rf{vnvn-1}. These equations tell us that
$\tilde V_n$ can be expressed in terms of $\tilde V_0$. Making use
of \rf{vnvn-1} and \rf{tv3} one can make sure that $\tilde V_n$
can be presented in the form
\be\label{vnv0} \tilde{V}_n =f_n
(M^{Lj}(\Lambda)M^{Lj}(\Lambda))^n \tilde{V}_0\,, \ee
which should be supplemented by obvious initial condition $f_0=1$.
Now making use of Eqs.\rf{mrltv0}-\rf{mijtv0} and commutation
relations
\be [M^{Ri},\,(M^{Lj} M^{Lj})^n] =( M^{Lj} M^{Lj})^{n-1} 2n(
M^{Lj} M^{ji} - M^{Li} M^{RL} -(\frac{N'}{2}-n) M^{Li}) \ee we get
\be\label{for11} M^{Ri} ( M^{Lj} M^{Lj})^n\tilde{V}_0
=-n(N'+2k-2n)( M^{Lj} M^{Lj})^{n-1}  M^{Li}\tilde{V}_0\,, \ee $N'
\equiv d-4$, where we use Eq.\rf{tv3} and for flexibility we keep
the spacetime dimension $d$ to be arbitrary. Making use of
\rf{for11} and \rf{vnv0} in \rf{vnvn-1} gives the following
equations for $f_n$
\be \frac{f_{n-1}}{f_n}=-n(N'+2k-2n)\,. \ee Solution to these
equations with $f_0=1$ is easily found to be
\be f_n=(-)^n\frac{\Gamma(\frac{N'}{2}+k-n)}{2^n
n!\Gamma(\frac{N'}{2}+k)}\,, \ee where $\Gamma$ is the Euler gamma
function. Collecting all steps of derivation we arrive at solution
\beq && \tilde{V}(\rho,\Lambda,\beta)
=E_\rho\tilde{V}_0(\Lambda,\beta)\,,
\\[6pt]
\label{erhod} && E_\rho\equiv\sum_{n=0}^{k}
(-\rho)^n\frac{\Gamma(\frac{d-4}{2}+k-n)}{2^n
n!\Gamma(\frac{d-4}{2}+k)} ( M^{Lj}(\Lambda) M^{Lj}(\Lambda))^n\,.
\eeq Restriction to eleven dimensions $d=11$ leads to the desired
relation \rf{tvrho}.

\appendix{Derivation of expression for density $j^{-I}$ \rf{j3p3}}

In this appendix we outline procedure of deriving the expression
for density $j^{-I}$ given in \rf{j3p3}. To simplify presentation
we focus on the calculation of $j_\sm3^{-L}$. For flexibility we
start with calculation of $n$-point commutators for arbitrary
value of $n$:
\be\label{appC01} [P_\smn^-,J_{(2)}^{-L}]\! =\!\int\! d\Gamma_n
\Phi_{\smn}\Bigl( \sum_{a=1}^n(J_a^{-L})^t + \frac{1}{n}
\sum_{b=1}^n p_b^- \partial_{p_a^R} \Bigr) p_\smn^-
+\frac{1}{n}\Bigl(\sum_{a=1}^n
\partial_{p_a^R} \sum_{b=1}^n p_b^- \Phi_\smn\Bigr)p_\smn^-\,,  \ee
\be\label{appC02} [J_\smn^{-L},P_{(2)}^-]\! =\!\int\! d\Gamma_n
\Phi_{\smn}\sum_{b=1}^n p_b^- \Bigl(j_\smn^{-L} + \frac{1}{n}
\sum_{a=1}^n \frac{\lambda_a}{\beta_a}q_\smn^{-L}\Bigr) +
\frac{1}{n}\Bigl(\sum_{a=1}^n
\partial_{p_a^R} \sum_{b=1}^n p_b^- \Phi_\smn\Bigr)p_\smn^-\,,  \ee
where notation $(J^{-L})^t$ is used for the operators obtainable
from \rf{jml} by applying of transposition that is defined to be
\be\label{appC03} \partial_{p^I}^t = -\partial_{p^I},\qquad
\theta^t=-\theta, \qquad (p^I)^t=p^I,\qquad
\lambda^t=\lambda\,.\ee Transposition on the product of bosonic
($B$) and fermionic ($F$) quantities is defined to be $(B_1B_2)^t
= B_2^t B_1^t$, $(BF)^t = F^t B^t$, $(F_1F_2)^t = - F_2^t F_1^t $.

In cubic approximation the relations \rf{appC01},\rf{appC02} and
commutator
\be \label{appC04} [P_\sm3^-,J_{(2)}^{-L}]=
[J_\sm3^{-L},P_{(2)}^-] \ee lead to the formula
\be\label{appC05} \sum_{b=1}^3 p_b^- \Bigl( j_\sm3^{-L} +
\frac{1}{3} \sum_{a=1}^3 \frac{\lambda_a}{\beta_a}
q_\sm3^{-L}\Bigr) = \Bigl(\sum_{a=1}^3 (J_a^{-L})^t + \frac{1}{3}
\sum_{b=1}^3 p_b^-\partial_{p_a^R}\Bigr) p_\sm3^-\,. \ee Taking
into account that $p_\sm3^-$ depends on $\Po^I$ and $\Lambda$ and
using $Li$ part of Eq.\rf{JIJp3} we can cast an action of
differential operator $\sum_{a=1}^3(J_a^{-L})^t$ on $p_\sm3^-$
into the form
\beq \label{jmltr1}\sum_{a=1}^3(J_a^{-L})^t p_\sm3^-
&=&\Bigl(-\frac{\Po^L}{3\hat{\beta}}\sum_{a=1}^3
\check{\beta}_a\beta_a\partial_{\beta_a} -\frac{1}{3}\sum_{a=1}^3
\frac{\lambda_a}{\beta_a} Q^{-L}(\Lambda)\Bigr)p_\sm3^-\,. \eeq
Plugging this relation into \rf{appC05} and using the second
relation in \rf{qQp} and formula
\be \sum_{a=1}^3 \partial_{p_a^R} p_\sm3^- =0\ee we get from
\rf{appC05} the relation
\be \sum_{a=1}^3 p_a^-  j_\sm3^{-L}=
-\frac{\Po^L}{3\hat{\beta}}\sum_{a=1}^3
\check{\beta}_a\beta_a\partial_{\beta_a}p_\sm3^-\,.\ee Taking into
account
\be \sum_{a=1}^3p_a^-=\frac{|\Po|^2}{2\hat{\beta}} \ee we arrive
at the relation \rf{j3p3} taken to be for transverse index $I=L$.
Above-given calculations can extended to the cases of
$j_\sm3^{-R}$, $j_\sm3^{-i}$ in a rather straightforward way. This
leads to the formula given in \rf{j3p3}.

\appendix{Derivation of representation for 4-point vertex \rf{p4fin}}

In this appendix we outline a procedure of solving the defining
equations for 4-point vertex \rf{defeqs0}-\rf{defeq3}. First we
express orbital momenta $L^{IJ}_{ab}$ in terms of the variables
given in \rf{newvar4}. The orbital momenta $L_{ab}^{Li}$ take then
the form as in \rf{Lli} and this allows us to write a solution of
$Li$ parts of Eqs.\rf{defeqs1} in the following form
\beq \label{p4vp} && p_{(4)}^-=E_{q_{13}}E_{q_{24}}V^\prime\,,
\\[6pt]
&& V^\prime \equiv V^\prime(q_{_L},\Lambda_{13},\Lambda_{24}, {\Po
}_{13}^L,{\Po }_{24}^L,\rho_{13},\rho_{24},\beta_a)\,, \eeq
where $E_{q_{ab}}$ and $q_{_L}^i$ are defined in
\rf{eqab},\rf{qi}. Now we have to reformulate remaining equations
in terms of $V^\prime$. Moving the operators $J^{RL}$ $J^{ij}$ and
$J^{Ri}$ throughout the operators $E_{q_{ab}}$ one can make sure
that Eqs.\rf{defeqs1} lead to the following equations for
$V^\prime$
\beq \label{rleq1} && (q_{_L}\partial_{q_{_L}} +
2\rho_{13}\partial_{\rho_{13}}+2\rho_{24}\partial_{\rho_{24}} -
{\Po }_{13}^L\partial_{{\Po }_{13}^L} -{\Po }_{24}^L\partial_{{\Po
}_{24}^L}+M_{13}^{RL}+M_{24}^{RL})V^\prime
=0\,,\\[6pt]
\label{jijp4} && (q_{_L}^i \partial_{q_{_L}^j} -
q_{_L}^j\partial_{q_{_L}^i} +M_{13}^{ij}+M_{24}^{ij})V^\prime
=0\,,
\\[10pt]
 && \frac{q_{_L}^i}{2}
(2\rho_{13}\partial_{\rho_{13}}-2\rho_{24}\partial_{\rho_{24}}
-{\Po }_{13}^L\partial_{{\Po }_{13}^L} + {\Po
}_{24}^L\partial_{{\Po }_{24}^L} +M_{13}^{RL}-M_{24}^{RL})V^\prime
\nonumber\\
\label{jrip42} && +\Bigl((\rho_{13}-\rho_{24})\partial_{q_{_L}^i}
+M^{Ri}_{13}+M^{Ri}_{24}-\rho_{13}M^{Li}_{13}-\rho_{24}M^{Li}_{24}
+\frac{q_{_L}^j}{2}(M^{ij}_{13}-M^{ij}_{24})\Bigr)V^\prime=0\,. \
\ \ \ \eeq Doing the same for supercharges we get from
\rf{defeqs2} the equations
\be\label{qri1324p4}
\Bigl(\frac{1}{\sqrt{2}}(\theta_{\Lambda_{13}}\bpline_{13}
+\theta_{\Lambda_{24}}\bpline_{24}) +\frac{\rho_{13}{\Po
}_{13}^L\Lambda_{13}}{\beta_1\beta_3\beta_{24}}
+\frac{\rho_{24}{\Po
}_{24}^L\Lambda_{24}}{\beta_2\beta_4\beta_{13}} \Bigr)V^\prime
=0\,, \ee
\be\label{qli1324p4} ({\Po }_{13}^L\theta_{\Lambda_{13}} +{\Po
}_{24}^L\theta_{\Lambda_{24}})V^\prime =0\,. \ee Thus the generic
equations \rf{defeqs1},\rf{defeqs2} are reduced to
\rf{jijp4}-\rf{qli1324p4}. We note that because Eqs.\rf{defeqs2}
are valid on the energy surface \rf{rhoman} we should also reduce
our equations to the energy surface, {\it i.e.} we should express
interaction vertex in terms of Mandelstam variable $u$ instead of
$\rho_{13}$, $\rho_{24}$. Before doing that we find solution to
the simple equation \rf{qli1324p4}
\be\label{solequ1} V^\prime=V^{\prime\prime}(q_{_L}, \Lambda^L,
{\Po }^L_{13},{\Po }_{24}^L, \rho_{13},\rho_{24},\beta_a)\,, \ee
where $\Lambda^L$ is given by \rf{defThe}. Taking into account
that on the energy surface the variables $\rho_{13}$ and
$\rho_{24}$ are expressible in terms of Mandelstam variable $u$
\rf{rhoman} we introduce the vertex $V^{\prime\prime\prime}$ by
relation
\be \label{solequ2}
V^{\prime\prime}=V^{\prime\prime\prime}(q_{_L}, \Lambda^L, {\Po
}_{13}^L, {\Po }_{24}^L,u,\beta_a)\,. \ee Now we have to repeat
our procedure and rewrite remaining Eqs.\rf{rleq1}-\rf{qri1324p4}
in terms of $V^{\prime\prime\prime}$. First let us consider
Eq.\rf{qri1324p4}. In terms of $V^{\prime\prime\prime}$ the
equation \rf{qri1324p4} takes the form
\be \label{solequ3} \Bigl(\frac{1}{\sqrt{2}}{\Po }_{13}^L{\Po
}_{24}^L\theta_{\Lambda^L} \qline +\frac{u
\Lambda^L}{2\beta_{24}{\Po }_{13}^L{\Po }_{24}^L}\Bigr)
V^{\prime\prime\prime}=0\,, \ee whose solution is easily found to
be
\be \label{solequ4} V^{\prime\prime\prime} =E_u V^{{\rm
iv}}(q_{_L},{\Po }_{13}^L,{\Po }_{24}^L,u,\beta_a)\,. \ee
where the operator $E_u$ is given in \rf{eu}.

Next step is to consider Eqs.\rf{rleq1},\rf{jrip42}. It is
straightforward to demonstrate that in terms of $V^{{\rm iv}}$
these equations take the form
\beq \label{solequ6} && (q_{_L}\partial_{q_{_L}} -
\Po_{13}^L\partial_{\Po_{13}^L}
-\Po_{24}^L\partial_{\Po_{24}^L}+4)V^{{\rm iv}}=0\,,
\\[8pt]
\label{solequ7} &&
\Bigl(\frac{q_{_L}^i}{2}(-\Po_{13}^L\partial_{\Po_{13}^L}
+\Po_{24}^L\partial_{\Po_{24}^L})
+u\frac{(\beta_1\beta_3\Po_{24}^{L2}
-\beta_2\beta_4\Po_{13}^{L2})}{2(\Po_{13}^L\Po_{24}^L)^2}
(\partial_{q_{_L}^i}-\frac{4q_{_L}^i}{q_{_L}^2})\Bigr)V^{{\rm
iv}}=0\,. \  \ \ \eeq In addition to these equations there are two
equations obtainable from Eqs.\rf{defeqs0},\rf{defeq3}
\beq \label{solequ5} && (\Po_{13}^L\partial_{\Po_{13}^L} +
\Po_{24}^L\partial_{\Po_{24}^L} + \beta_{13}\partial_{\beta_{13}}
+ y_{13}\partial_{y_{13}} + y_{24} \partial_{y_{24}} - 4)V^{\rm
iv} = 0\,,
\\[8pt]
\label{solequ8}  && \Bigl(\frac{\Po_{13}^L}{\beta_{13}}
\partial_{y_{13}}
+ \frac{\Po_{24}^L}{\beta_{24}}
\partial_{y_{24}}\Bigr) V^{\rm iv}=0\,,\eeq where we use the
notation
\be y_{13}\equiv \beta_1 - \beta_3\,,\qquad y_{24}\equiv
\beta_2-\beta_4\,.\ee In Eqs.\rf{solequ6}-\rf{solequ8} and below
the vertex $V^{\rm iv}$ is considered to be function depending on
three independent light cone momenta $\beta_{13}$, $y_{13}$,
$y_{24}$ instead of four momenta $\beta_a$ subject to the
conservation low $\sum_{a=1}^4 \beta_a=0$. Eq.\rf{solequ5} is
obtainable from Eq.\rf{defeqs0}, while Eq.\rf{solequ8} is
obtainable from commutator $[P_\smf^-, J_{(2)}^{-L}]=0$. Helpful
relation to analyze this commutator is given in \rf{appC01}. Now
we focus on the equations \rf{solequ6}-\rf{solequ8}.

Solution to \rf{solequ8} is easily fixed to be
\be \label{solequ9} V^{\rm iv} =  V^{\rm v}(u,q_{_L},
\Po_{13}^L,\Po_{24}^L, Y, \beta_{13} )\,,\qquad Y \equiv
\Po_{13}^Ly_{24} + \Po_{24}^Ly_{13}\,.\ee Plugging this
representation into \rf{solequ7} we get the following equation
\beq \label{solequ10} && \Bigl(\frac{q_{_L}^i}{2}( -
\Po_{13}^L\partial_{\Po_{13}^L} +\Po_{24}^L\partial_{\Po_{24}^L})
+ u \beta_{13}^2 \frac{(\Po_{24}^{L2} -
\Po_{13}^{L2})}{8(\Po_{13}^L\Po_{24}^L)^2}
(\partial_{q_{_L}^i}-\frac{4q_{_L}^i}{q_{_L}^2})\Bigr)V^{{\rm v}}
\nonumber \\[8pt]
&& - y^L \Bigl(- \frac{q_{_L}^i}{2}\partial_Y + \frac{u
Y}{8(\Po_{13}^L\Po_{24}^L)^2}(\partial_{q_{_L}^i}-\frac{4q_{_L}^i}{q_{_L}^2})
\Bigr) V^{\rm v} = 0\,, \eeq where
\be \label{solequ11} y^L \equiv y_{13} \Po_{24}^L - y_{24}
\Po_{13}^L\,. \ee Because the vertex $V^{\rm v}$ does not depend
on $y^L$ equation \rf{solequ8} implies that this vertex should
satisfy the following two equations
\beq\label{solequ12} &&
\Bigl(\frac{q_{_L}^i}{2}(-\Po_{13}^L\partial_{\Po_{13}^L}
+\Po_{24}^L\partial_{\Po_{24}^L}) + u \beta_{13}^2
\frac{(\Po_{24}^{L2} - \Po_{13}^{L2})}{8(\Po_{13}^L\Po_{24}^L)^2}
(\partial_{q_{_L}^i}-\frac{4q_{_L}^i}{q_{_L}^2})\Bigr)V^{{\rm
v}}=0\,,
\\[8pt]
\label{solequ13} && \Bigl(-\frac{q_{_L}^i}{2}\partial_Y + \frac{u
Y}{8(\Po_{13}^L\Po_{24}^L)^2}(\partial_{q_{_L}^i}-\frac{4q_{_L}^i}{q_{_L}^2})
\Bigr) V^{\rm v} = 0\,. \eeq
Solution to these equations is found to be\footnote{Instead of $v$
\rf{solequ14} we could use another function of the Mandelstam
variables, which is not constant on the surface $u=const$. We
prefer to exploit the variable $v$ as this variable has simple
transformation rule upon cyclic permutation of four external line
indices 1,2,3,4. Namely, upon cyclic permutations of 1,2,3,4 we
get $v\rightarrow -v$.}
\be \label{solequ14} V^{\rm v} = (q_{_L}^2)^2 V^{\rm vi}(u,\,
v,\,\omega,\,\beta_{13})\,,\qquad  \omega \equiv
\Po_{13}^L\Po_{24}^L\,,\qquad v\equiv \frac{t-s}{2}\,,\ee where we
use the following helpful relation for the variables $u$, $v$ and
$Y$:
\be v = \frac{\Po_{13}^L\Po_{24}^L}{\beta_{13}^2}q_{_L}^2 +
\frac{Y^2u}{4\beta_{13}^2\Po_{13}^L\Po_{24}^L} -
\frac{\Po_{13}^{L2} + \Po_{24}^{L2}}{4\Po_{13}^L\Po_{24}^L}u\,.
\ee
{}Formula \rf{solequ14} implies the following representation for
$V^{\rm iv}$ (see \rf{solequ9}):
\be \label{solequ16} V^{\rm iv} = (q_{_L}^2)^2 V^{\rm vi}( u,\,
v,\,\omega,\,\beta_{13})\,. \ee
Plugging the $V^{\rm iv}$ \rf{solequ16} into \rf{solequ6} we get
solution for $V^{\rm vi}$:
\be \label{solequ17}  V^{\rm vi}(u, v, \omega, \beta_{13}) =
\omega^4 V^{\rm vii}(u, v,\beta_{13})\,,  \ee which implies
\be \label{solequ18} V^{\rm iv} = \omega^4 (q_{_L}^2)^2 V^{\rm
vii}( u,\, v,\,\beta_{13})\,. \ee Plugging \rf{solequ18} into
\rf{solequ5} we get solution for vertex $V^{\rm vii}$:
\be \label{solequ19} V^{\rm vii}( u,\, v,\,\beta_{13}) =
\frac{1}{\beta_{13}^4} V^{\rm viii}( u,\, v)\,.\ee {}For the case
of supergravity theory the vertex $V^{\rm viii}$ should be
symmetric with respect to Mandelstam variables $s$, $t$, $u$. To
respect this requirement we take into account the relation
\rf{stu0} which implies that dependence on $u$, $v$ can be
replaced by dependence on $s$, $t$, $u$  and therefore we can
simply rewrite the vertex $V^{\rm viii}$ in the form $V^{\rm viii}
= g(s,t,u)$, where the function $g(s,t,u)$ is considered to be
symmetric in $s$, $t$, $u$.\footnote{Note that for the case of YM
theories the vertex should be symmetric only upon cyclic
permutations of 1,2,3,4. This is reason why for the case of YM
theory we use representation $V^{\rm viii}=g(s,t)$ and impose
constraint \rf{gdua} in (see \rf{p4finym}).} Collecting all steps
of derivation we get the following solution to $V^{\rm iv}$:
\be \label{solequ20}  V^{\rm iv} =
\frac{(\Po_{13}^L\Po_{24}^L)^4}{\beta_{13}^4}(q_{_L}^2)^2
g(s,t,u)\,.\ee The dependence on Mandelstam variables $s$, $t$,
$u$ cannot be defined by commutation relations Poincar\'e
superalgebra and this is freedom of our solution. Taking into
account formulas \rf{p4vp},\rf{solequ1},\rf{solequ2},\rf{solequ4},
\rf{solequ20} we arrive at formula \rf{p4fin}.

\appendix{ $R^4$ and $F^4$ terms in light cone basis}\label{r4APP}

In this appendix we explain how various $R^4$ (and $F^4$) terms
given in \rf{r42def}-\rf{r46def} (and \rf{symL}) can be cast into
light cone basis. This will allow us to relate normalization of
covariant Lagrangian and corresponding supersymmetric light cone
Hamiltonian. We begin our discussion with the gravitational $R^4$
terms.

Making Fourier transformation to momentum space for all
coordinates except for the time  $x^+$
\be\label{r4ter0}  \Phi(x) = \int \frac{d^{d-1}
p}{(2\pi)^{(d-1)/2}} e^{{\rm i}( x^- \beta + x^I p^I) }
\Phi(x^+,p) \ee (by setting $d=11$ for fields of $11d$
supergravity) we cast covariant action corresponding to 4-point
Lagrangian \rf{covlagr4} into the form
\be \label{r4ter1}  \int d^{11}x\, {\cal L}_4(x) = \int dx^+
P_{(4)}^- \,,\ee
where we introduce an appropriate 4-point Hamiltonian
\be \label{r4ter2}  P_{(4)}^- = \int d\Gamma_4(p) {\cal L}_4(p)\,,
\ee and ${\cal L}_4$ indicates 4-point approximation of covariant
Lagrangian \rf{covlagr4}. The measure $d\Gamma_4(p)$ is given by
formula \rf{delfun01} in which we set $n=4$. In what follows we
assume: 1) massless fields are on mass-shell $({\rm
i}\partial_{x^+}  + p^-) \Phi(x^+,p)=0$, where $p^-$ is given in
\rf{dyn1}; 2) momenta of fields are restricted to the energy
surface \rf{enesur}. To find ${\cal L}_4(p)$ we should find
expressions for $W_1$ and $W_2$ in the 4-point approximation,
which we shall denote as $W_1(p)$ and $W_2(p)$ (see
\rf{covlagr4})\footnote{Note that beyond of establishing relation
\rf{g0k4} we confirmed ourselves that gravitational body of our
Hamiltonian \rf{pm122} is indeed related with Lagrangian
\rf{covlagr4} in which the coefficients in front of various $R^4$
terms \rf{r42def}-\rf{r46def} should be equal to those of
Eq.\rf{covlagr4}. To keep discussion from becoming unwieldy here
we do not discuss these relative coefficients in front of various
$R^4$ terms and use the coefficients that are evident from
Eq.\rf{covlagr4}.}
\be\label{r4ter2a} {\cal L}_4(p) =
\kappa_{_{(4)}}W_{R^4}(p)\,,\qquad W_{R^4}(p)= W_1(p)
+\frac{1}{16}W_2(p)\,. \ee We start our discussion with $W_1(p)$
term that is obtainable from \rf{w1def}.

{}Following Fourier transform \rf{r4ter0} we introduce Fourier
modes for linearized Riemann tensor and Lorentz connection
\beq \label{r4ter3} && R_{\mu\nu}(p) \equiv  p_\mu \omega_\nu(p)
-p_\nu\omega_\mu(p)\,,
\\[8pt]
\label{r4ter4}  &&  \omega^{\mu \, AB}(p) \equiv  -p^A \bar h^{B
\mu} + p^B \bar h^{A\mu}\,,\qquad \bar h^{A\mu} \equiv \sqrt{2}\,
\kappa h^{A\mu}\,, \eeq  where $h^{A \mu} \equiv \delta_\nu^A
h^{\nu\mu}$ and we keep a dependence on the gravitational constant
$\kappa$ to respect expansion \rf{gmnexp}. Indices $A,B=0,1,\ldots
,10$ are flat Lorentz indices. In \rf{r4ter3} and below the
quantities $R_{\mu\nu}$ and $\omega^\mu$ stand for respective
matrices $R_{\mu\nu}^{AB}$ and $\omega^{\mu\,AB}$. Taking into
account \rf{w1def} and representation for Riemann tensor
\rf{r4ter3} we get the following expression for $W_1(p)$:
\beq\label{E12} W_1(p)  &   =  &  \Tr -
\frac{ut}{2}\omega_{12}\omega_{34} - \frac{st}{4}\omega_1^\mu
\omega_2^\nu \omega_3^\mu \omega_4^\nu + 2t
\omega_{12}b_{13}b_{24} + 2u \omega_{12}b_{23}b_{14}
\nonumber\\[7pt]
&+& t \omega_1^\mu b_{12} \omega_3^\mu b_{34} + s \omega_1^\mu
b_{32} \omega_3^\mu b_{14}\,, \eeq where we use the notation
\be \label{r4ter7}  \omega_{ab} \equiv \omega_a^\mu
\omega_b^\mu\,,\qquad b_{ab}\equiv p_a^\mu \omega_b^\mu\,, \qquad
\omega_a^\mu \equiv \omega^\mu(p_a)\,. \ee $W_1(p)$ in \rf{E12}
can be easily cast into light cone basis by noticing that in light
cone gauge \rf{lcgau} we have the relations
\be \label{r4ter8}  \omega^+(p) = 0\,,\qquad \omega^{-}(p)=
-\frac{p^I}{\beta}\omega^I(p)\,,\ee which lead to helpful formula
\be  \label{r4ter9}  p_a^\mu \omega_b^\mu = \frac{1}{\beta_b}
\Po_{ab}^I \omega^I_b \,.\ee Making use of this formula in
\rf{E12} gives representation for $W_1(p)$ in light cone basis
\beq\label{r4ter10} W_1(p) &   = &\Tr   -
\frac{ut}{2}\omega_{12}\omega_{34} - \frac{st}{4}\omega_1^I
\omega_2^J \omega_3^I \omega_4^J
\nonumber\\[7pt]
& + & \frac{2}{\beta_3\beta_4} ( t \Po_{13}^I\Po_{24}^J + u
\Po_{23}^I\Po_{14}^J ) \omega_{12} \omega_3^I \omega_4^J
\nonumber\\[7pt]
& + & \frac{1}{\beta_2\beta_4} ( t \Po_{12}^I\Po_{34}^J - s
\Po_{23}^I \Po_{14}^J ) \omega_1^M \omega_2^I \omega_3^M
\omega_4^J\,. \eeq We note that because of the first relation
given in \rf{r4ter8} the Lorentz invariant scalar products
$\omega_{ab}$ are reduced to transverse scalar products
$\omega_{ab} = \omega_a^I \omega_b^I$.

Now we turn to $W_2$. Transformation of $W_2$ to the light cone
basis is simplified by using the relation
\be\label{r4ter14} R_{\mu\nu}^{A_1B_1}(p_1)
R_{\mu\nu}^{A_2B_2}(p_2) = -
\frac{g^{IJ}(\Po_{12})}{\beta_1\beta_2} \omega^{I\, A_1B_1}(p_1)
\omega^{J\, A_2B_2}(p_2)\,, \ee where we use the notation
\be \label{r4ter15}  g^{IJ}(x) \equiv |x|^2 \delta^{IJ} - 2 x^I
x^J \,, \qquad |x|^2 \equiv x^I x^I \,. \ee
{}Formula \rf{r4ter14} can be proved by using
\rf{r4ter3},\rf{r4ter9} and representation for Mandelstam
variables given in \rf{stulcdef}. Making use of formula
\rf{r4ter14} all $R^4$- terms in  $W_2$ \rf{r43def}-\rf{r46def}
can be cast into light cone basis in a rather straightforward way
\beq \label{r4ter16}  && R_{43}(p) = \frac{g^{IJ}(\Po_{12})
g^{MN}(\Po_{34})}{\beta_1\ldots \beta_4} \Tr \omega_1^I \omega_3^J
\Tr  \omega_2^M\omega_4^N\,,
\\[8pt]
\label{r4ter17}  &&  R_{44}(p) = \frac{g^{IJ}(\Po_{12})
g^{MN}(\Po_{34})}{\beta_1\ldots \beta_4} \Tr \omega_1^I \omega_2^J
\Tr \omega_3^M\omega_4^N\,,
\\[8pt]
\label{r4ter18}  &&  R_{45}(p) = \frac{g^{IJ}(\Po_{12})
g^{MN}(\Po_{34})}{\beta_1\ldots \beta_4} \Tr \omega_1^I \omega_2^J
\omega_3^M\omega_4^N\,,
\\[8pt]
\label{r4ter19}  && R_{46}(p) = \frac{g^{IJ}(\Po_{13})
g^{MN}(\Po_{24})}{\beta_1\ldots \beta_4} \Tr \omega_1^I \omega_2^M
\omega_3^J\omega_4^N\,. \eeq Note that $\Tr$ in formulas
\rf{r4ter10},\rf{r4ter16}-\rf{r4ter19} still indicates trace over
all Lorentz indices $A,B=0,1,\ldots, 10$. In fact formulas
\rf{r4ter10},\rf{r4ter16}-\rf{r4ter19} give a desired
representation of $R^4$ terms in the light cone basis. These
representations can be reformulated in terms of $h^{\mu\nu}$
\rf{r4ter4} by using the following helpful formulas
\be \Tr (\omega_a^\mu \omega_b^\nu) = \frac{h_a^{\mu I}
g^{IJ}(\Po_{ab})h_b^{J\nu}}{\beta_a\beta_b}\,,\qquad
h_a^{\mu\nu}\equiv h^{\mu\nu}(p_a)\,, \ee
\be \omega_1^{\mu AC} \omega_2^{\nu CB} =-p_{12} h_1^{\mu
A}h_2^{\nu B} - p_1^A p_2^B h_1^{\mu\rho} h_2^{\rho\nu} + h_1^{\mu
A} \Po_{12}^I h_2^{I\nu} \frac{p_2^B}{\beta_2} - \Po_{12}^I
h_1^{I\mu} \frac{p_1^A}{\beta_1} h_2^{\nu B}\,, \ee where
$p_{12}\equiv p_1^\mu p_2^\mu$.

Now we would like to demonstrate how these results can be used to
relate Lagrangian in covariant and light cone bases. It turns out
that it suffices to consider terms proportional to $h_1^{LL}
h_2^{RR} h_3^{RR} h_4^{RR}$. To this end we evaluate contribution
of such terms to $W_1$ and $W_2$. We introduce
\be W_1(p)|, W_2(p)|  =\frac{\bar h_1^{LL} \bar h_2^{RR} \bar
h_3^{RR} \bar h_4^{RR}}{(\beta_1\ldots \beta_4)^2} \widetilde
W_1(p), \widetilde W_2(p), \ee where  notation $W_1(p)|$,
$W_2(p)|$ indicates that we keep only those parts of $W_1(p)$,
$W_2(p)$ which are proportional to $h_1^{LL} h_2^{RR} h_3^{RR}
h_4^{RR}$. Making use of formulas \rf{w2def}, \rf{r4ter10},
\rf{r4ter16}-\rf{r4ter19} we get
\beq && \widetilde W_1(p)  =  \beta_1^3 su \Po_{14}^L\Po_{23}^L
\Bigl (-2\beta_1 \Po_{14}^L\Po_{23}^L -\beta_3 (\Po_{24}^L)^2
+\beta_2 (\Po_{34}^L)^2\Bigr)\,,
\\[10pt]
&& \widetilde W_2(p)  =  16 \beta_1^3 su \Po_{14}^L\Po_{23}^L
\Bigl (\beta_3 (\Po_{24}^L)^2 - \beta_2 (\Po_{34}^L)^2\Bigr)\,.
\eeq Taking into account the second formula in \rf{covlagr4} we
get
\be  \label{r4ter30} W_{R^4}(p)  = \frac{\bar h_1^{LL} \bar
h_2^{RR} \bar h_3^{RR} \bar h_4^{RR}}{(\beta_1\ldots \beta_4)^2}
\Bigl(- 2 \beta_1^4 su (\Po_{14}^L\Po_{23}^L)^2\Bigr)\,.\ee

To relate this result to the light cone Hamiltonian
\rf{pm122},\rf{p4fin} we should find contribution of $h_1^{LL}
h_2^{RR} h_3^{RR} h_4^{RR}$- terms to the Hamiltonian \rf{pm122}.
To this end we note the relation
\be \Phi_1|_{_{h_1^{LL}}} \Phi_2 \Phi_3
\Phi_4|_{_{h_2^{RR}h_3^{RR}h_4^{RR}}}
=\frac{\beta_1^4(\epsilon\lambda_2^8) (\epsilon\lambda_3^8)
(\epsilon\lambda_4^8)}{4(\beta_1 \ldots
\beta_4)^2}h_1^{LL}h_2^{RR}h_3^{RR}h_4^{RR}\ee and this implies
the following relation
\be \int d\Gamma_4(\lambda) \prod_{a=1}^4 \Phi(p_a,\lambda_a)\,
p_{(4)}^-|  = g_0
\frac{h_1^{LL}h_2^{RR}h_3^{RR}h_4^{RR}}{(\beta_1\ldots
\beta_4)^2}\, \beta_1^4\,
\frac{(\Po_{13}^L\Po_{24}^L)^4}{\beta_{13}^4}(q_{_L}^2)^2\,.\ee
where the measure $d\Gamma_4(\lambda)$ is given by formula
\rf{delfun02} in which we set $n=4$. Making use then the formula
\rf{q4stu} we get the following contribution of
$h_1^{LL}h_2^{RR}h_3^{RR}h_4^{RR}$ terms to the 4-point
Hamiltonian
\be P_{(4)}^-| =  g_0   \int d\Gamma_4(p)\,
\frac{h_1^{LL}h_2^{RR}h_3^{RR}h_4^{RR}}{(\beta_1\ldots
\beta_4)^2}\,( - 3su \beta_1^4\,(\Po_{14}^L\Po_{23}^L)^2)\,. \ee
Comparing this formula with \rf{r4ter2a},\rf{r4ter30}(see also the
second formula in \rf{r4ter4}) we arrive at \rf{g0k4}.

The analysis above-given is extended to the case of higher
derivative Lagrangian \rf{covlagr4f} straightforwardly because the
factor $f(s,t,u)$ is symmetric with respect to Mandelstam
variables and therefore this factor does not affect derivation.

These considerations can be easily extended to the case of $10d$
SYM theory. In this case the covariant Lagrangian ${\cal
L}_{_{F^4}}$ given by \rf{symL} leads to the 4-point Hamiltonian
\be \label{f4ter2}  P_{(4)}^- = \int d\Gamma_4(p)\, {\cal
L}_4(p)\,,\qquad  {\cal L}_4(p) = g_{_{F^4}}W_{_{F^4}}(p) \,,\ee
where the measure $d\Gamma_4(p)$ is given by formula \rf{delfun01}
in which we set $n=4$, $d=10$. It easy to see that if in
expression for $W_1$ \rf{w1def} we replace the Lorentz connection
$\omega_\mu$ by the gauge field $\phi_\mu$ then we get $W_{F^4}$
\rf{symW}. Therefore Lorentz covariant representation for
$W_{F^4}(p)$ in terms of gauge field $\phi^\mu$ can be obtained
from \rf{E12} by making there just mentioned replacement
\beq W_{_{F^4}}(p)&   = & \Tr - \frac{ut}{2} \phi_{12} \phi_{34} -
\frac{st}{4} \phi_1^\mu \phi_2^\nu \phi_3^\mu \phi_4^\nu
 +  2t \phi_{12}b_{13}b_{24} + 2u  \phi_{12}b_{23}b_{14}
\nonumber\\[7pt]
&+& t \phi_1^\mu b_{12} \phi_3^\mu b_{34} + s \phi_1^\mu b_{32}
\phi_3^\mu b_{14}\,, \eeq where we use the notation
\be \phi_{ab} \equiv \phi_a^\mu \phi_b^\mu\,,\qquad b_{ab} \equiv
p_a^\mu \phi_b^\mu\,, \qquad  \phi_a^\mu \equiv
\phi^\mu(p_a)\,.\ee Making use of light cone gauge and the
relations similar to the ones in \rf{r4ter8},\rf{r4ter9} adopted
for gauge fields we get representation $W_{F^4}(p)$ in the light
cone basis (cf. \rf{r4ter10}):
\beq W_{_{F^4}}(p) &   = &  \Tr - \frac{ut}{2}\phi_{12}\phi_{34} -
\frac{st}{4}\phi_1^I \phi_2^J \phi_3^I \phi_4^J
\nonumber\\[7pt]
& + & \frac{2}{\beta_3\beta_4} ( t \Po_{13}^I\Po_{24}^J + u
\Po_{23}^I\Po_{14}^J ) \phi_{12} \phi_3^I \phi_4^J
\nonumber\\[7pt]
& + & \frac{1}{\beta_2\beta_4} ( t \Po_{12}^I\Po_{34}^J - s
\Po_{23}^I \Po_{14}^J ) \phi_1^M \phi_2^I\phi_3^M \phi_4^J\,. \eeq
To fix normalization it suffices to analyze terms proportional to
$\phi_1^L \phi_2^R\phi_3^R\phi_4^R$. On the one hand
$W_{_{F^4}}(p)$ gives the following contribution to $\phi_1^L
\phi_2^R\phi_3^R\phi_4^R$- term:
\be\label{WLRRR10} W_{_{F^4}}(p)|_{_{\phi_1^L
\phi_2^R\phi_3^R\phi_4^R}}= \Tr\ \frac{\phi_1^L
\phi_2^R\phi_3^R\phi_4^R}{\beta_1\beta_2\beta_3\beta_4}\,
\frac{(\Po_{13}^L\Po_{24}^L)^2}{\beta_{13}^2} q_{_L}^2 (-
2\beta_1^2 )\,. \ee On the other hand our Hamiltonian \rf{Ham4fin}
gives the following contribution to $\phi_1^{LL}
\phi_2^{RR}\phi_3^{RR} \phi_4^{RR}$ - term:
\be P_{(4)}^-| =  g_0  \Tr \int d\Gamma_4(p)\, \frac{\phi_1^L
\phi_2^R\phi_3^R \phi_4^R}{\beta_1\ldots
\beta_4}\,\frac{(\Po_{13}^L\Po_{24}^L)^2}{\beta_{13}^2}q_{_L}^2( -
4 \beta_1^2)\,. \ee
Comparison of this formula with Eqs.\rf{f4ter2},\rf{WLRRR10} leads
to the relationship for the coefficients $g_0$ and $g_{_{F^4}}$
given in Eq.\rf{g0grel} (see \rf{gstexp}). Generalization of this
consideration to the case of the Lagrangian \rf{LfF4} to get the
relationship \rf{gstfstrel} is straightforward.

\end{document}